\documentclass[12pt]{article}
\usepackage{amsmath,amssymb,graphicx} 
\usepackage{epsf}
\usepackage{subfigure}
\usepackage{cite}

\def \be  {\begin{equation}}
\def \ee  {\end{equation}}
\def \baa {\begin{eqnarray}}
\def \eaa {\end{eqnarray}}

\def \Tr {\mbox{Tr\,}}

\newcommand \ket [1] {|{#1}\rangle}
\newcommand \bra [1] {\langle {#1}|}
\def \e {\mbox{e}}

\newcommand \da {\Delta{\eta}}
\newcommand \ad {\Gamma_S}
\newcommand \f {\mbox{\scriptsize{f}}}
\newcommand \tf {\mbox{\tiny{f}}}

\newcommand \bea{\begin{eqnarray}}
\newcommand \eea{\end{eqnarray}}

\newcommand \de {\Delta \eta}
\newcommand \nn{\nonumber}

\def\hepph  #1 {{\tt hep-ph/#1}}

\begin{document}
\begin{flushright}
YITP-SB-09-24\\
\end{flushright}
\baselineskip=18pt 

\vskip.5cm
\begin{center}
\bf{\Large Probing the Gauge Content of Heavy Resonances \\
\vskip 1.5mm with Soft Radiation}
\vskip.2cm
\end{center}

\begin{center}
{\bf  {Ilmo Sung}}\\
\vskip.8mm
 {\it C.N. Yang Institute for Theoretical Physics,
Stony Brook University\\ Stony Brook, New York 11794-3840, USA}\\
\end{center}

\begin{abstract}
The use of energy flow is investigated 
as a diagnostic tool for 
determining the color SU(3) representation of new resonances.
It is found that the pattern of soft gluon radiation into a rapidity gap 
depends on color flow in the hard scattering, 
and reflects the gauge content of new physics.
The massive soft anomalous dimension matrix for rapidity gap events is 
introduced for describing soft gluon emission analytically 
in heavy quark pair production.
A gap fraction is used for 
quantifying the amount of soft radiation into the gap region.
In general, the results illustrate that radiation is greater for 
a singlet resonance than for an octet.
Especially, it is found that the quantitative difference is quite distinguishable 
for spin-1 resonances, depending on the gauge content in the new sector.
\end{abstract}

\section{Introduction}
\label{sec:intro}

QCD has been successfully tested in many hadronic processes at high energy. 
Current efforts in perturbative QCD have moved toward precision, 
to help gain accurate predictions for Standard Model processes 
in collider experiments as backgrounds to new physics. 
Precise prediction is a key to find new physics, 
because QCD events should be distinguished from signals 
in order to claim the discovery of new phenomena at colliders. 
In this paper, we show how to use factorized cross sections 
as a tool for analyzing the gauge content of new heavy particles.

Factorization is a statement of the quantum mechanical incoherence 
of short and long distance physics, 
and plays an important role in the use of perturbative QCD.
Perturbative calculations of factorized partonic cross sections 
depend on the separation of regions of momentum space. 
The three basic regions describe off-shell ``hard" partons at short distances, 
energetic, on-shell ``collinear" partons near the light cone, and long-wavelength, 
``soft" partons~(see Ref.~\cite{Collins:1989gx} for reviews and references therein).
Once we obtain a factorized form of the cross section for a certain process, 
resummation can be achieved 
from the independence of physical quantities from factorization scales, 
which leads to evolution equations~\cite{Contopanagos:1997nh}. 
The evolution equations can be solved to resum large logarithmic corrections 
in terms of perturbative anomalous dimensions. 
Especially, the soft anomalous dimension matrix 
introduced in Refs.~\cite{Sen:1981sd,Botts:1989kf}, 
which is the main interest in this paper, 
allows us to describe soft gluon radiation systematically in hadronic collisions.

In the mid 1990's, dijet rapidity gap events 
with anomalously low radiation in a wide interjet rapidity region 
for electron-proton collision were observed at HERA~\cite{hera1,hera2,hera3} 
and for proton-antiproton scattering at the Tevatron~\cite{tevatron}.
Such events were predicted 
from the exchange of two or more gluons in a color singlet configuration, 
which avoids color recombination between jets~\cite{Bjorken:1992er}. 
A quantitative explanation of these events 
in terms of factorizable cross sections was studied in Ref.~\cite{Oderda:1998en},
which discussed the dependence of the dijet cross section 
on energy flow, $Q_c$, into the central region, 
in terms of a soft function 
that is a matrix in the space of the possible color exchange at the hard scattering.
A ``gap" here refers to low energy flow $Q_c$.
Since then, there have been various studies of gap events from this point of view. 
The soft anomalous dimension matrix, 
denoted by $\Gamma_S$ below, for rapidity gap events, 
was calculated in Ref.~\cite{Oderda:1998en}, 
and used to resum the leading logarithmic contributions 
in Refs.~\cite{Berger:2001ns,Appleby:2003sj}. 
The soft anomalous dimension matrix for three jet rapidity gap processes 
was calculated in Refs.~\cite{Kyrieleis:2006dt,Sjodahl:2008fz,Sjodahl:2009wx}.

Cross sections computed from the soft anomalous dimension matrix 
organize ``global" logarithms, 
following the terminology introduced in Ref.~\cite{Dasgupta:2001sh}, 
and as such do not organize all logarithms of the gap energy.
Non-global logarithmic contributions, 
mis-cancellations from secondary emission, 
discovered by Dasgupta and Salam~\cite{Dasgupta:2001sh}, 
were resummed in the large $N_c$ approximation~\cite{Dasgupta:2002bw}.
More recently, Forshaw, Kyrieleis and Seymour 
have discovered new logarithmic contributions 
which they name ``super-leading logarithms", 
which do not seem possible to exponentiate by 
current techniques~\cite{Forshaw:2006fk,Forshaw:2008cq,Keates:2009dn}.  
The contributions of these non-global logarithms, however, 
are generally not dominant~\cite{Appleby:2002ke,Forshaw:2009fz}.

In this paper, we study the pattern of soft gluon radiation 
into a rapidity gap for the production of heavy particles 
in hadron-hadron scattering at leading order in the soft matrix, $\Gamma_S$, 
which resums global logarithms. 
The pattern depends on color flow in the hard scattering, 
and reflects the gauge content of new particles
that participate in the process, for example, as $s$-channel resonances.

The results of this paper illustrate the use of energy flow, treated by perturbative QCD 
and factorization, as a diagnostic tool for 
determining the color SU(3) representation of new resonances, 
and the use of resummation for more precise cross sections. 
Previous work on soft gluon radiation for the study of new physics 
includes Ref.~\cite{Forshaw:2007vb} 
for Higgs boson production in association with two hard jets, 
and Refs.~\cite{Kulesza:2008jb,Kulesza:2009kq} 
for gluino and squark pair production, 
which describe resummed cross sections 
to provide more accurate predictions for signals.

There are many proposals for extensions of the Standard Model. 
These could provide solutions to the hierarchy problem, 
dark matter explanation, the origin of mass, etc. 
The most common feature of the models is 
that they include new heavy particles 
transforming under definite gauge groups.
In many scenarios, we expect that 
they are created with $\mathcal{O}$(TeV) mass 
via $s$-channel processes in hadron-hadron scattering. 
Possible processes with various spins and gauge groups of new heavy particles 
are studied and summarized in Ref.~\cite{Frederix:2007gi}, 
which considers new sectors that decay into a top quark pair.
There have been studies for identifying signals, 
including hadronic or leptonic decays 
from gauge Kaluza-Klein~(KK) resonance states 
and from QCD background~\cite{Agashe:2006hk,Fitzpatrick:2007qr,Lillie:2007yh,Agashe:2007zd,Agashe:2007ki,Baur:2008uv,Agashe:2008jb,Davoudiasl:2009jk}. 
These studies show 
that the ability to distinguish signals especially from hadronic decays, 
which have much larger branching ratio than leptonic decays, 
depends on both how strongly the new sector is coupled to the Standard Model 
and to what extent signals and backgrounds are separated~\cite{Butterworth:2008iy, Thaler:2008ju,Kaplan:2008ie,Almeida:2008yp,Almeida:2008tp,Godfrey:2008vf,FileviezPerez:2008ib,Ellis:2009su}.
Also, various methods for spin measurements of the intermediate new particles 
have been suggested 
and discussed~(see Ref.~\cite{Wang:2008sw} for reviews and references therein).

Our study begins from this point. 
Once we succeed in distinguishing a resonance from backgrounds, 
a remaining task is to determine 
the SU(3) color gauge content of the resonance particle, denoted as $V'$.
For simplicity we consider below two possibilities; 
color-singlet~($Z'$) and color-octet~($G$).
Since we consider QCD radiation 
with total energy much less than resonance energy, 
our considerations do not depend on the details of the model.
In this paper, we assume that the decay width is large enough that a hard function is 
an effective vertex. 
We will, however, discuss the consequences of a narrow resonance at the end of 
Secs.~\ref{sec:sadm} and \ref{sec:spins}.

In many models there are gauge interactions 
whose coupling with a top quark pair is enhanced in particular.
These include models with a KK excited graviton, 
weak and strong gauge bosons
in extra dimensions~\cite{ArkaniHamed:1998rs,Randall:1999ee},
as well as heavy spin-0 particles in MSSM and 
two-higgs-doublet model~(2HDM)~\cite{Dicus:1994bm,Bernreuther:1997gs} and 
spin-1 coloron and axigluon~\cite{Choudhury:2007ux}.
Such new particles~$V'$ of mass $M$ and spin~$(sp)$ 
could show up as resonances 
in $pp(\bar{p}) \to V'(M,sp) \to Q \bar{Q}$ at the Tevatron and the LHC.
We may expect larger branching ratios for heavy quark pair production 
than for other channels such as massless dijets or dileptons, 
due to the small couplings of $V'$ to light particles.
This motivates the calculation of an analytic form 
describing soft gluon emission in heavy quark pair production. 
At the level of global logarithms, 
this is determined by the soft anomalous dimension matrices 
for rapidity gap processes involving massive particles.
The ``massive" soft anomalous dimension matrix for heavy pair production 
was studied in Refs.~\cite{Kidonakis:1996aq,Kidonakis:1997gm,Kidonakis:1998bk,Kidonakis:2009ev,Mitov:2009sv,Becher:2009kw,Beneke:2009rj,Czakon:2009zw,Ferroglia:2009ep} 
for threshold resummation.
In this paper, we extend this work to introduce 
the massive soft anomalous dimension matrix for rapidity gap cross sections, 
which will be used as a tool for our analysis.

Reference~\cite{Oderda:1998en} studied 
rapidity gaps for a process in which $t$-channel exchange was dominant. 
The result indicates that, in the limit of a very large interjet region, 
the color singlet component dominates, 
with more radiation into the gap when the exchange is a color-octet.
In this study, however, we might expect a different pattern of soft gluon radiation 
due to new heavy particle resonances in the $s$-channel 
rather than $t$-channel. The quantity of interest is the gap fraction, 
the ratio of the number of events 
for heavy quark pair production with a gap energy up to some value $Q_0$ 
to the total number of events for the process.
We will find a quantitative difference that is quite distinguishable 
and can be used for the study of the gauge content in the new sector.
This difference, especially caused by the SU(3) color content, 
 is related to ``drag" effects in $e^+ e^-$ three-jet~($q \bar{q} g$) events, compared to
$q \bar{q} \gamma$ events~\cite{Dokshitzer:1988bq,Dokshitzer:1991wu,Ellis:1991qj}.
The color dipole configurations 
explain a surplus of radiation in the $qg$ and $\bar{q}g$ interjet regions
for three-jet events,
while radiation is enhanced in the $q\bar{q}$ region for $q \bar{q} \gamma$ events.

In this study, we do not give a detailed study of non-global~\cite{Dasgupta:2001sh} 
or super-leading logarithms~\cite{Forshaw:2006fk}, 
arising when gluons radiated outside the gap emit back into the gap.
The latter are sometimes called secondary emissions.
We can in principle introduce correlations 
between energy flow and event shapes 
that suppress such secondary emissions naturally. 
This technique was introduced in Ref.~\cite{Berger:2003iw} 
for rapidity gap $e^+ e^-$ dijet events. 
Such correlations control secondary radiation, 
by suppressing states with radiation outside the gap at intermediate energy scales, 
leading to perturbatively computable measures insensitive to non-global logarithms.
In our study, the rapidity gap we choose does not allow 
radiation of high transverse momentum outside the gap by definition, 
which suppresses the non-global logarithms from final-state radiation.
On the other hand, another issue arises for hadronic collisions, 
where there is secondary emission nearly parallel to the initial partons.
Several classes of observables studied 
in Refs.~\cite{Banfi:2004nk,Sterman:2005bf,Berger:2005ct} 
could be used to suppress exponentially forward radiation, 
which would then ensure the suppression of non-global, 
including super-leading, logarithms. 

In the following section, 
we study the definition and kinematics of rapidity gap processes
and review the relevant factorization procedure. 
Here, we introduce the energy flow defined in Ref.~\cite{Oderda:1998en}, 
and define the gap fraction. 
After studying the gap fraction in terms of energy flow, 
we study its relation to color flow at short distances in Secs.~\ref{sec:sadm} and \ref{sec:hard}.
We argue that the gap fraction exposes the pattern of soft gluon radiation, 
depending on the color SU(3) gauge content of the resonance in 
$q \bar{q} (gg) \to Z'/G \to Q \bar{Q}$. 
In Sec.~\ref{sec:spin1}, the gap fraction for heavy quark pair production 
induced by a heavy spin-1 resonance is studied. 
Finally, in Sec.~\ref{sec:spins}, we apply our formalism for 
a study of spin-0 or 2 resonance particles 
associated with different partonic processes.

\section{Heavy Quark Pair Rapidity Gap Cross Sections}
\label{sec:def}
\subsection{Definitions}
In this paper, we will focus on rapidity gap cross sections 
for heavy quark pair production
through a new physics particle resonance 
denoted by $V'(M, sp)$ of mass $M$ and spin~$sp$.
The partonic scattering process for proton and (anti)-proton collisions,
creating $s$-channel resonance $V'$, is
\bea
\mbox{f}:\, \, f_1(p_1)\, + f_2(p_2)\, \to \, V'(M, sp) \, \to \, Q(p_a) \, 
+ \bar{Q}(p_b) \, + \Omega_{gap}(Q_c) + X \, ,
\label{eq:process}
\eea
for the production of a heavy quark pair at fixed rapidity difference, 
$\de=\eta_a - \eta_b$.
Here the $f_i$ refer to partons that participate in the collision, 
and we have labeled process~(\ref{eq:process}) ``f".
We sum inclusively over final states,
while measuring the energy flow, $Q_c$, 
into the rapidity region, $\Omega_{gap}$, in Eq.~(\ref{eq:process}).
The transverse momentum $p_T$ of a parton $Q(p_a)$ 
or $\bar{Q}(p_b)$ of mass $m_Q$ in Eq.~(\ref{eq:process}) 
is related to the rapidity difference $\de$ according to
\bea
{p}_T = \sqrt{ \frac{M^2}{4 \cosh^2{\frac{\de}{2}}}- m_Q^2 } \, .
\label{eq:pt}
\eea
As discussed in the introduction, 
we may expect different patterns of radiation 
from intermediate resonances 
if they are distinguished by different gauge content.

\begin{figure}[fhptb]
\begin{center}
\includegraphics[width=.5\hsize]{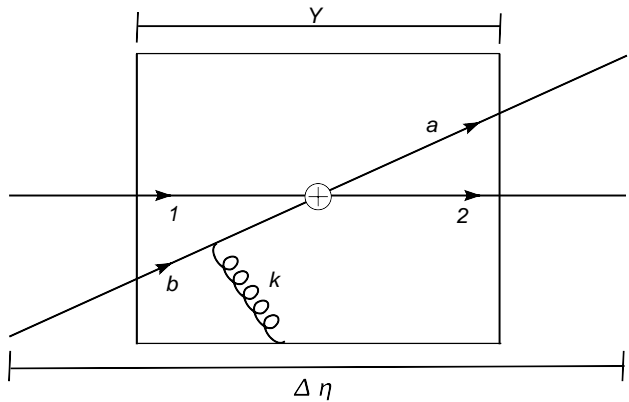}
\end{center}
\caption{Light to heavy process with gluon radiation of momentum~$k^{\mu}$ into a rapidity gap. 
The figure describes $q\bar{q} \to Q \bar{Q}$ process. 
$\de$ is the rapidity difference of a heavy quark pair, 
$Q(p_a)$ and $\bar{Q}(p_b)$, 
and $Y$ is the rapidity range of the gap region. 
1 and 2 denote the initial partons while $a$ and $b$ denote the final partons. }
\label{fig:gap}
\end{figure}

The geometry for the gap at a collider is presented schematically 
in Fig.~\ref{fig:gap}, 
where the gap region is determined by rapidity range $Y$.
The factorization that we will discuss in Sec.~\ref{sec:factorization} 
enables us to work perturbatively.
The partonic cross section at leading order for process~(\ref{eq:process})
involves five external particles.
The kinematics of this analysis are described in the Appendix.

\subsection{Factorized and refactorized cross sections}
\label{sec:factorization}
The inclusive cross section 
for heavy quark pair production 
at rapidity difference $\de$ from a resonance at mass $M$ 
with fixed energy flow $Q_c$
is given in standard factorized form by
\bea
\frac{ d \sigma_{AB}}{d \de \, d Q_c} 
=\sum_{f_1,f_2} \int d x_1 d x_2 \, 
\phi_{f_1/A} (x_1, \mu_F) \phi_{f_2/B}  (x_2, \mu_F)
\frac{ d \hat{\sigma}^{(\f)}}{d \de  \, d Q_c} \, . 
\label{eq:pdf}
\eea
Here $\phi_{f_1/A}$ and $\phi_{f_2/B}$ are parton distribution functions~(PDFs), 
evaluated at the factorization scale $\mu_F$. 
Below, we use the scale $\mu_F=p_T$, 
the transverse momentum $p_T$ in Eq.~(\ref{eq:pt}).
The partonic scattering cross section 
$ d \hat{\sigma}^{(\f)} / d \de d Q_c$ 
can be expanded in $\alpha_s$, 
starting from the lowest order~(LO) Born cross section
\bea
\frac{ d \hat{\sigma}^{(\f)}}{d \de  \, d Q_c}
=\frac{ d \hat{\sigma}^{(\f,LO)}}{d \de } \delta(Q_c) + \cdots \, ,
\label{eq:pcross}
\eea
where corrections may include potentially large logarithms of $p_T/Q_c$.
We recall that the index f denotes the partonic process in Eq.~(\ref{eq:process}).
Following Ref.~\cite{Berger:2001ns}, 
we consider the partonic cross sections 
integrated over the transverse energy, $Q_c$, 
radiated into a symmetric, central rapidity gap region, 
$Y$ in Fig.~\ref{fig:gap}, up to a fixed value $Q_0$, 
\bea
\frac{ d \hat{\sigma}^{(\f)} (M, m_Q, Q_0, \mu_F, \de, Y, \alpha_s(\mu_F)) }
{d \de } 
=\int^{Q_0}_0 d Q_c 
\frac{ d \hat{\sigma}^{(\f)}(M, m_Q, Q_c, \mu_F, \de, Y, \alpha_s(\mu_F))}
{d \de  \, d Q_c} 
 \, .
\label{eq:p2}
\eea
When Eq.~(\ref{eq:p2}) is divided by the total cross section, 
the ratio can be interpreted as 
the probability for heavy quark pair production 
with soft gluon radiation into the gap region up to energy flow $Q_0$. 
For simplicity of notation, we keep dependence on the resonance width $\Gamma$ 
implicit but recall that we treat $\Gamma/M$ as a number of order unity. 

The partonic cross sections of Eq.~(\ref{eq:p2}) 
can be further refactorized into a hard scattering matrix $H_{IL}^{(\f)}$ 
and a soft matrix $S^{(\f)}_{LI}$,
\bea
\frac{ d \hat{\sigma}^{(\f)}}{d \de } 
(M, m_Q, Q_0, \mu_F, \de, Y, \alpha_s(\mu_F))
&=& \sum_{L,I} H^{(\f)}_{IL} (M, m_Q, \mu_F, \mu,  \de,  \alpha_s(\mu))
\nn \\
 &&\hspace{7mm} \times S^{(\f)}_{LI} \left(\de, Y, \frac{Q_0}{\mu}, \alpha_s(\mu), m_Q \right) \, ,
 \label{eq:hs}
 \eea
where matrix $H^{(\f)}_{IL}$ represents interactions at short distances, 
independent of soft gluon radiation, 
and contains the dynamics of resonance $V'(M,sp)$. 
Here we have introduced a new refactorization scale $\mu$.
The soft function $S^{(\f)}_{LI}$ describes 
the radiation of soft gluons up to the scale $Q_0$, 
which decouples from the dynamics of the hard scattering. 
The soft function is written in terms of 
products of path-ordered exponentials, $w^{(\f)}_L (x)_{\{b_i\}}$, 
of the gluon field as~\cite{Berger:2001ns},
\bea
S^{(\f)}_{LI} 
= \int^{Q_0}_0 d Q'_c \sum_n \sum_{\{b_i\}} \delta(Q^{(n)}_0 - Q'_c) 
\bra{0} \bar{T} [(w^{(\f)}_L(0))^{\dag}_{\{b_i\}} ] \ket{n}
\bra{n} T [(w^{(\f)}_I(0))_{\{b_i\}} ] \ket{0} \, ,
\label{eq:soft}
\eea
where we sum over all states $n$ 
whose transverse energy flow into the gap is restricted to 
equal $Q'_c$ and where color tensors labeled by $L$ and $I$ 
account for the color flow at the hard scattering. 
Eikonal multipoint operators, $(w^{(\f)}_L(x))_{\{b_i\}}$, 
with color indices $\{b_i\}$ are written as 
products of ordered exponentials, 
tied together by color tensors, $c_L$,
\bea
(w^{(\f)}_L(x))_{\{b_i\}} &=&
\sum_{\{d_i\}} \Phi^{(f_b)}_{\beta_b} (\infty, 0;x)_{b_b,d_b} 
\Phi^{(f_a)}_{\beta_a} (\infty, 0;x)_{b_a,d_a}
\nonumber \\
&&\hspace{7mm} \times (c_L^{(\f)})_{d_b, d_a ; d_1, d_2}  
\Phi^{(f_1)}_{\beta_1} (0, -\infty;x)_{d_1,b_1} 
\Phi^{(f_2)}_{\beta_2} (0, -\infty;x)_{d_2,b_2} \, ,
\eea
where the non-Abelian path-ordered phase operators (Wilson lines), 
$\Phi^{(f)}_{\beta}$, are given by
\bea
\Phi^{(f)}_{\beta}(\infty,0;x) = 
P \exp \left[ -i g \int^{\infty}_0 d \lambda \beta \cdot A^{(f)} (\lambda \beta +x ) \right]\, .
\eea
Here, $\beta$ is a four-velocity, 
and the vector potentials $A^{(f)}$ are 
in the color representation appropriate to flavor $f$.
At the tree level, 
Eq.~(\ref{eq:soft}) reduces to the trace of the product of color tensors.

The product of hard and soft functions in Eq.~(\ref{eq:hs}) is 
independent of the refactorization scale, $\mu$. 
From this, we easily derive for $S^{(\f)}_{LI}$ 
the evolution equation~\cite{Contopanagos:1997nh}
\bea
\left(\mu \frac{\partial}{\partial \mu} + \beta(g_s) \frac{ \partial}{\partial g_s} \right) 
S^{(\f)}_{LI}
= - (\Gamma_S^{(\f)})^{\dag}_{LJ} S_{JI}^{(\f)}
-S_{LJ}^{(\f)} (\Gamma_S^{(\f)})_{JI} \, ,
\label{eq:evolution}
\eea
in terms of a matrix of anomalous dimensions, 
$(\Gamma^{(\f)}_S)_{JI}$. 
The matrix $\Gamma_S^{(\f)}(\de, Y, \rho)$ depends on kinematics, 
including the geometric information of the gap region and the heavy quark mass. 
Solving this equation will enable us to resum 
the leading logarithms of the soft scale $Q_0$.
In heavy quark pair production, dependence on 
quark mass and the partonic center of mass energy $M$ 
in soft functions will always appear
through the parameter $\rho$, given by
\bea
\rho \equiv \sqrt{ 1+\left( \frac{m_Q}{p_T}\right)^2 }= 
\frac{1}{\sqrt{1-4 \frac{m_Q^2}{M^2}
\cosh^2 \frac{\de}{2}}} \, .
\label{eq:rho}
\eea 
We will find that the soft function depends on the velocities $\beta_i$ 
only through $\de$ and $\rho$.

To obtain the solution to Eq.~(\ref{eq:evolution}), 
we treat above equation in a basis that diagonalizes $\Gamma_S^{(\f)}$,
\bea
(\Gamma_S^{(\f)}( \de, Y, \rho) )_{\gamma \beta} 
&\equiv& \lambda_{\beta}^{(\f)} (\de, Y, \rho) \delta_{\gamma \beta}
\nonumber \\
&=& (R^{(\f)})_{\gamma I} \, 
(\Gamma_S^{(\f)}(\de, Y, \rho ))_{IJ} \, 
({R^{(\f)}}^{-1})_{J \beta} \, ,
\label{eq:gdiagonal}
\eea
where the eigenvalues $\lambda_{\beta}^{(\f)}$ are given by a series in $\alpha_s$,
\bea
\lambda_{\beta}^{(\f)}(\de, Y, \rho)
 = \frac{ \alpha_s}{\pi} \lambda_{\beta}^{(\f,1)}(\de, Y, \rho) + \cdots \, .
 \eea
We can transform the soft and hard matrices in the diagonalized basis
\bea
S^{(\f)}_{\gamma \beta} &=&
[(({R^{(\f)}})^{-1})^{\dag} ]_{\gamma L} \, 
S^{(\f)}_{LK} \,  
[({R^{(\f)}})^{-1} ]_{K \beta} \, ,
\nonumber \\
H^{(\f)}_{\gamma \beta} &=&
[{R^{(\f)}} ]_{\gamma K} \, 
H^{(\f)}_{KL} \, 
[{R^{(\f)}}^{\dag} ]_{L \beta} \, .
\label{eq:shdiagonal}
\eea
This change of basis uses the transformation matrix 
$[({R^{(\f)}})^{-1}]_{K\beta}=(e_{\beta})_K$, 
where $e_{\beta}$ are eigenvectors of the soft anomalous dimension matrix.

The solution to Eq.~(\ref{eq:evolution}) gives  
the dependence of $S^{(\f)}$ on the ratio $\mu/Q_0$,
\bea
S^{(\f)}_{\gamma \beta} \left( \de, Y, \frac{Q_0}{\mu}, \alpha_s(\mu), \rho \right)
&=& S^{(\f)}_{\gamma \beta} ( \de, Y, 1, \alpha_s(Q_0), \rho)
\exp\left[ - E^{(\f)}_{\gamma \beta} \int^{\mu}_{Q_0} \frac{d \mu'}{\mu'} 
\left( \frac{ \beta_0}{2 \pi} \alpha_s (\mu') \right) \right] 
\nonumber \\
&=& S^{(\f)}_{\gamma \beta} ( \de, Y, 1, \alpha_s(Q_0), \rho)
\left[  \ln \left( \frac{Q_0}{\Lambda} \right) \right]^{E^{(\tf)}_{\gamma \beta}}
\left[ \ln \left( \frac{\mu}{\Lambda} \right) \right]^{-E^{(\tf)}_{\gamma \beta}}  .
\label{eq:sd2}
\eea
We will choose below $\mu$ and $\mu_F$ to be the hard scale $p_T$ 
for use in the refactorized partonic cross section~(\ref{eq:hs}).
In Eq.~(\ref{eq:sd2}) the exponents $E^{(\f)}_{\alpha \beta}$ are given 
in terms of sums of the eigenvalues of $\Gamma_S$
by 
\bea
E^{(\f)}_{\gamma \beta} (\de, Y,\rho) =
\frac{2}{\beta_0} [ \lambda_{\gamma}^{(\f,1)*} (\de, Y,\rho) 
+  \lambda_{\beta}^{(\f,1)} (\de, Y, \rho) ] \, ,
\label{eq:exponents}
\eea
where $\beta_0$ is the lowest-order coefficient 
in the expansion of the QCD beta function, 
$\beta_0 = (11N_c-2n_f)/3$, for $N_c$ colors and $n_f$ quark flavors. 
Equation~(\ref{eq:exponents}) shows that 
the diagonal exponents, $E^{(\f)}_{\alpha \alpha}$, are real.
In the second form of Eq.~(\ref{eq:sd2}),
we have used the one-loop QCD running coupling, 
\bea
\alpha_s(\mu) = 
\frac{2\pi}{\beta_0 \ln\left( \mu / \Lambda \right)} \, .
\eea
Combining Eqs.~(\ref{eq:hs}) and (\ref{eq:sd2}), 
the partonic cross section, valid to leading logarithm, is then
\bea
\frac{ d \hat{\sigma}^{(\f)}}{d \de } 
&=& \sum_{\beta, \gamma} H^{(\f,LO)}_{\beta \gamma}
(M, m_Q, \de, \alpha_s (p_T)) 
\, S^{(\f,0)}_{\gamma \beta}
\left[  \ln \left( \frac{Q_0}{\Lambda} \right) \right]^{E^{(\tf)}_{\gamma \beta}}
\left[ \ln \left( \frac{p_T}{\Lambda} \right) \right]^{-E^{(\tf)}_{\gamma \beta}}  .
\label{eq:qc3}
\eea
In the following section,
we will use Eq.~(\ref{eq:qc3}) to evaluate the gap fraction.

\subsection{The gap fraction}
\label{sec:gapfraction}
We define the gap fraction $f_{gap}$ as the ratio of the number of events 
for heavy quark pair production with a specified rapidity gap 
to the total number of the pair production events. 
A gap event was originally identified experimentally by 
the lack of particle multiplicity in the interjet region~\cite{hera1,tevatron}. 
In that case, the multiplicity is determined from 
the number of calorimeter cells 
which measure energy deposition above a threshold. 
A gap event is then defined by the absence of such cells in the rapidity region. 
Our formulation of the problem is in terms of the transverse energy flow, $Q_c$, 
of hadronic radiation carried by the particles~\cite{Oderda:1998en}. 
We introduce a variable, energy  ``gap threshold" $Q_0$, 
which is different in principle from the experimental calorimeter threshold, 
and identify a gap event from the condition of interjet radiation less than $Q_0$. 
Then, to get the partonic gap cross section, 
we evaluate Eq.~(\ref{eq:p2}) at the value $Q_0$ 
for the maximum of interjet energy flow $Q_c$. 
The resummed result is given by Eq.~(\ref{eq:qc3}).
We note that values of $Q_0$ at the order of the hard scale $p_T$
would violate the requirement of factorization, 
since the emission into the gap region would not be soft any more. 

To define the gap fraction, 
we approximate the total cross sections from 
the LO~(Born) partonic cross section, related to Eq.~(\ref{eq:hs})  
\bea
\frac{ d \hat{\sigma}^{(\f,LO)}}{d \de } 
&=&
 \sum_{\beta, \gamma} H^{(\f,LO)}_{\beta \gamma}(M, m_Q, \de, \alpha_s (p_T)) 
\, S^{(\f,0)}_{\gamma \beta}\, .
\label{eq:LO}
\eea
We may consider the LO and the gap differential partonic cross sections 
with respect to $M$ and $\eta_{V'} = \frac{1}{2} \ln \frac{x_1}{x_2}$,
given for a resonance by, respectively, 
\bea
\frac{ d \hat{\sigma}^{(\f,LO)} 
\left( M,\, m_Q,  \,  \de, \, \alpha_s(p_T), \, Y \right)
}{d \de d M^2 d \eta_{V'} }
&=&
\frac{ d \hat{\sigma}^{(\f,LO)}}{d \de } \delta \left(M^2- x_1 x_2 S\right) \delta\left( \eta_{V'} - \frac{1}{2} \ln \frac{x_1}{x_2}\right) \, ,\nn \\
\frac{ d \hat{\sigma}^{(\f)}
\left( M,\, m_Q, \,  Q_0, \,  \de, \, \alpha_s(p_T), \, Y \right)
}{d \de d M^2 d \eta_{V'} }
&=&
\frac{ d \hat{\sigma}^{(\f)}}{d \de } \delta \left(M^2- x_1 x_2 S\right) \delta\left( \eta_{V'} - \frac{1}{2} \ln \frac{x_1}{x_2}\right) \, ,
\label{eq:df}
\eea
which are distinguished by the argument $Q_0$ in the latter, and
where $S$ is the hadronic center of mass energy.
Here, the differential with respect to $\eta_{V'}$ 
corresponds to fixing the center of mass rapidity of the final-state pair in the lab frame.
We emphasize that the above condition does not change 
the description of soft QCD radiation in the soft functions, 
which depend on the rapidity difference $\de$.

We now define the gap fraction at leading order 
in terms of Eqs.~(\ref{eq:qc3}), (\ref{eq:LO}) and (\ref{eq:df}), by
\bea
f_{gap}^{(LO)} = \frac{\sum_{f_1, f_2}  \int d x_1 dx_2 
\phi_{f_1/A}(x_1)  \phi_{f_2/B}(x_2) 
\frac{ d \hat{\sigma}^{(\tf)} 
\left( M,\, m_Q, \,  Q_0, \,  \de, \, \alpha_s(p_T), \, Y \right)}
{d \de \, d M^2 \, d \eta_{V'}  } }
{ \sum_{f_1, f_2}  
\int d x_1 dx_2 \phi_{f_1/A}(x_1)  \phi_{f_2/B}(x_2) 
\frac{ d \hat{\sigma}^{(\tf,LO)}
\left( M,\,  m_Q,  \,  \de, \, \alpha_s(p_T), \, Y \right) }
{d \de \, d M^2 \, d \eta_{V'} }  } \, .
\label{eq:fraction}
\eea 
An advantage of using the differential partonic cross sections~(\ref{eq:df}) 
to obtain the gap fraction becomes clear 
when there is only one partonic initial state for a resonance process. 
The gap fraction at leading order in Eq.~(\ref{eq:fraction}) 
for a single partonic channel process is given by
\bea
f_{gap}^{(LO)} 
\left( M,\, m_Q, \, Q_0, \, \de, \, \alpha_s(p_T), \, Y,  \, \eta_{V'} \right)
 =
 \frac{  \frac{ d \hat{\sigma}^{(\tf)} 
 \left( M,\,  m_Q, \, Q_0, \,  \de, \, \alpha_s(p_T), \, Y \right)}
 {d \de }  }
{  \frac{ d \hat{\sigma}^{(\tf,LO)}
\left( M,\,  m_Q,  \,  \de, \, \alpha_s(p_T), \, Y \right)}
{d \de} 
  } \, ,
\label{eq:fraction2}
\eea 
where the PDFs cancel in the ratios of the gap fraction in this case.

\section{Massive Gap Soft Anomalous Dimension Matrices}
\label{sec:sadm}
The one-loop soft anomalous dimension matrix, $\Gamma_S$, 
for a rapidity gap process is calculated from 
the mismatch between real and virtual corrections, 
generated by imposing a rapidity gap. 
The soft function is calculated by taking 
single ultraviolet pole parts from virtual and real corrections, 
which we will label as  
${\omega_V}_{(ij)}$ and ${\omega_R}_{(ij)}$, respectively. 
The real part, ${\omega_R}_{(ij)}$, comes with a phase space integral 
outside the gap,
where it cancels ${\omega_V}_{(ij)}$. 
The coefficient of the pole results from the phase space integration 
inside the rapidity gap region. 
Here, the single ultraviolet pole arises from 
the limitation on the energy integration of soft gluons. 

In practice, following \cite{Berger:2001ns},
we construct the gap soft anomalous dimension matrix 
from a set of integrals, $\omega_{(ij)}$,
\bea
{\gamma^{(1)}_S}_{(ij)} 
= 
\frac{\omega_{(ij)}(-2\varepsilon)}
{(\alpha_s / \pi)} \, ,
\label{eq:ad}
\eea 
where we use dimensional regularization, with $D= 4 - 2\varepsilon$. 
The indices $i$ and $j$ label partons of momenta $p_i$ and $p_j$, 
and the $\omega_{(ij)}$ are defined by
\bea
\omega_{(ij)} 
&=& 
{\omega_V}_{(ij)} + {\omega_R}_{(ij)} 
\nn \\
&=& 
-(4 \pi \alpha_s) \delta_i \delta_j\Delta_i \Delta_j 
\int_{P.P.} \frac{ d^{D} k }{(2 \pi)^{D-1}} 
\delta_+(k^2)\Theta(\vec{k}) 
\frac{ (p_i \cdot p_j) }{(p_i \cdot k)(p_j \cdot k )}  
\nn \\
&& 
+ \delta_i \delta_j\Delta_i \Delta_j  
\frac{\alpha_s}{2\pi} 
\frac{i \pi}{2\varepsilon}(1-\delta_i \delta_j) \, .
\label{eq:ss}
\eea
Here the subscript $P.P.$ indicates that 
the integral is defined by its ultraviolet pole part. 
We also define $\delta_i =1 (-1)$ for momentum $k$ 
flowing in the same (opposite) direction as the momentum flow of line $i$, 
and $\Delta_i=1 (-1)$ for $i$ a quark (antiquark) line. 
The function $\Theta(\vec{k})=1$ 
when the vector $\vec{k}$ is directed into a rapidity gap.
Equation (\ref{eq:ss}) describes the incomplete cancellation 
between real and virtual gluons from the integration over 
the geometric phase space of the gap region. 
Only the part of the phase integration inside the gap survives, 
and contributes to the soft anomalous dimension matrix for the process.

We express the above integrals in terms of 
transverse momentum, rapidity, and azimuthal angle coordinates, 
with the help of the kinematics described in the Appendix, 
and use the relations
\bea
\int_{P.P.} d^{D } k \delta_+ (k^2)
&=& 
\int_{P.P.}  \frac{k_T^{D-3}}{2} d k_T d y d \phi \, , 
\nn \\
\int_{P.P} \frac{d k_T}{k_T^{1+2\varepsilon}} 
&=& 
\frac{1}{2\varepsilon} \, .
 \eea
In these terms, the $\omega_{(ij)}$ become
\bea
\omega_{(ij)} 
&=& 
\left(\frac{\alpha_s}{\pi}\right) 
\Bigg( 
- \delta_i \delta_j\Delta_i \Delta_j \frac{1}{2} \frac{1}{2\varepsilon} 
\int \frac{d y d \phi}{2\pi}\Theta(\vec{k}) \Omega_{ij} 
+ \delta_i \delta_j\Delta_i \Delta_j  \frac{i \pi}{2\varepsilon}
\frac{(1-\delta_i \delta_j)}{2} 
\Bigg)\, ,
\label{eq:w}
\eea
where we define $\Omega_{ij}$ as
\bea
\Omega_{ij} 
= 
\frac{ (p_i \cdot p_j) k_T^2}
{(p_i \cdot k)(p_j \cdot k )} \, .
\eea
In these coordinates, the $\Omega_{ij}$ are given by
\bea
\Omega_{12} 
&=& 
2 \, , 
\nn \\
\Omega_{ab} 
&=& 
\frac{ \rho^2 \cosh \da + 1}
{\left( \rho \cosh\left( \frac{\da}{2}-y \right) - \cosh \phi \right)
\left( \rho \cosh\left( \frac{\da}{2}+y \right)+ \cos \phi \right) } \, , 
\nn \\
\Omega_{1a} 
&=& 
\frac{ \rho \, \e^{-\left( \frac{\da}{2} - y \right) }}
{\left( \rho  \cosh\left( \frac{\da}{2}- y \right)-  \cos \phi \right) } \, , 
\nn \\
\Omega_{1b} 
&=& 
\frac{ \rho \, \e^{\left( \frac{\da}{2} + y \right) }}
{\left( \rho  \cosh\left( \frac{\da}{2}+ y \right)+  \cos \phi \right) } \, , 
\nn \\
\Omega_{2a} 
&=& 
\frac{ \rho \, \e^{\left( \frac{\da}{2} - y \right) }}
{\left( \rho  \cosh\left( \frac{\da}{2}- y \right)-  \cos \phi \right) } \, , \nn \\
\Omega_{2b} 
&=& 
\frac{ \rho \, \e^{-\left( \frac{\da}{2} + y \right) }}
{\left( \rho  \cosh\left( \frac{\da}{2} + y \right) +  \cos \phi \right) } \, .
\eea
We recall that indices 1 and 2 are for initial-state partons,
and $a$ and $b$ for final-state partons, 
as described in Eq.~(\ref{eq:process}).
The parameter $\rho$ has been defined in Eq.~({\ref{eq:rho}), 
and is always larger than one for massive partons 
and equal to one for the massless case.
We will see that all the results we derive here go back to 
the previous calculation for massless dijet events, 
$\rho=1$~\cite{Oderda:1998en, Appleby:2003sj}.

For complete integrations over the rapidity gap geometry, 
it is convenient to integrate $\Omega_{ij}$ over an azimuthal angle $\phi$ first. 
It is then straightforward to evaluate 
rapidity integrals over a gap region $-Y/2$ to $Y/2$. 
The final expressions for the ${\gamma_S^{(1)}}_{(ij)}$ of 
Eq.~(\ref{eq:ad}) are then
\bea
{\gamma_S^{(1)}}_{(12)} 
&=& 
Y - i \pi \, , 
\nn \\
{\gamma_S^{(1)}}_{(ab)} 
&=& 
- i \pi 
+ \frac{\rho^2 \cosh \da + 1}
{ 2\rho \cosh \left( \frac{\da}{2} \right) 
\sqrt{1+\rho^2 \sinh^2\left( \frac{\da}{2} \right)}} 
\nn \\
&&\hspace{-20mm}
\times 
\ln \left(  
\frac{
\rho^2 \cosh \left( \frac{\da+Y}{2} \right) \sinh \left( \frac{\da}{2} \right) 
+ \sinh \left( \frac{Y}{2} \right) 
+ \sqrt{ \left(\rho^2 \sinh^2 \left( \frac{\da}{2} \right) 
+1 \right) 
\left( \rho^2  \cosh^2 \left( \frac{\da+Y}{2} \right)   -1   \right)   }   
}
{
\rho^2 \cosh \left( \frac{\da-Y}{2} \right)\sinh \left( \frac{\da}{2} \right) 
- \sinh \left( \frac{Y}{2} \right) 
+ \sqrt{\left(\rho^2 \sinh^2 \left( \frac{\da}{2} \right) +1 \right) 
\left( \rho^2 \cosh^2 \left( \frac{\da-Y}{2} \right) -1   \right)     }  
}  
\right)
 \, , 
 \nn \\
{\gamma_S^{(1)}}_{(1a)} 
&=& 
\frac{1}{4} \ln \left( 
\frac{ 
2\rho \left( \rho \cosh^2 \left( \frac{ \da+Y}{2} \right) 
+  \sinh \left( \frac{ \da+Y}{2} \right) 
\sqrt{\rho^2  \cosh^2 \left( \frac{ \da+Y}{2} \right) -1 }  \right) - (\rho^2 +1)
}
{
2\rho \left( \rho \cosh^2 \left( \frac{ \da-Y}{2} \right) 
+ \sinh \left( \frac{ \da-Y}{2} \right) 
\sqrt{\rho^2  \cosh^2 \left( \frac{ \da-Y}{2} \right) -1 }  \right) 
- (\rho^2 +1)
}  
\right) 
\nn \\
&& 
- \frac{1}{2} \ln \left( 
\frac{  
\rho \cosh \left( \frac{ \da+Y}{2} \right) 
+ \sqrt{\rho^2  \cosh^2 \left( \frac{ \da+Y}{2} \right) -1 } 
}
{ 
\rho \cosh \left( \frac{ \da-Y}{2} \right) 
+ \sqrt{\rho^2  \cosh^2 \left( \frac{ \da-Y}{2} \right) -1 }
}  
\right) 
 \, , 
 \nn \\
{\gamma_S^{(1)}}_{(1b)} 
&=& 
- \frac{1}{4} \ln \left( 
\frac{
2\rho \left( \rho \cosh^2 \left( \frac{ \da+Y}{2} \right) 
+  \sinh \left( \frac{ \da+Y}{2} \right) 
\sqrt{\rho^2  \cosh^2 \left( \frac{ \da+Y}{2} \right) -1 }  \right) 
- (\rho^2 +1)
}
{
2\rho \left( \rho \cosh^2 \left( \frac{ \da-Y}{2} \right) 
+  \sinh \left( \frac{ \da-Y}{2} \right) 
\sqrt{\rho^2  \cosh^2 \left( \frac{ \da-Y}{2} \right) -1 }  \right) 
- (\rho^2 +1)
}  
\right) 
\nn \\
&& 
- \frac{1}{2} \ln \left( 
\frac{  
\rho \cosh \left( \frac{ \da+Y}{2} \right)
+ \sqrt{\rho^2  \cosh^2 \left( \frac{ \da+Y}{2} \right) -1 } 
}
{ 
\rho \cosh \left( \frac{ \da-Y}{2} \right)
+ \sqrt{\rho^2  \cosh^2 \left( \frac{ \da-Y}{2} \right) -1 }
}  \right) 
 \, , 
 \nn \\
{\gamma_S^{(1)}}_{(2a)} 
&=& 
{\gamma_S^{(1)}}_{(1b)}  \, , 
\nn \\
{\gamma_S^{(1)}}_{(2b)} 
&=&
{\gamma_S^{(1)}}_{(1a)}  \, .
\label{eq:ga3}
\eea
We take the massless limit, $\rho=1$, in Eq.~(\ref{eq:ga3}) 
to cross-check these results  
with the previous calculation for massless dijet events,
and find~\cite{Appleby:2003sj}
\bea
{\gamma_S^{(1)}}_{(12)} 
&=& 
Y - i \pi \, , 
\nn \\
{\gamma_S^{(1)}}_{(ab)} 
&=& 
\ln \left( 
\frac{
\sinh \left( \frac{\da+Y}{2} \right)
}
{
\sinh \left( \frac{\da-Y}{2} \right) 
} 
\right) - i \pi \, , 
\nn \\
{\gamma_S^{(1)}}_{(1a)} 
&=&  
{\gamma_S^{(1)}}_{(2b)} 
= 
\frac{1}{2} 
\left( 
\ln \left( 
\frac{\sinh \left( \frac{\da+Y}{2} \right)}
{\sinh \left( \frac{\da-Y}{2} \right) } 
\right) - Y 
\right) \, , 
\nn \\
{\gamma_S^{(1)}}_{(1b)} 
&=& 
{\gamma_S^{(1)}}_{(2a)}  
= 
- \frac{1}{2} \left( 
\ln \left( 
\frac{\sinh \left( \frac{\da+Y}{2} \right)}
{\sinh \left( \frac{\da-Y}{2} \right) } 
\right) +Y 
\right) \, . 
\label{eq:ga4}
\eea

We can now construct the soft anomalous dimension matrices 
which depend on the color basis. 
We use an $s$-channel singlet-octet basis~\cite{Kidonakis:1998bk} for 
quark-antiquark annihilation to a heavy quark pair,
\bea
c_{singlet} 
= 
c_1, \, \,\,\, c_{octet}
= 
-\frac{1}{2N_c}c_1+\frac{c_2}{2} \, ,
\label{eq:basis}
\eea
where $c_1$ and $c_2$ are color tensor basis for 
singlet exchange in the $s$ and $t$ channels,
\bea
(c_1)_{\{r_i\}} 
= 
\delta_{r_1 r_2} \delta_{r_a r_b}, \, \, (c_2)_{\{r_i\}}
= 
\delta_{r_1 r_a} \delta_{r_2 r_b} \, .
\label{eq:basis1}
\eea
Here, we denote by $r_i$ the color index 
associated with the incoming or outgoing parton $i$.
The lowest-order soft matrix is  
the trace of the color basis, $S_{LI}^{(\f,0)}= \Tr(c_L^{\dag} c_I)$,
given in singlet-octet basis by
\bea
S^{(\f,0)} 
&=&
\left( 
\begin{array}{cc} 
N_c^2 & \hspace{3mm} 0 \vspace{3mm}  
\\ 
0 & \hspace{3mm} \frac{1}{4}(N_c^2-1) 
\end{array}
\right) \, ,
\label{eq:s0qq}
\eea
which will be used in the resummed cross sections, Eq.~(\ref{eq:qc3}).

To construct the soft anomalous dimension matrix 
in the color basis~(\ref{eq:basis}), 
we rewrite the various one-loop diagrams, 
using the identity shown in Fig.~\ref{fig:c}:
\bea
(T^a_F)_{ji} (T^a_F)_{kl} 
= 
\frac{1}{2} \left( 
\delta_{ki} \delta_{jl} - \frac{1}{N_c}  \delta_{ji} \delta_{kl} 
\right) \, ,
\label{eq:c}
\eea
where the $T^a_F$'s are the generators of SU($N_c$) 
in the fundamental representation.
\begin{figure}[fhptb]
\begin{center}
\includegraphics[width=.75\hsize]{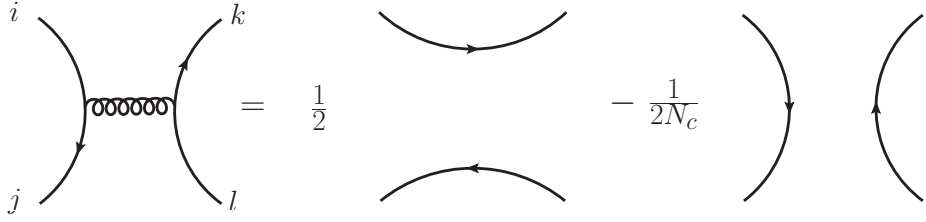}
\end{center}
\caption{Color identity corresponding to Eq.~(\ref{eq:c}).}
\label{fig:c}
\end{figure}
In the basis of Eq.~(\ref{eq:basis}) for the process $q \bar{q} \to Q \bar{Q}$, 
we derive in this way the soft anomalous dimension matrix 
in terms of the integrals ${{\gamma_S^{(1)}}}_{(ij)}$ in Eq.~(\ref{eq:ga3}),
\bea
\ad^{(1)(q\bar{q} \to Q\bar{Q})} 
&=&
\left( 
\begin{array}{cc} 
C_F \alpha & 
\hspace{5mm} \frac{C_F}{2N_c}\left( \chi + \tau \right) \vspace{3mm}  
\\  
\chi + \tau  & 
\hspace{5mm} C_F \chi - \frac{1}{2N_c} 
\left( 2 \tau  +\alpha+ \chi  \right)
\end{array}
\right) \, ,
\label{eq:so}
\eea
where $\alpha$, $\chi$ and $\tau$ are defined by
\bea
\alpha 
&\equiv& 
{\gamma_S^{(1)}}_{(12)}+{\gamma_S^{(1)}}_{(ab)} \, , 
\nn \\
\chi 
&\equiv& 
{\gamma_S^{(1)}}_{(1a)}+{\gamma_S^{(1)}}_{(2b)} \, , 
\nn \\
\tau 
&\equiv& 
{\gamma_S^{(1)}}_{(1b)}+{\gamma_S^{(1)}}_{(2a)} \, .
\label{eq:variables}
\eea

The eigenvectors of $\ad^{(1)}$ in Eq.~(\ref{eq:so}) may be chosen as
\bea
e_1 
&=& 
\left(
\begin{array}{c} 
1 \\
\frac{4N_c \xi}{\Delta_1 + \sqrt{ \Delta_1^2 + 8 C_F N_c \xi^2}}
\end{array}
\right) \, ,
\nonumber \\
e_2 
&=& 
\left(
\begin{array}{c}
\frac{\Delta_1 - \sqrt{ \Delta_1^2 +  8 C_F N_c \xi^2}}{4N_c \xi}
 \\
1
\end{array}
\right) \, ,
\label{eq:e1e2}
\eea
and the corresponding eigenvalues are given by
\bea
\lambda_1^{(\f,1)} 
&=& 
\frac{ \Delta_2 + \sqrt{ \Delta_1^2 + 8 C_F N_c \xi^2}}{4N_c} \, ,
\nonumber \\
\lambda_2^{(\f,1)} 
&=& 
\frac{ \Delta_2 - \sqrt{ \Delta_1^2 + 8 C_F N_c \xi^2}}{4N_c} \, ,
\label{eq:exponents2}
\eea
where we have defined
\bea
\Delta_1 
&\equiv& 
(\alpha + \chi + 2 \tau) + 2C_F N_c (\alpha-\chi) \, ,
\nonumber \\
\Delta_2 
&\equiv&  
-(\alpha + \chi + 2 \tau) + 2C_F N_c (\alpha+\chi) \, ,
\nonumber \\
\xi 
&\equiv& 
(\tau+ \chi) \, .
\eea
Finally, the transformation matrix $R$ in Eq.~(\ref{eq:gdiagonal}), 
which diagonalizes the soft anomalous dimension matrix, 
is given from Eq.~(\ref{eq:e1e2}) by
\bea
 ({R^{(\f)}}^{-1})_{J \beta} 
 = 
 \left(
 \begin{array}{cc}
 1 & \frac{\Delta_1 - \sqrt{ \Delta_1^2 + 8 C_F N_c \xi^2}}{4N_c \xi} 
 \\ 
 \frac{4N_c \xi}{\Delta_1 + \sqrt{ \Delta_1^2 + 8 C_F N_c \xi^2}} & 1 
 \end{array}
 \right) \, .
 \label{eq:trans}
 \eea
It was observed in Ref.~\cite{Oderda:1998en} that 
for massless partons the above eigenvectors are 
independent of the rapidity difference $\de$, 
and depend only on the rapidity gap range, $Y$.
However, we find that this is not the case any more 
for massive quark pair production, $\rho>1$.
As an example, we take the values $\de=2.5$ and $\de=4$ 
at fixed $m_Q=m_t$, $M=1.5~$TeV and $Y=1.5$, 
and confirm a mild rapidity dependence of the eigenvectors, 
\bea
e_1 
&=& 
\left(
\begin{array}{c} 
1 \\ -0.098-0.269 i 
\end{array}
\right) \, 
\mbox{for $\de=2.5$} , \, \, \,\,
e_1 
= 
\left(
\begin{array}{c} 
1 \\ -0.102 -0.274 i 
\end{array}
\right) \, 
\mbox{for $\de=4$} ,
\nonumber \\
e_2 
&=& 
\left(
\begin{array}{c}
0.022+0.060 i \\ 1
\end{array}
\right) \,
\mbox{for $\de=2.5$} , \, \, \,\,
e_2 
= 
\left(
\begin{array}{c} 
0.023+0.061 i \\ 1 
\end{array}
\right) \, 
\mbox{for $\de=4$} \, .
\label{eq:ddd}
\eea
For these configurations, 
the first eigenvector, $e_1$, is close to a color singlet, 
while the second, $e_2$, is close to a color octet. 
Following the terminology in Ref.~\cite{Oderda:1998en}, 
we denote the eigenvectors, $e_1$ and $e_2$, of $\ad$ by 
``quasi-singlet" and ``quasi-octet", respectively.

We will see below that 
the process $q \bar{q} \to Q \bar{Q}$ is the only relevant partonic channel 
involving a spin-1 particle resonance, $V'(M, sp=1)$.
For other resonances $V'(M,sp \neq 1)$, 
we can consider the gluon-induced partonic process
\bea
g(p_1)\, + g(p_2)\, 
\to \, 
V'( M , sp \neq 1) \, 
\to \, 
Q(p_a) \, + \bar{Q}(p_b) \, + \Omega_{\mbox{gap}}(Q_c) \, .
\label{eq:processgg}
\eea
To describe the color flow in the latter process, 
a suitable $s$-channel basis of color tensors has been defined 
in Ref.~\cite{Berger:2001ns},
\bea
(c_1)_{\{r_i\}}
&=& 
\delta_{r_1 r_2} \delta_{r_a r_b} \, , 
\nn \\
(c_2)_{\{r_i\}} 
&=& 
d_{r_1 r_2 c} (T^c_F)_{r_a r_b} \, , 
\nn \\
(c_3)_{\{r_i\}} 
&=& 
if_{r_1 r_2 c} (T^c_F)_{r_a r_b} \, ,
\label{eq:basis2}
\eea
where $c_1$ is again the $s$-channel singlet tensor, 
and $c_2$ and $c_3$ are the symmetric and antisymmetric octet tensors, 
respectively.
In this basis, the anomalous dimension matrix for the process $gg \to Q \bar{Q}$ 
is found to be~\cite{Appleby:2003sj}
\bea
\ad^{(1)(gg \to Q\bar{Q})} 
=  
\left(
\begin{array}{ccc}
C_F \,  {\gamma_S^{(1)}}_{(ab)}+N_c \, {\gamma_S^{(1)}}_{(12)}
& 0 
& \frac{1}{2} ( \chi+ \tau) \vspace{3mm}
\\
0 
& N_c\, \zeta 
& \frac{N_c}{4} (\chi + \tau)
\vspace{3mm}
\\
\chi+\tau 
& \frac{N_c^2-4}{4N_c}(\chi+\tau) 
& N_c\, \zeta 
\end{array}
\right) \, ,
\label{eq:adg}
\eea
where 
$\zeta
\equiv 
\frac{1}{4} (\chi- \tau) 
- \frac{1}{2N_c^2}  {\gamma_S^{(1)}}_{(ab)} 
+ \frac{1}{2} {\gamma_S^{(1)}}_{(12)}$, 
and where $\chi$ and $\tau$ have been defined in Eq.~(\ref{eq:variables}).
In analogy to Eq.~(\ref{eq:s0qq}), 
we derive the zeroth order soft matrix for this process 
in basis~(\ref{eq:basis2}) as
\bea
S^{(\f,0)} 
=  
\left(
\begin{array}{ccc}
N_c(N_c^2-1)
& 0 
& 0 \vspace{3mm}
\\
0 
& \frac{(N_c^2 -4 ) (N_c^2-1)}{2N_c}
& 0
\vspace{3mm}
\\
0
& 0
& \frac{N_c(N_c^2-1)}{2} 
\end{array}
\right) \, .
\label{eq:s0gg}
\eea

One may now find the corresponding eigenvalues and transformation 
matrices ${R^{(\f)}}^{-1}$ for $gg \to Q\bar{Q}$. 
The results are too cumbersome to present here, but are straightforward to derive and evaluate numerically.
In large $N_c$ limit, however, 
the eigenvectors and the corresponding eigenvalues can be obtained readily, since Eq.~(\ref{eq:adg}) 
becomes 
\bea
\ad^{(1)(gg \to Q\bar{Q})} 
{\simeq } 
\, \, N_c \left(
\begin{array}{ccc}
\frac{1}{2} \,  {\gamma_S^{(1)}}_{(ab)}+ {\gamma_S^{(1)}}_{(12)}
& 0 
& 0 \vspace{3mm}
\\
0 
&  \zeta 
& \frac{1}{4} (\chi + \tau)
\vspace{3mm}
\\
0
& \frac{1}{4}(\chi+\tau) 
&  \zeta 
\end{array}
\right) \, ,
\label{eq:adgN}
\eea
for large $N_c$.
Normalized eigenvectors of $\ad^{(gg\to Q\bar{Q})}$ in large $N_c$ limit 
are then
\bea
e_1 
&=& 
\left(
\begin{array}{c} 
1 \\
0\\
0
\end{array}
\right) \, , \, \, \, \,\,
e_2 
= 
\frac{1}{\sqrt{2}} \left(
\begin{array}{c} 
0 \\
1\\
1
\end{array}
\right) \, , \, \, \, \,\, 
e_3 
= 
\frac{1}{\sqrt{2}} \left(
\begin{array}{c} 
0 \\
-1\\
1
\end{array}
\right) \, ,
\label{eq:e1e2e3}
\eea
and the corresponding eigenvalues are given by
\bea
\lambda_1^{(\f,1)} 
&=& 
\left(\frac{1}{2}   {\gamma_S^{(1)}}_{(ab)}+ {\gamma_S^{(1)}}_{(12)} \right) N_c \, ,
\nonumber \\
\lambda_2^{(\f,1)} 
&=& 
\left(\zeta+ \frac{1}{4}(\chi+\tau) \right) N_c \, ,
\nonumber \\
\lambda_3^{(\f,1)} 
&=& 
\left(\zeta- \frac{1}{4}(\chi+\tau) \right) N_c \, .
\label{eq:exponentsNc}
\eea
We emphasize that, in Eq.~(\ref{eq:e1e2e3}), $e_1$ is an exact color singlet and that $e_2$ and $e_3$ are octet basis vectors.
We will not restrict ourselves the above approximation in following sections. 
We will see, however, in Sec.~{\ref{sec:spins} that the explicit results of eigenvalues and eigenvectors for $gg \to Q\bar{Q}$ in the kinematics and the gap geometry we study in this paper  
are close to the results of the large $N_c$ limit. 

The analysis in this section is based on 
wide resonances with $\Gamma \sim M$ 
of decay width, $\Gamma$, and resonance mass, $M$.
We note, however, that 
the energy scale $\mu$ of gluon radiation into a gap 
should be separated into two different regions, 
depending on the decay widths of heavy resonances. 
This may lead to different prescriptions for describing soft gluon radiation, 
depending on $\mu$.
For gluon radiation at scale $\mu$, with $Q_0 < \mu < \sqrt{\Gamma M}$, 
the hard function can be treated as an effective vertex.
This is the condition we have used for describing soft gluon radiation in this section, where we assume that the hard scale $p_T$ is of the same order as  
$\sqrt{\Gamma M}$. 
For the case of gluon radiation in the range $ \sqrt{\Gamma M} < \mu < M$, 
direct radiation from the resonance, 
which was not considered in this section, 
has to be included in the soft function. 
In addition to this, the contributions from soft gluon interactions between initial and final state partons are suppressed for both octet and singlet resonance processes 
if the hard scale falls into the regime $ \sqrt{\Gamma M} < \mu < M$. 
In this paper, we concentrate on the case that $\sqrt{\Gamma M}$ is 
of the order of $p_T$.\footnote{
In principle, a hard scale for the specific model could be chosen as
$\sqrt{\Gamma M}$, the minimum off-shellness of the resonance. 
The hard scale $p_T$ is suitable in our study since we are interested in a model independent approach 
and since we consider broad resonances 
that satisfy $\sqrt{\Gamma M} \sim p_T$ by assumption.}
In general, the decay width of an octet resonance would be large 
enough to satisfy this condition. 
For example, a KK gluon above $1~$TeV~(as required by precision tests) 
has decay width of about $M_G /6$~\cite{Agashe:2006hk,Lillie:2007yh} in basic RS models, 
and $0.215 M_G$~\cite{Djouadi:2007eg,Baur:2008uv} in RS models based on
 extended electroweak gauge symmetry with specific fermion charges and localization that explain the LEP anomaly of the forward-backwardy asymmetry for $b$ quarks.  
Axigluons or universal colorons have decay width of about $\alpha_s M_G$~\cite{Choudhury:2007ux}, resulting in 
$\sqrt{\Gamma M_G} \sim p_T$ where $p_T$ is the hard scale we have used. 
On the other hand, 
the widths of singlet resonances are sensitive to the model parameters 
for the $Z'$.
A $Z'$ can be very narrow, ${\cal{O}} (10^{-3} M_{Z'})$, 
or very broad, ${\cal{O}} (M_{Z'})$, in both Topcolor $Z'$ models~\cite{Harris:1999ya} 
and in the Little Higgs models~\cite{ArkaniHamed:2002qy,Schmaltz:2005ky,Boersma:2006hw,Boersma:2007fd}, 
depending on parameters such as the $SU(2)_1 \otimes SU(2)_2$ mixing angle.  
In the following sections, our study will be based primarily on 
the broad widths of heavy resonances, $ Q_0 < \mu < \sqrt{\Gamma M}$. 
We have observed that for narrow resonances we have to evolve $\mu$ both from $Q_0$ to $ \sqrt{\Gamma M} $ with the exponents we have obtained in this section and from $ \sqrt{\Gamma M} $ to the hard scale with a different analysis. 
We will discuss briefly how such a narrow singlet resonance can change our results at the end of Sec.~\ref{sec:spins}.

\section{Hard Functions}
\label{sec:hard}
The remaining piece to complete constructing 
a partonic cross section in Eq.~(\ref{eq:hs}) is a hard function.
Hard functions describe physics at the large momentum scale.
In our study, the mechanism of new heavy particle resonances is 
thus contained in  the hard functions, 
decoupled from soft QCD radiation. 
The direct radiation 
from an octet resonance can thus be neglected.
In the previous section, 
we have constructed soft functions in the space of color flow at short distances.
Thus, it is convenient to construct hard functions in the same basis.
In Fig.~\ref{fig:feyn}, we represent several possible diagrams for the hard part, 
involving $s$-channel heavy resonances at lowest order.
In the color basis~(\ref{eq:basis}), 
the corresponding hard functions for $q \bar{q} \to G/Z' \to Q \bar{Q}$ 
in Figs.~\ref{fig:qqGQQ} and \ref{fig:qqZQQ} are given by, respectively, 
 \bea
 (H_{G}^{(\f,LO)})_{KL} 
 = 
 h_{G} (M,s,t,u) 
 \left(
 \begin{array}{cc} 
 0 & 0
 \\
 0 & 1
 \end{array}
 \right) \, ,
 \label{eq:hg}
 \eea
\bea
 (H_{Z'}^{(\f,LO)})_{KL} 
 = 
 h_{Z'} (M,s,t,u) 
 \left(
 \begin{array}{cc} 
 1 & 0
 \\
 0 & 0
 \end{array}
 \right) \, .
 \label{eq:hz}
 \eea
A hard function for  $gg \to G \to Q \bar{Q}$ in Fig.~\ref{fig:GGGQQ} 
such as a KK gluon resonance process~\cite{ArkaniHamed:1998rs,Randall:1999ee} 
 would take the form    
 \bea
 (H_{G}^{'(\f,LO)})_{KL} 
 = 
 h'_{G} (M,s,t,u)  
 \left(
 \begin{array}{ccc} 
 0 & 0 & 0 
 \\ 
 0&0&0 
 \\
 0 & 0&1
 \end{array}
 \right) \, ,
 \label{eq:hg2}
 \eea
 and for $gg \to Z' \to Q \bar{Q}$ in Fig.~\ref{fig:GGZQQ} 
 a hard function is given by
 \bea
 (H_{Z'}^{'(\f,LO)})_{KL} 
 = 
 h'_{Z'} (M,s,t,u)  
 \left(
 \begin{array}{ccc} 
 1 & 0 & 0 
 \\ 
 0&0&0 
 \\
 0 & 0&0
 \end{array}
 \right) \, ,
 \label{eq:hz2}
 \eea
in the color basis~(\ref{eq:basis2}).
The overall functions $h_{G/Z'} (M,s,t,u)$ and $h'_{G/Z'} (M,s,t,u)$ 
depend on hard scale kinematic variables, 
$M$, $s$, $t$ and $u$, independent of the gap variables, 
$Q_0$ and $Y$.
\begin{figure}[fhptb]
\centering
\subfigure[]{
\includegraphics[scale=0.4]{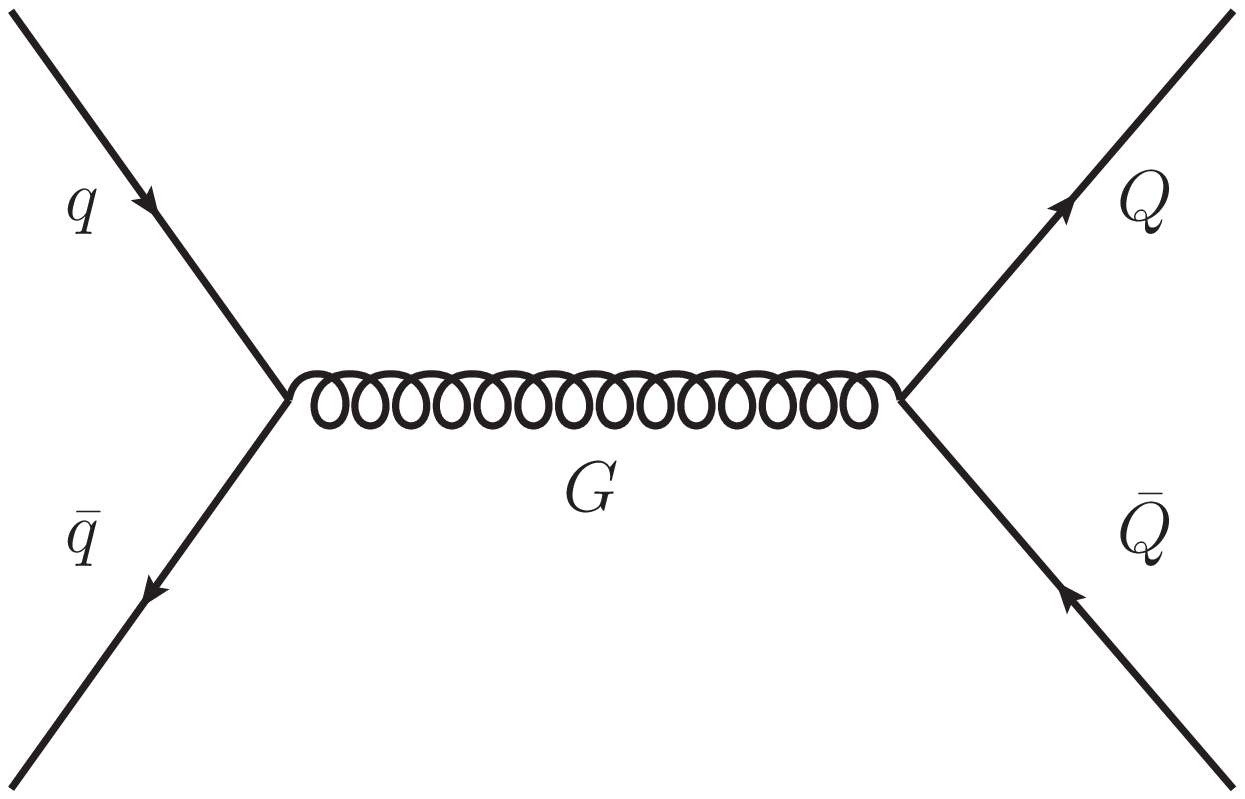}
\label{fig:qqGQQ}
}
\subfigure[]{
\includegraphics[scale=0.4]{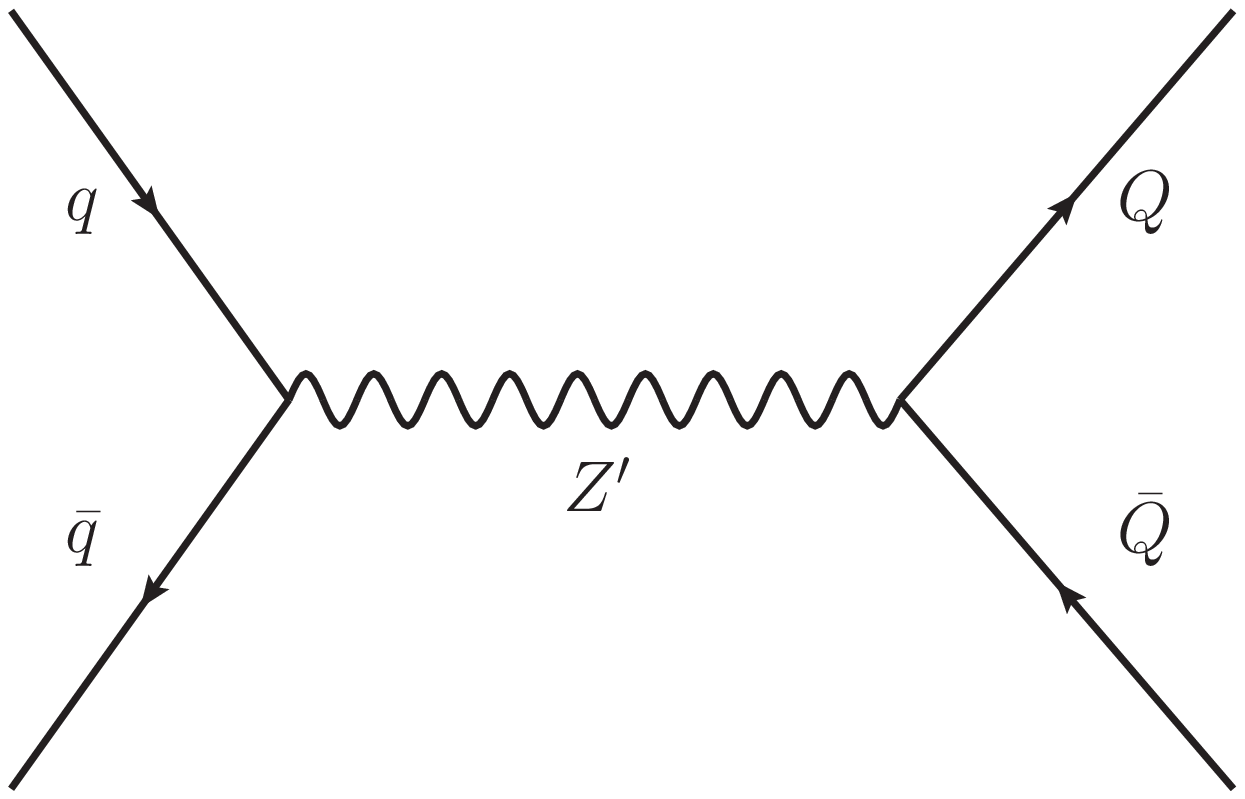}
\label{fig:qqZQQ}
}\\
\subfigure[]{
\includegraphics[scale=0.4]{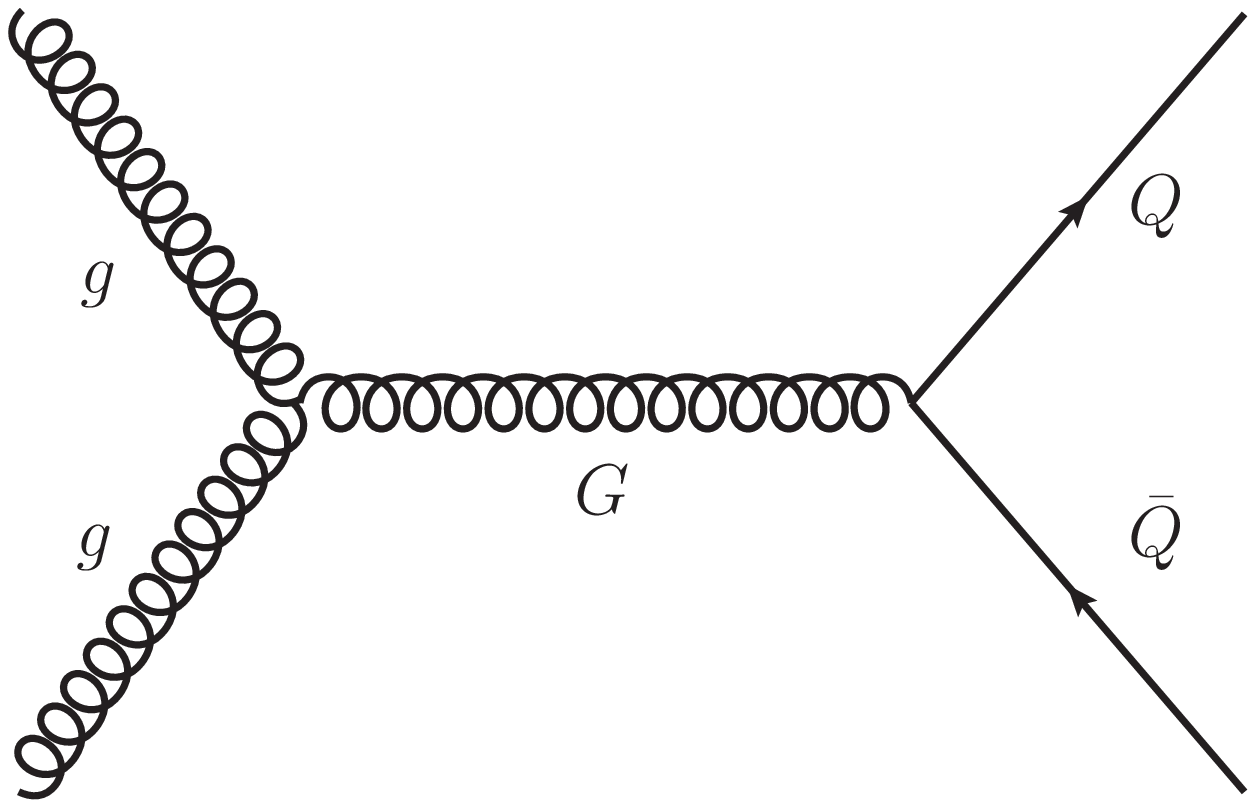}
\label{fig:GGGQQ}
}
\subfigure[]{
\includegraphics[scale=0.4]{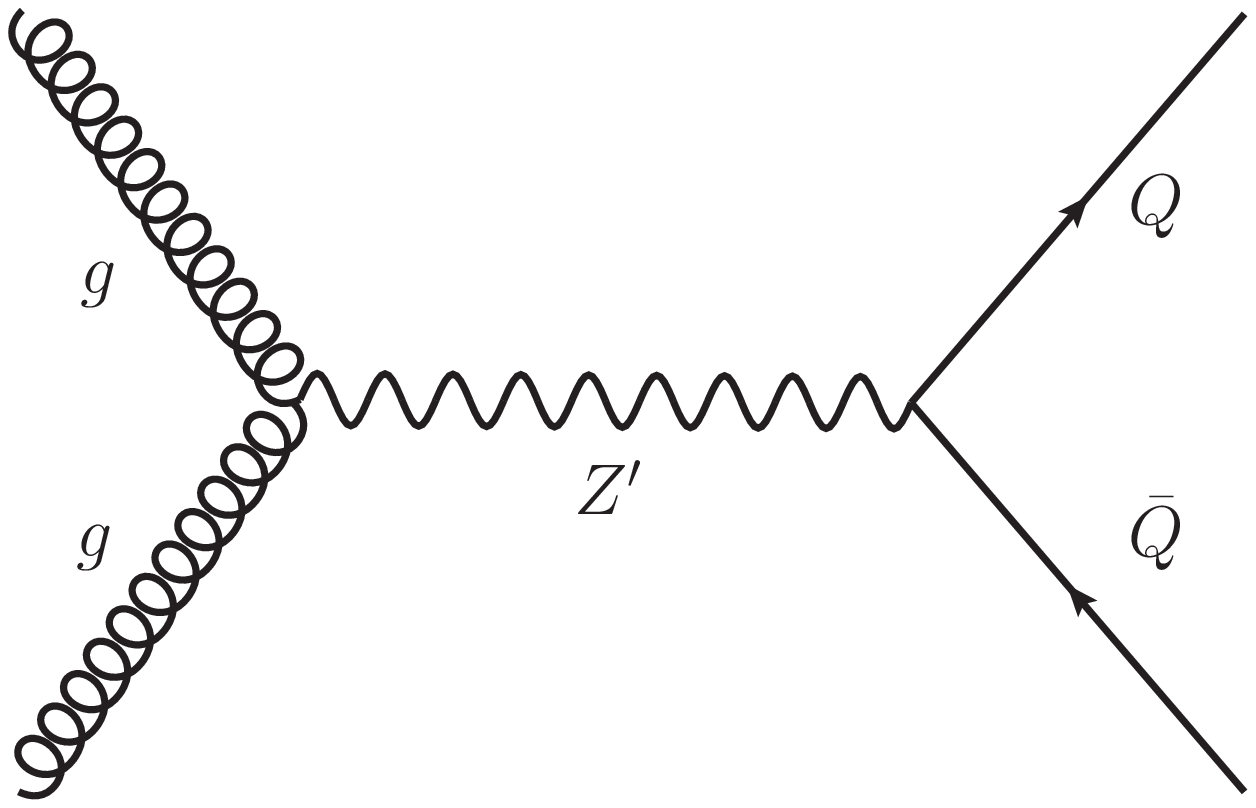}
\label{fig:GGZQQ}
}
\caption[ad]{
Figures (a)-(d) represent the resonance mechanisms for hard functions in 
Eqs.~(\ref{eq:hg}), (\ref{eq:hz}), (\ref{eq:hg2}) and (\ref{eq:hz2}), respectively. 
}
\label{fig:feyn}
\end{figure}
We note that the spin of the particle $V'$ limits possible partonic cross sections.
Specifically, as noted in Refs.~\cite{Pasechnik:2009bq,Hagiwara:2009wt}, 
the Landau-Yang theorem~\cite{Yang} forbids 
the creation of a massive state of total angular momentum $J=1$ 
from two massless spin-1 particles.  
Therefore, the gluon-induced production of a spin-1 massive state 
in Fig.~\ref{fig:GGGQQ} or \ref{fig:GGZQQ} is not allowed 
if the spin of the resonance is one.
Spin-0 singlet or octet resonances could arise, however, 
in split SUSY scenarios via $gg$ annihilation. 
In such cases, $gg \to G/Z'(sp=0)$ could be the dominant production channel, 
giving hard functions of the forms, Eqs.~(\ref{eq:hg2}) and (\ref{eq:hz2}). 
Also, spin-2 particles~$V'(sp=2)$ such as gravitons and reggeons~\cite{Perelstein:2009qi} 
could be created in $s$-channel resonances 
through $gg \to V'(sp=2) \to Q \bar{Q}$, as well as 
$q\bar{q} \to V'(sp=2) \to Q \bar{Q}$.
The cross sections for resonances in these cases 
depend crucially on the model we study. 
For spin-1 resonance processes, 
 we do not need to know specific coupling strength to the SM particles, 
because spin-1 $G$ and $Z'$ involve {\it{only}} the quark-induced partonic process, 
$q \bar{q} \to G/Z' \to Q\bar{Q}$. 
This simplifies the study of the gap fraction.
According to the argument in Sec.~\ref{sec:gapfraction}, 
the gap fraction at leading logarithm results in 
the ratio of partonic cross sections~(\ref{eq:fraction2}), 
free from the PDFs that cancel in the ratios.
Also, $h_G$ or $h_{Z}$ of the hard-scattering functions 
cancels in the ratios~(\ref{eq:fraction2}). 
We can thus study the gap fraction analytically.

In Sec.~\ref{sec:spins}, we will discuss how gluon-induced resonances 
contribute to the gap fraction in massive spin-0 or 2 resonance processes. 
For a complete study of the gap fraction in this case, 
we should combine all the possible partonic processes 
that depend on the specific model.
In addition to these, we must convolute the partonic level results with PDFs, 
following Eq.~(\ref{eq:pdf}).
We leave this study for a further project.

To obtain the partonic cross section in Eq.~(\ref{eq:qc3}), 
hard functions in Eqs.~(\ref{eq:hg}), (\ref{eq:hz}), (\ref{eq:hg2}), 
and (\ref{eq:hz2}) are transformed to $(H^{(\f,LO)})_{\gamma \beta}$ 
through Eq.~(\ref{eq:shdiagonal}) in the color basis 
that diagonalizes $\ad^{(\f)}$ in Eq.~(\ref{eq:gdiagonal}). 
We will study explicitly how partonic cross sections 
and gap fractions are evaluated in the following sections.

\section{Gap Fractions for Spin-1 Resonances}
\label{sec:spin1}
From the previous sections, 
we have all the tools necessary for the evaluation of the partonic cross section 
in Eq.~(\ref{eq:qc3}). 
As mentioned above, the main quantity we focus on is 
the gap fraction~(\ref{eq:fraction}).
For spin-1 resonances, 
we can limit ourselves to quark-induced resonances, 
shown in Figs.~\ref{fig:qqGQQ} and \ref{fig:qqZQQ}, 
following the argument in Sec.~\ref{sec:hard}.
The partonic cross section for heavy quark production via $G$ or $Z'$ resonances is
\bea
\frac{ d \hat{\sigma}^{(\f)}_{G/Z'}}{d \de  } 
&=& 
\sum_{\beta, \gamma} (H_{G/Z'}^{(\f,LO)})_{\beta \gamma}
(M, m_Q, \de, \alpha_s (p_T)) 
\, S^{(\f,0)}_{\gamma \beta}
\left[ 
\frac{
 \ln \left( \frac{Q_0}{\Lambda} \right) 
 }
 {
 \ln \left( \frac{p_T}{\Lambda} \right) 
 }
 \right]^{E^{(\tf)}_{\gamma \beta}} ,
\label{eq:qct}
\eea
where the $(H_{G/Z'}^{(\f,LO)})_{\beta \gamma}$ are hard functions 
written in the basis that diagonalizes the soft anomalous dimension matrix. 
We will implement Eq.~(\ref{eq:qct}) in numerical studies below.
Following Eq.~(\ref{eq:shdiagonal}), to obtain $(H_{G/Z'}^{(\f,LO)})_{\beta \gamma}$, 
we transform the hard functions in Eqs.~(\ref{eq:hg}) and (\ref{eq:hz}), 
through matrix $R$ given explicitly in Eq.~(\ref{eq:trans}).
The soft matrix $S^{(\f,0)}_{\gamma \beta}$ is obtained 
by transforming Eq.~(\ref{eq:s0qq}) through the same matrix $R$.
Finally, the exponents $E^{(\f)}_{\alpha \beta}$ 
that appear in the cross sections (\ref{eq:qct}) are given by Eq.~(\ref{eq:exponents}).

In Eq.~(\ref{eq:qct}), we observe that
the $Q_0$-dependent factor, $\left[ 
 \ln \left( \frac{Q_0}{\Lambda} \right) 
/
 \ln \left( \frac{p_T}{\Lambda} \right) 
 \right]^{E^{(\tf)}_{\gamma \beta}}$, 
 is larger at fixed $Q_0$ for smaller $E^{(\f)}_{\gamma \beta}$,
 because $\left[ 
 \ln \left( \frac{Q_0}{\Lambda} \right) 
/
 \ln \left( \frac{p_T}{\Lambda} \right) 
 \right]$
 is less than one for $Q_0 < p_T$.
 Note that we apply Eq.~(\ref{eq:qct}) only for $Q_0 < p_T$.
 A constraint for the maximum value of $\de$ 
 is obtained from $\rho$ in Eq.~(\ref{eq:rho}),
\bea
Y< \de<2 \cosh^{-1}\left(\frac{M}{2 m_Q}\right) \, .
\eea

Recall that the gap fraction measures the fraction of events 
for which the radiated transverse energy is less than or equal to 
the gap threshold $Q_0$. 
For any $E_{\alpha \beta}^{(\f)}>0$, 
the fraction vanishes for $Q_0=\Lambda$, 
and approaches unity for $Q_0=p_T$, 
where we choose $\mu = p_T$ in numerical calculations.
The soft anomalous dimension matrix for this analysis is given by Eq.~(\ref{eq:so}).
Representative numerical values of the real 
and imaginary parts of the eigenvalues are shown in Fig.~\ref{fig:eigen}
as a function of gap size $Y$.
The corresponding exponents at fixed $Y$
are shown in Fig.~\ref{fig:exp} as a function of $\de$. 
Only the real parts enter the cross section. 
Imaginary parts are shown for completeness. 
Qualitatively, in Figs.~\ref{fig:eigen} and \ref{fig:exp}, 
we see that the eigenvalues and the corresponding exponents are larger
for larger $Y$ and for smaller $\de$.
\begin{figure}[fhptb]
\centering
\subfigure[]{
\includegraphics[scale=0.61]{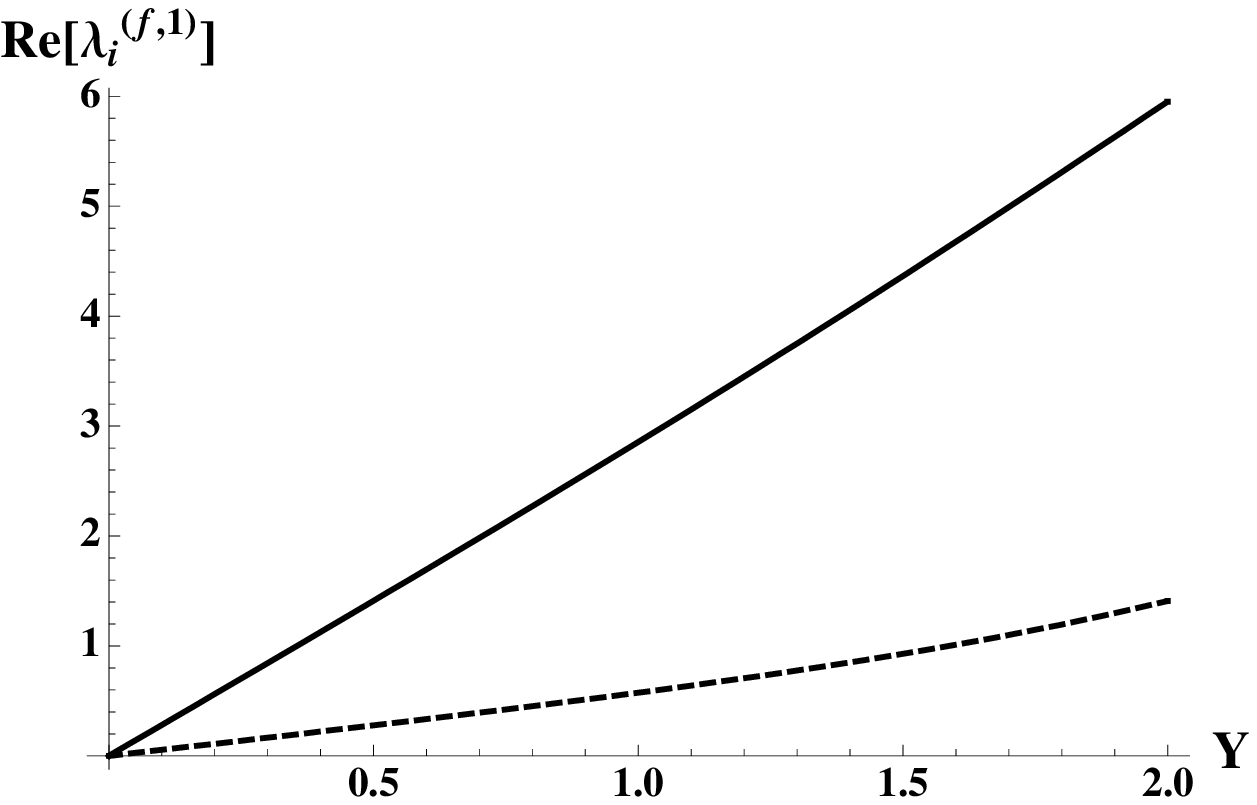}
\label{fig:real}
}
\subfigure[]{
\includegraphics[scale=0.61]{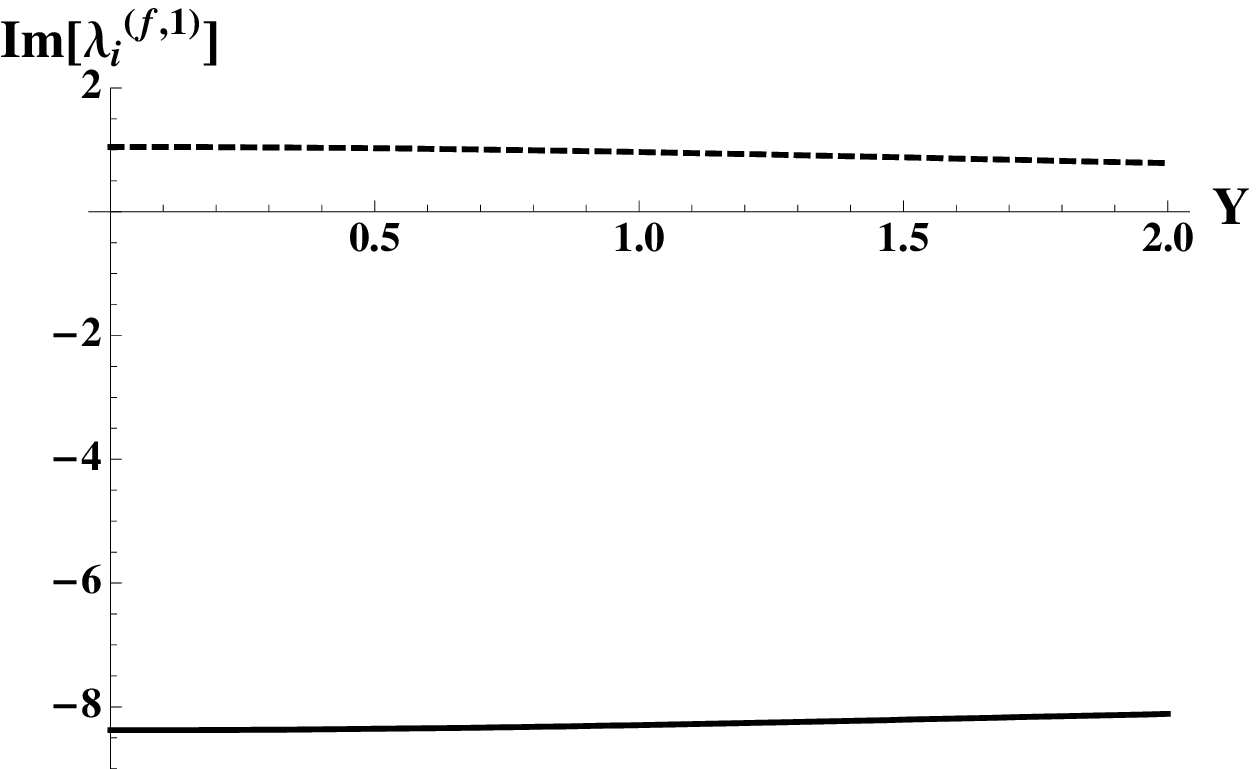}
\label{fig:im}
}
\caption[ad]{
Plot of the real~(a) and imaginary~(b) parts of 
the eigenvalues of the soft anomalous dimension matrix in Eq.~(\ref{eq:so}) for $q\bar{q} \to Q\bar{Q}$, 
as a function of $Y$
for $M=1.5~$TeV, $\de=2.5$ and $m_Q=m_t$. 
The solid line identifies the quasi-singlet eigenvalue, $\lambda_1^{(\f,1)}$, 
the dashed line, quasi-octet,  $\lambda_2^{(\f,1)}$, in Eq.~(\ref{eq:e1e2}).}
\label{fig:eigen}
\end{figure}
\begin{figure}[fhptb]
\centering
\subfigure[]{
\includegraphics[scale=0.6]{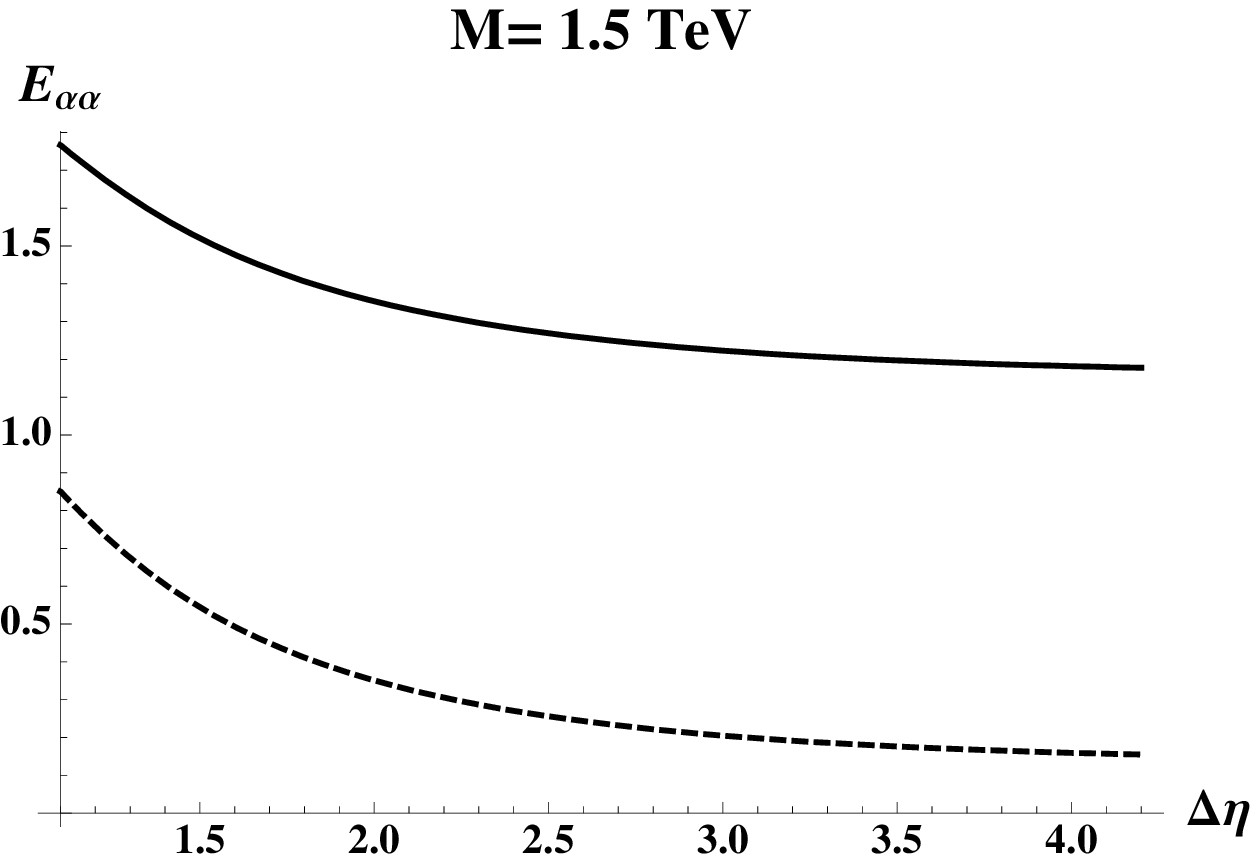}
\label{fig:exp2000}
}
\subfigure[]{
\includegraphics[scale=0.6]{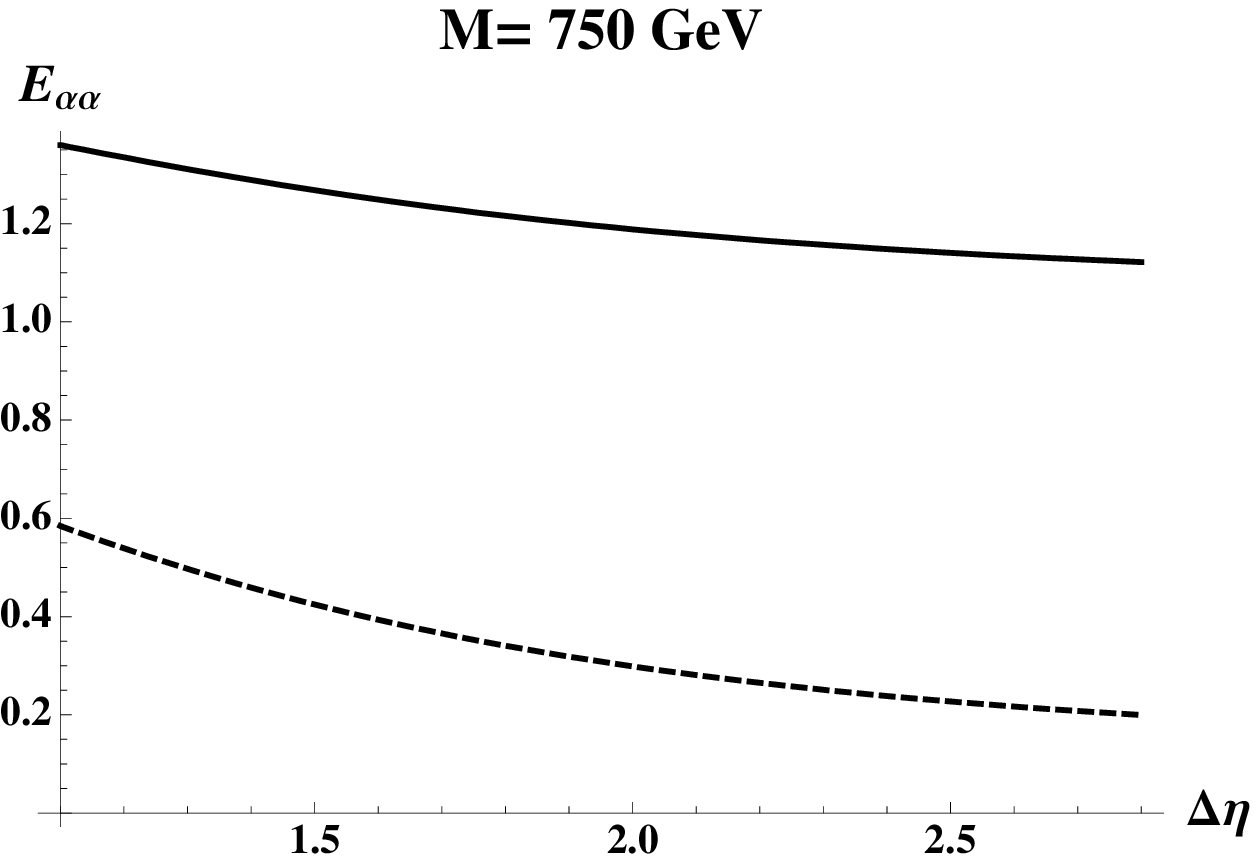}
\label{fig:exp500}
}
\caption[ad]{
Plot of the exponents, $E_{\alpha \alpha}^{(\f)}$, 
of the soft anomalous dimension matrix for $q\bar{q} \to Q\bar{Q}$, 
as a function of $\de$
for $Y=1$ and $m_Q=m_t$
with resonance mass $M=1.5~$TeV and $M=750~$GeV
for (a) and (b), respectively.
The solid lines identify the quasi-singlet exponent, $E^{(\f)}_{11}$, 
the dashed lines, quasi-octet,  $E^{(\f)}_{22}$, obtained from Eq.~(\ref{eq:exponents}).
}
\label{fig:exp}
\end{figure}
In Fig.~\ref{fig:eigen}, we observe that for $Y$ of order unity, 
Re$[\lambda_1]$  is much larger than Re$[\lambda_2]$, 
leading to $E^{(\f)}_{22} < 1 < E^{(\f)}_{11}$, as shown in Fig.~\ref{fig:exp}.
Following the discussion below Eq.~(\ref{eq:qct}), 
a smaller value of  
the exponent $E^{(\f)}_{22}$ enhances the factor  
 $\left[ 
 \ln \left( \frac{Q_0}{\Lambda} \right) 
/
 \ln \left( \frac{p_T}{\Lambda} \right) 
 \right]^{E^{(\tf)}_{22}}$ 
in the partonic cross section~(\ref{eq:qct}) 
relative to a larger value. 
We note that off-diagnal contributions are small.

The resulting gap fractions for two different spin-1 resonances, $Z'$ and $G$, of mass $2~$TeV, $1.5~$TeV, $750~$GeV, and $550~$GeV  
are illustrated as functions of $Q_0$ 
in Fig.~\ref{fig:fractiont} as the solid and the dashed curves 
for a top pair, and  
as the dot-dashed and the dotted curves~(blue) for a bottom pair
at fixed $Y=1.5$ and $\de=2$.
The ratios of gap fractions, $f^G_{gap}/f^{Z'}_{gap}$,
are shown in Fig.~\ref{fig:a} at fixed $Q_0=5~$GeV 
in terms of $Y$ and $\de$ for $M=1.5~$TeV and $750~$GeV.
We observe that for the ratios the dependence on $\de$ is negligible at fixed $Y$.
Also, Fig.~\ref{fig:a} shows that these ratios get larger at fixed $Q_0$ 
as the gap range increases.

In Figs.~\ref{fig:150035} and \ref{fig:75025}, 
the gap fractions and their ratios are presented
as functions of $Q_0$ and $Y$ for $M=1.5~$TeV and $750~$GeV, respectively.
The gap fraction for an octet resonance increases rapidly 
for small values of $Q_0$
in comparison to a singlet resonance for any gap size.
Many fewer events are accumulated at low $Q_0$ 
through a color-singlet resonance than for a color-octet resonance.
For instance, at $Q_0=3$ GeV and $V'( \mbox{1.5 TeV}, sp=1)$ in Fig.~\ref{fig:1600_1}, 
we obtain gap fractions of 12\% and 54\%
for color-singlet and -octet resonances, respectively.
The figures also show that the ratios are larger for larger $M$ and $Y$.
\begin{figure}[fhptb]
\centering
\subfigure[]{
\includegraphics[scale=0.61]{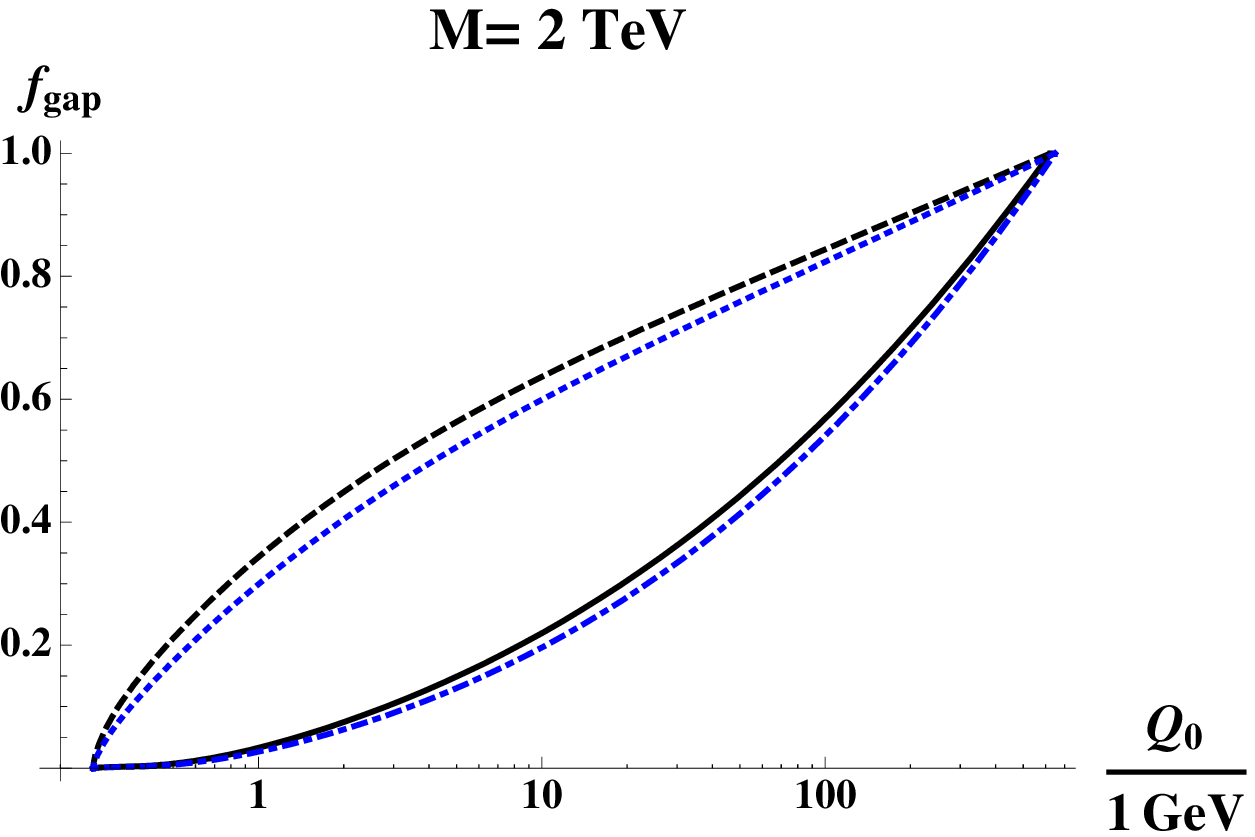}
\label{fig:2000}
}
\subfigure[]{
\includegraphics[scale=0.61]{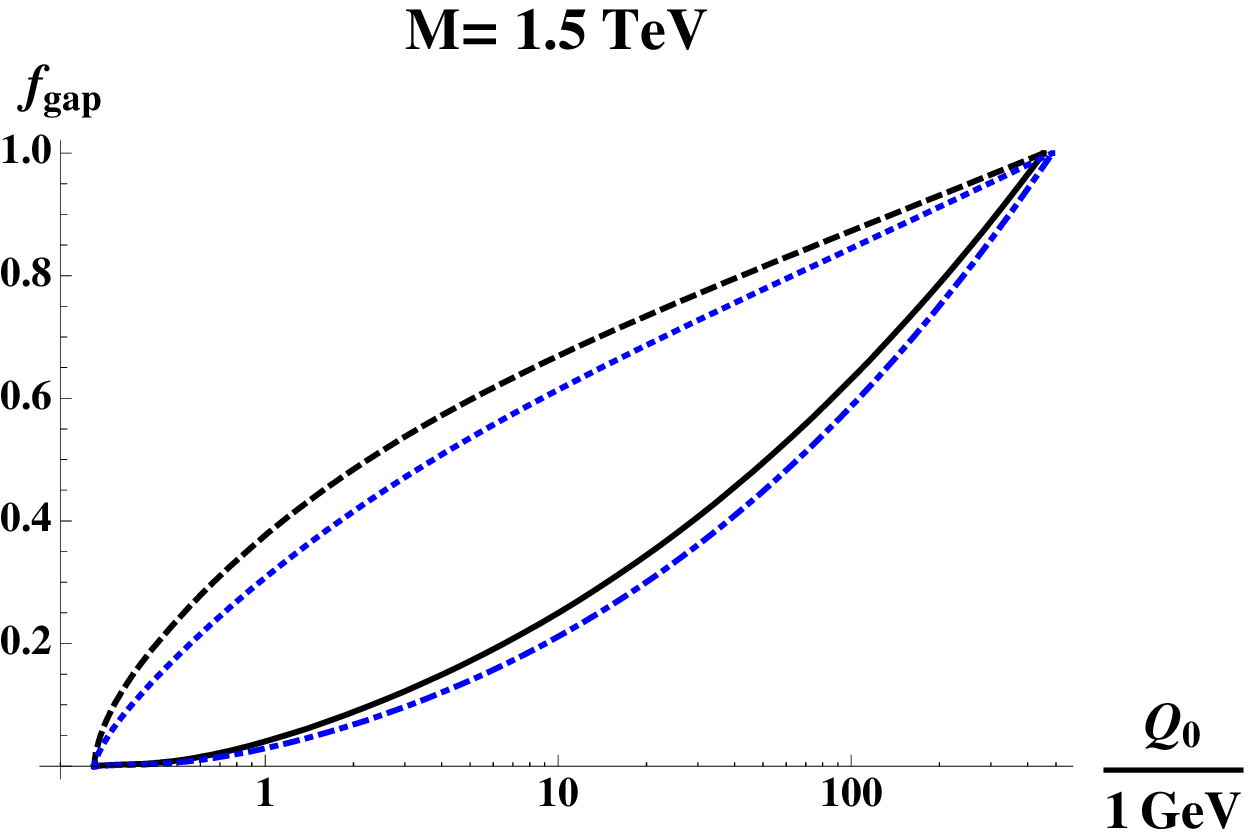}
\label{fig:1600_1}
}
\subfigure[]{
\includegraphics[scale=0.61]{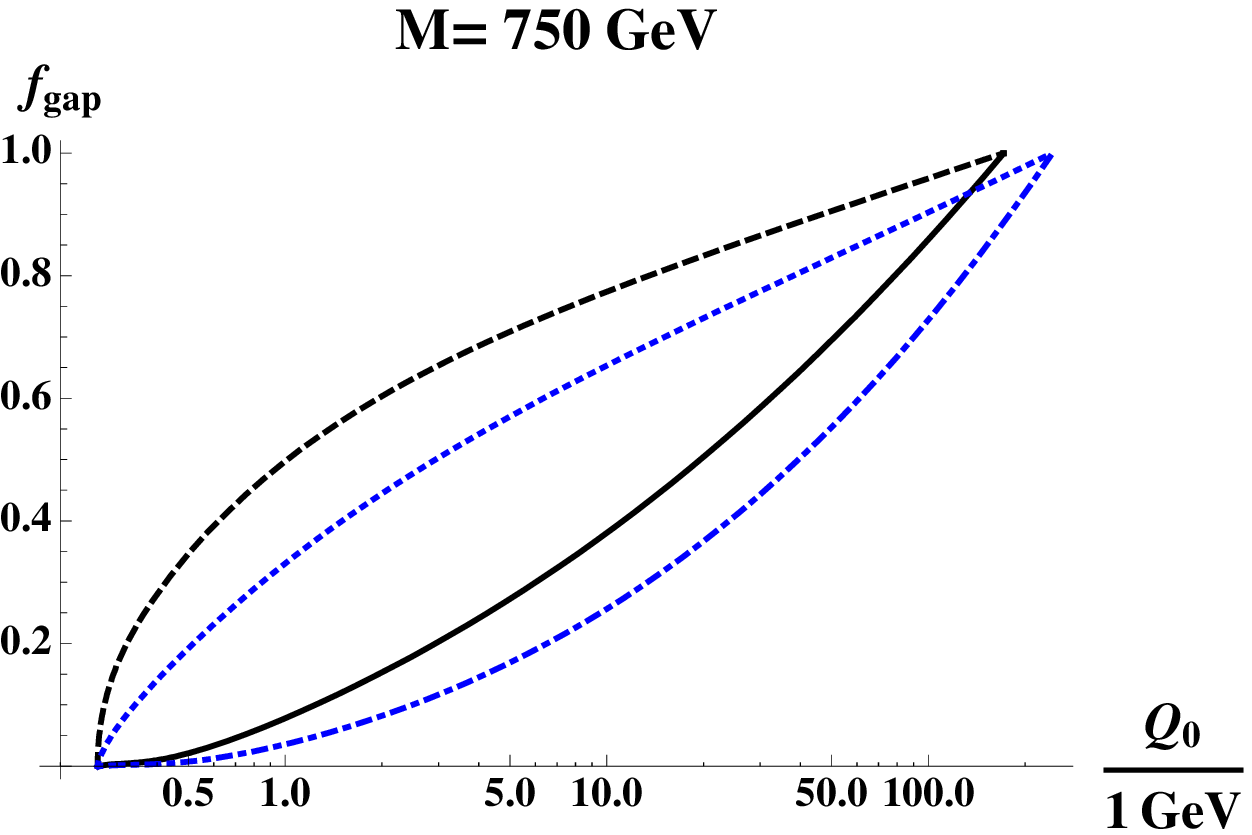}
\label{fig:800_1}
}
\subfigure[]{
\includegraphics[scale=0.61]{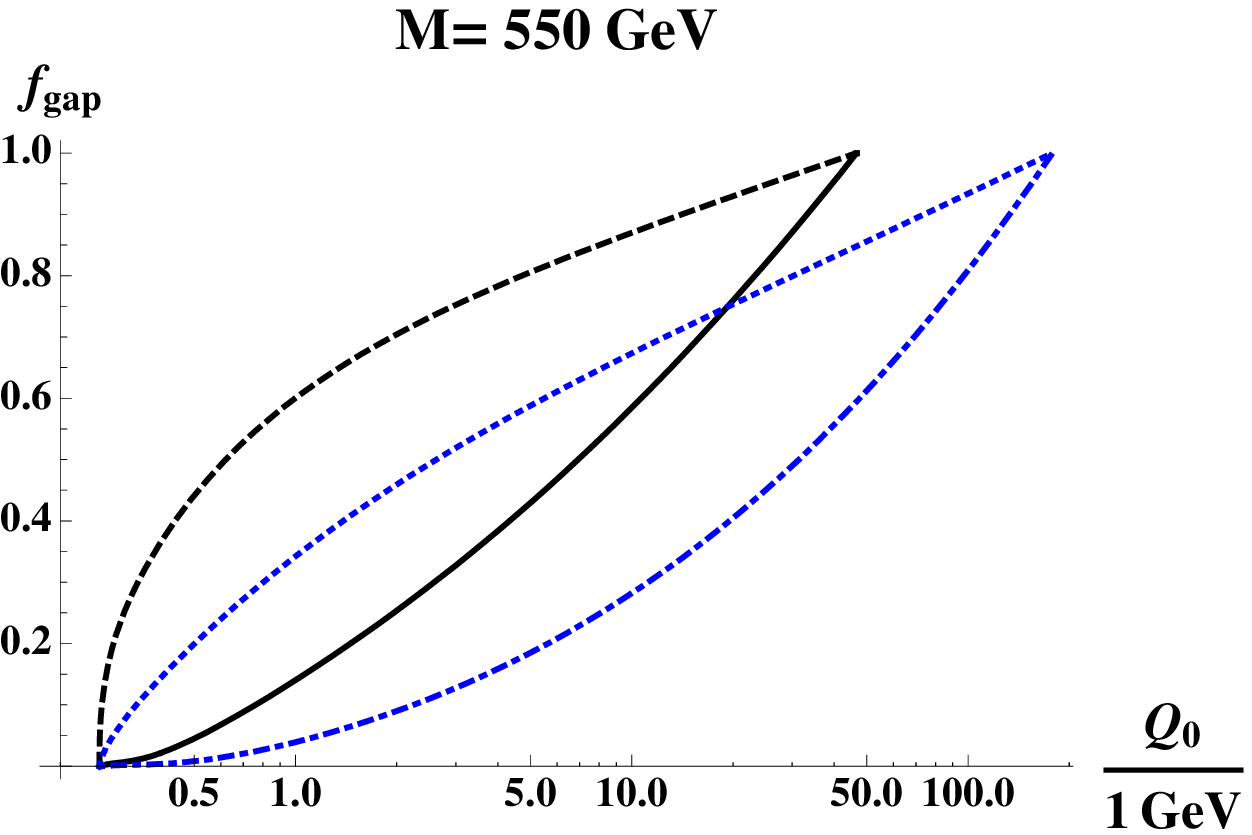}
\label{fig:550a}
}
\caption[ad]{
The fractions for gaps identified by 
the energy threshold $Q_0$ at $\de=2$ and $Y=1.5$ 
for resonance masses $M=2$, $1.5$, $0.75$, and $0.55~$TeV. 
In the above figures, the solid curves describe the gap fraction 
through a $Z'$ resonance (color-singlet), and 
the dashed curves throughout a $G$ resonance (color-octet) into a {\it{top}} quark pair.
The dot-dashed curves~(blue) describe the gap fraction 
through a $Z'$ resonance (color-singlet), and
the dotted curves~(blue) throughout a $G$ resonance (color-octet) into a {\it{bottom}} pair.
 }
\label{fig:fractiont}
\end{figure}

\begin{figure}[fhptb]
\centering
\subfigure[]{
\includegraphics[scale=0.52]{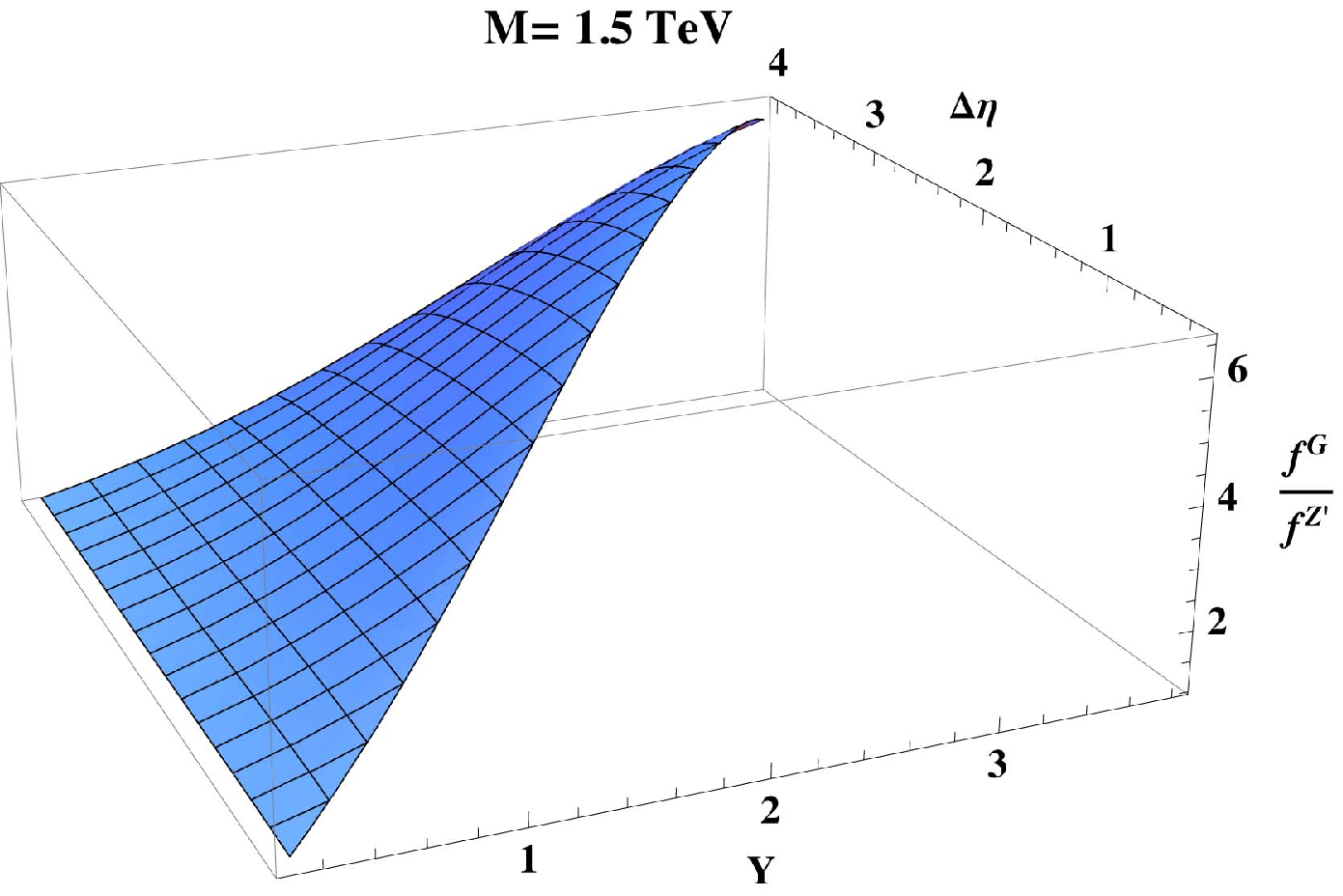}
\label{fig:1500etaY}
}
\subfigure[]{
\includegraphics[scale=0.52]{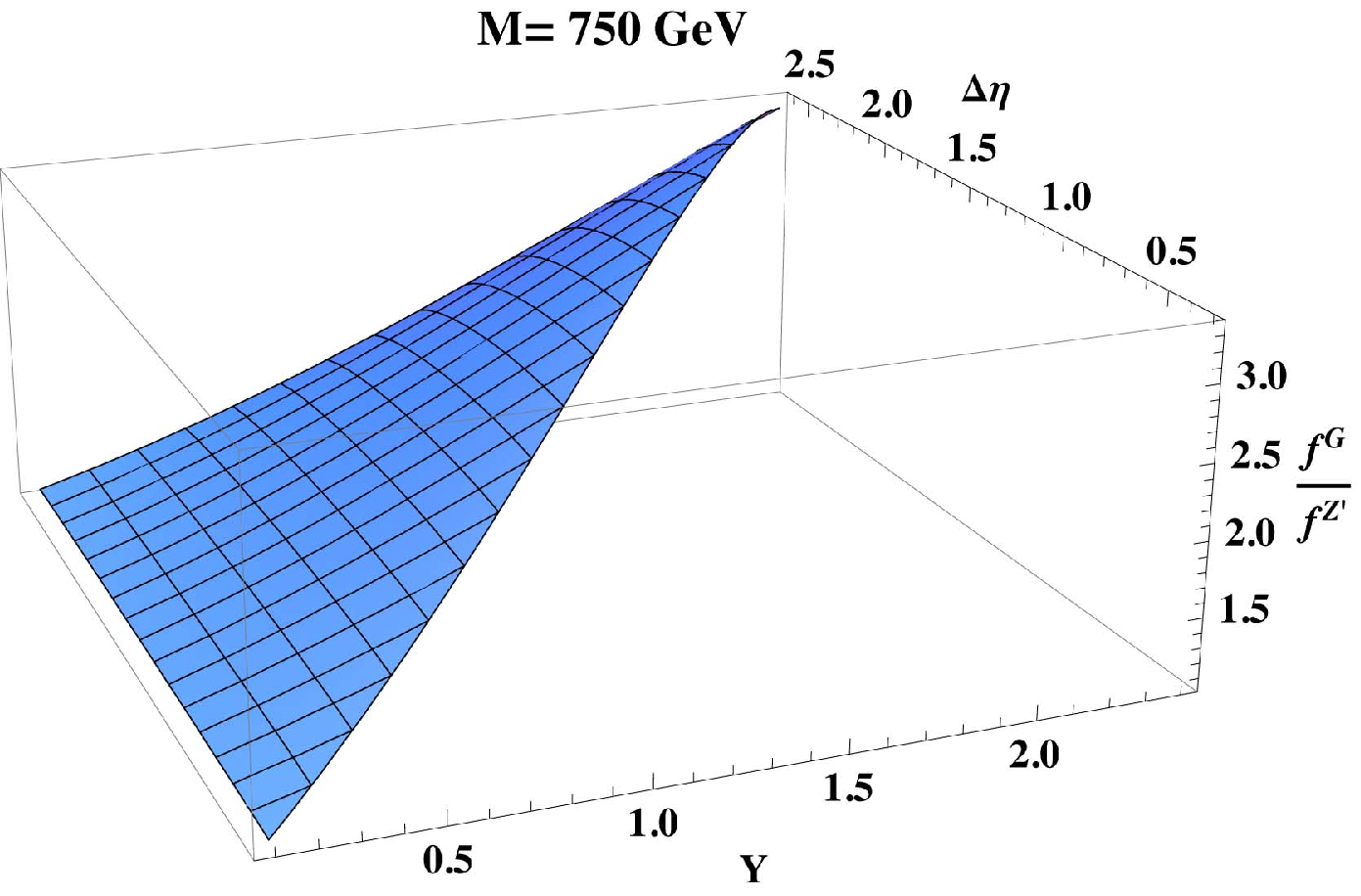}
\label{fig:750etaY}
}
\caption[ad]{
The ratios of two different gap fractions, $f^G_{gap}/f^{Z'}_{gap}$, 
as functions of $Y$ and $\de$ at fixed $Q_0=5~$GeV
are shown (a) for $M=1.5~$TeV and (b) $M=750~$GeV, respectively. 
 }
\label{fig:a}
\end{figure}

\begin{figure}[fhptb]
\centering
\subfigure[]{
\includegraphics[scale=0.55]{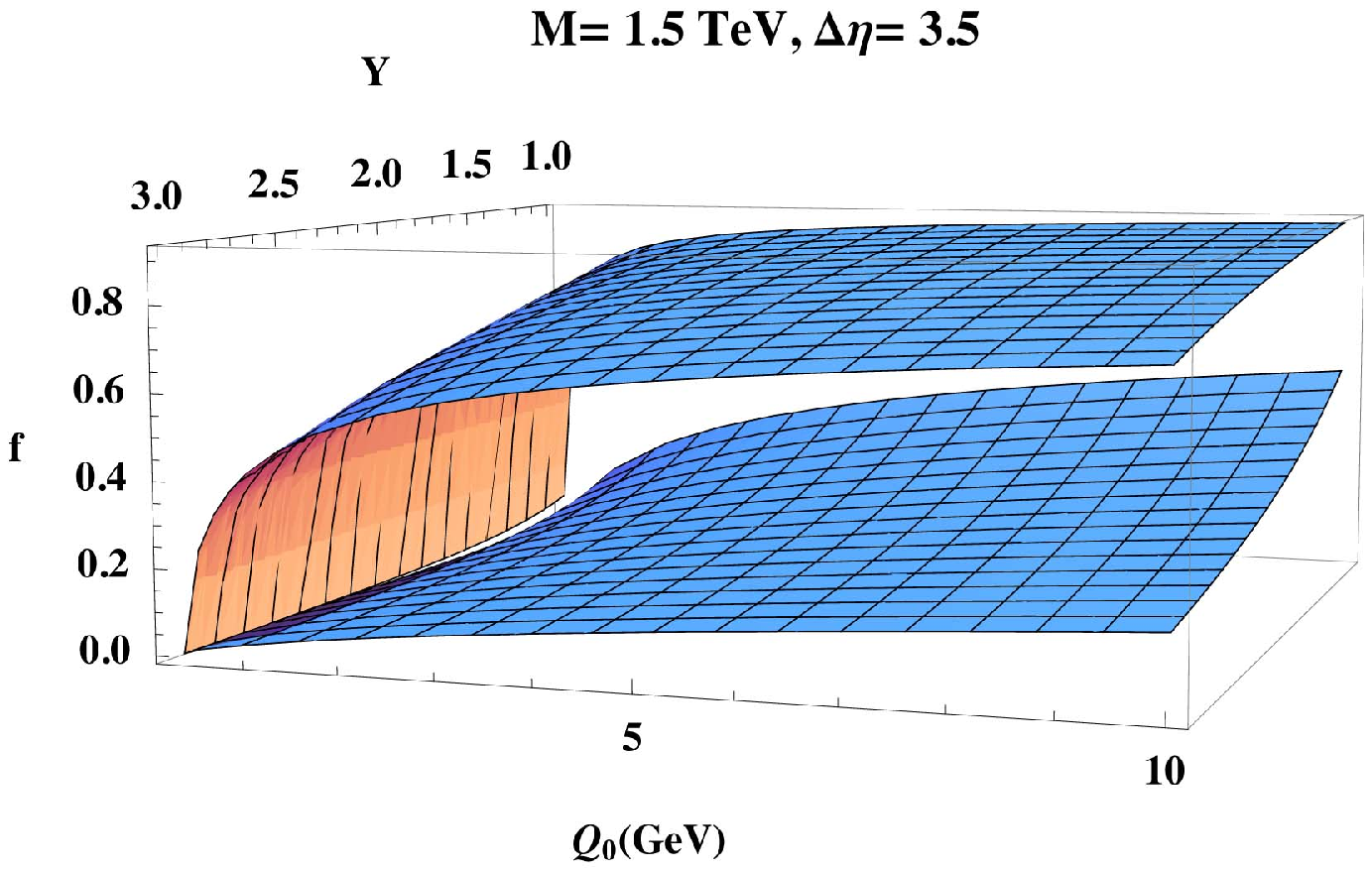}
\label{fig:1500f35}
}
\subfigure[]{
\includegraphics[scale=0.55]{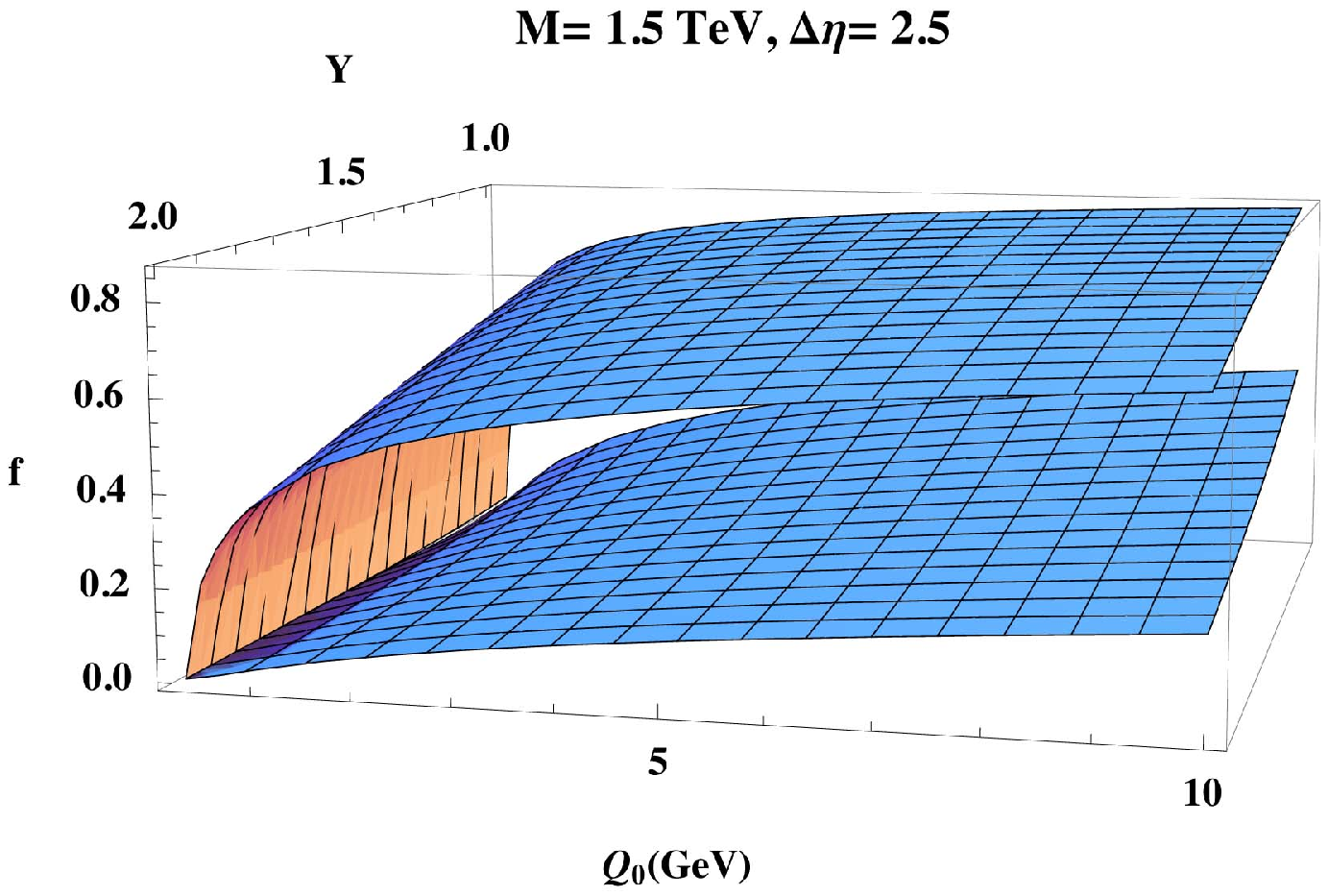}
\label{fig:1500f25}
}
\subfigure[]{
\includegraphics[scale=0.54]{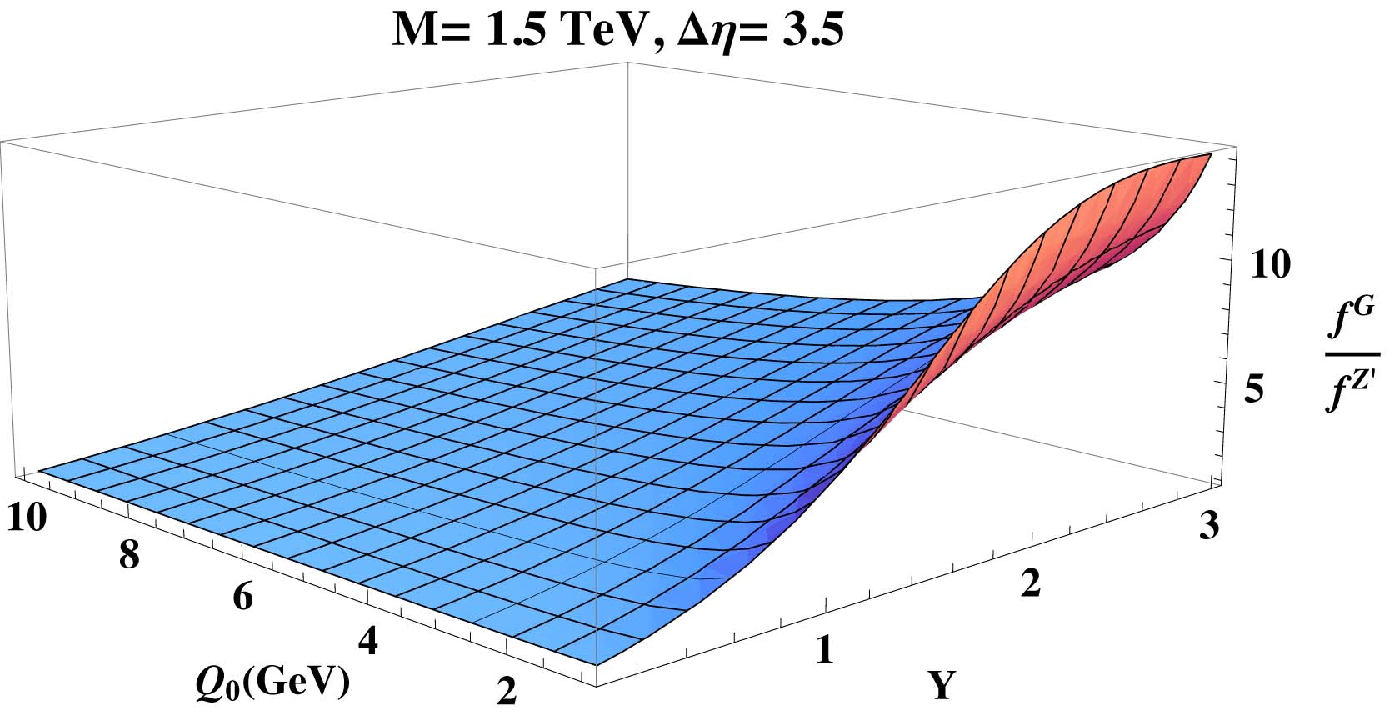}
\label{fig:150035QY}
}
\subfigure[]{
\includegraphics[scale=0.54]{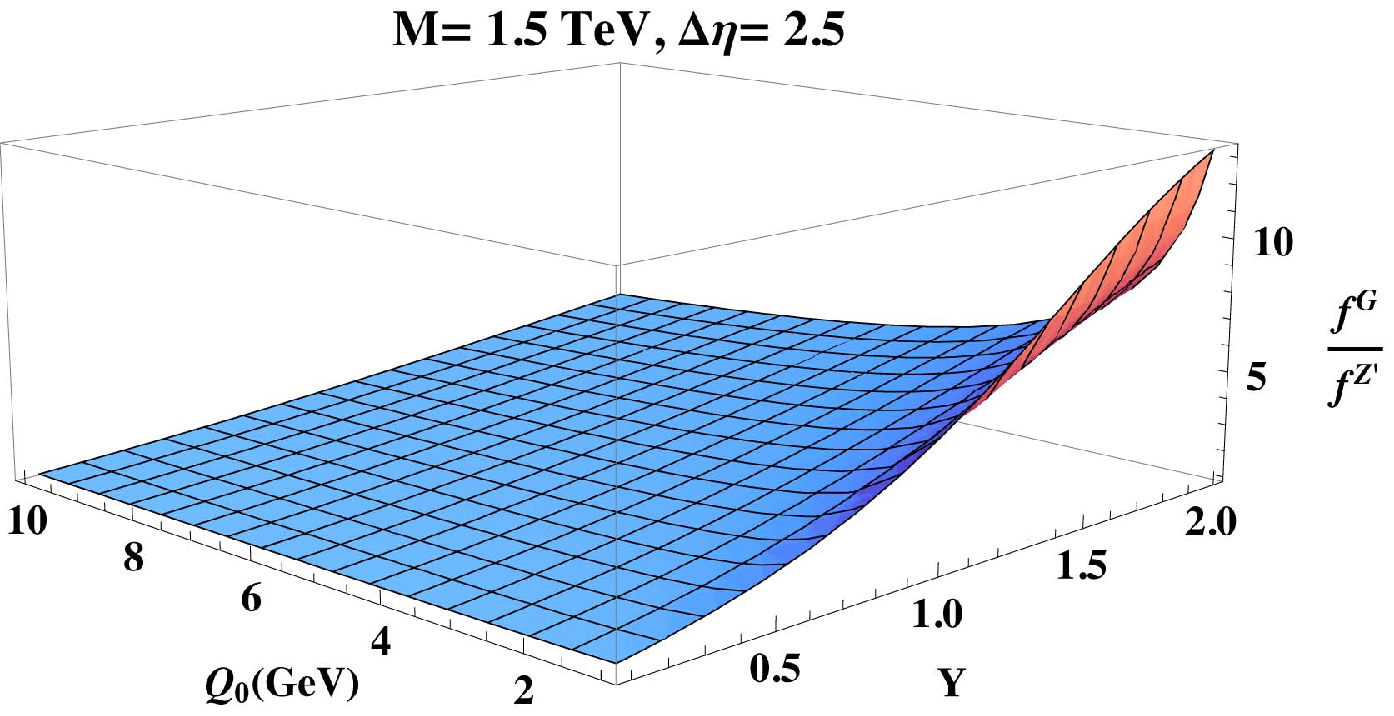}
\label{fig:150025QY}
}
\caption[ad]{
Gap fractions as functions of energy threshold $Q_0$ and
gap range $Y$ 
from a resonance of $M=1.5~$TeV, (a) for $\de=3.5$ and (b) for $\de=2.5$.
The lower surfaces in (a) and (b) describe the gap fractions 
through a $Z'$ resonance (color-singlet), 
the upper surfaces throughout a $G$ resonance (color-octet) decaying into a top quark pair.
The ratios of the gap fraction for an octet resonance to the gap fraction for a singlet resonance are illustrated in (c) and (d) as functions of $Q_0$ and $Y$.
}
\label{fig:150035}
\end{figure}

\begin{figure}[fhptb]
\centering
\subfigure[]{
\includegraphics[scale=0.55]{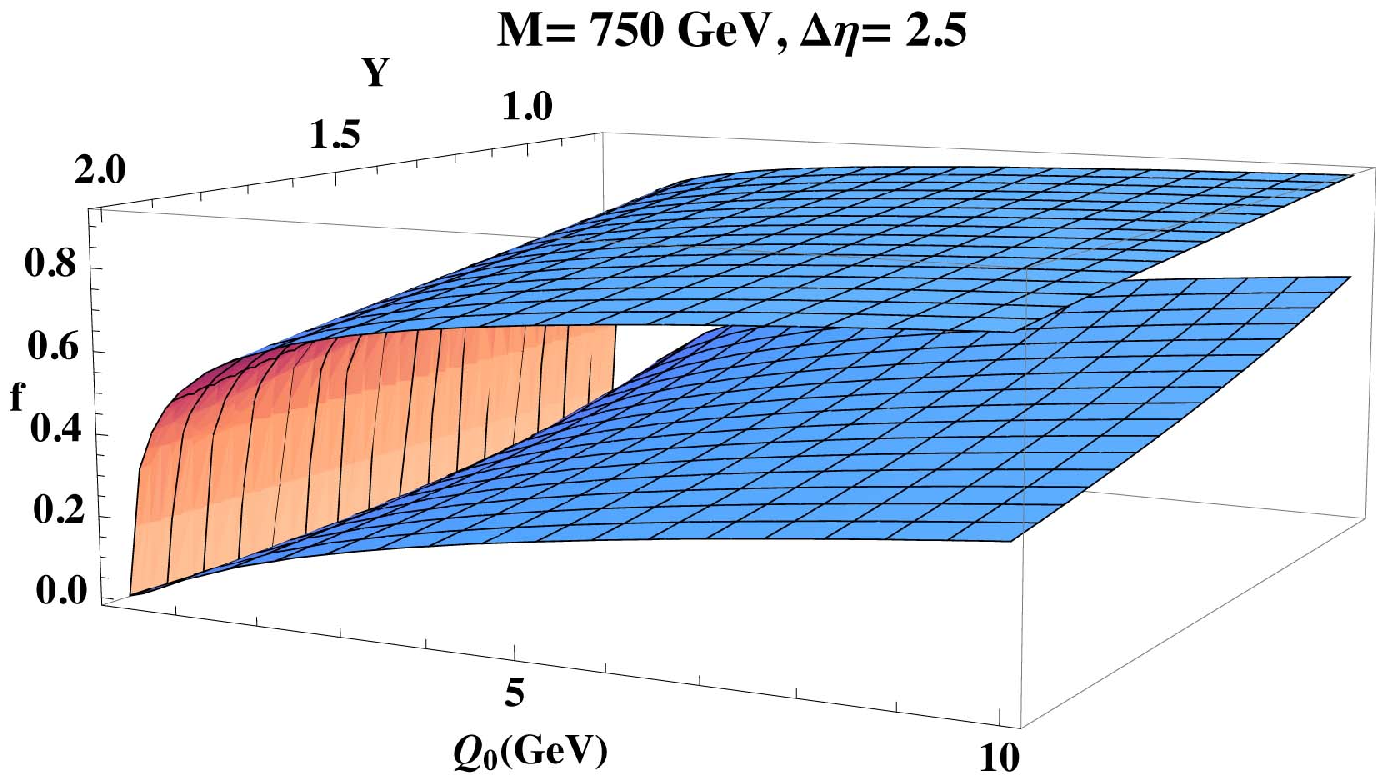}
\label{fig:750f25}
}
\subfigure[]{
\includegraphics[scale=0.55]{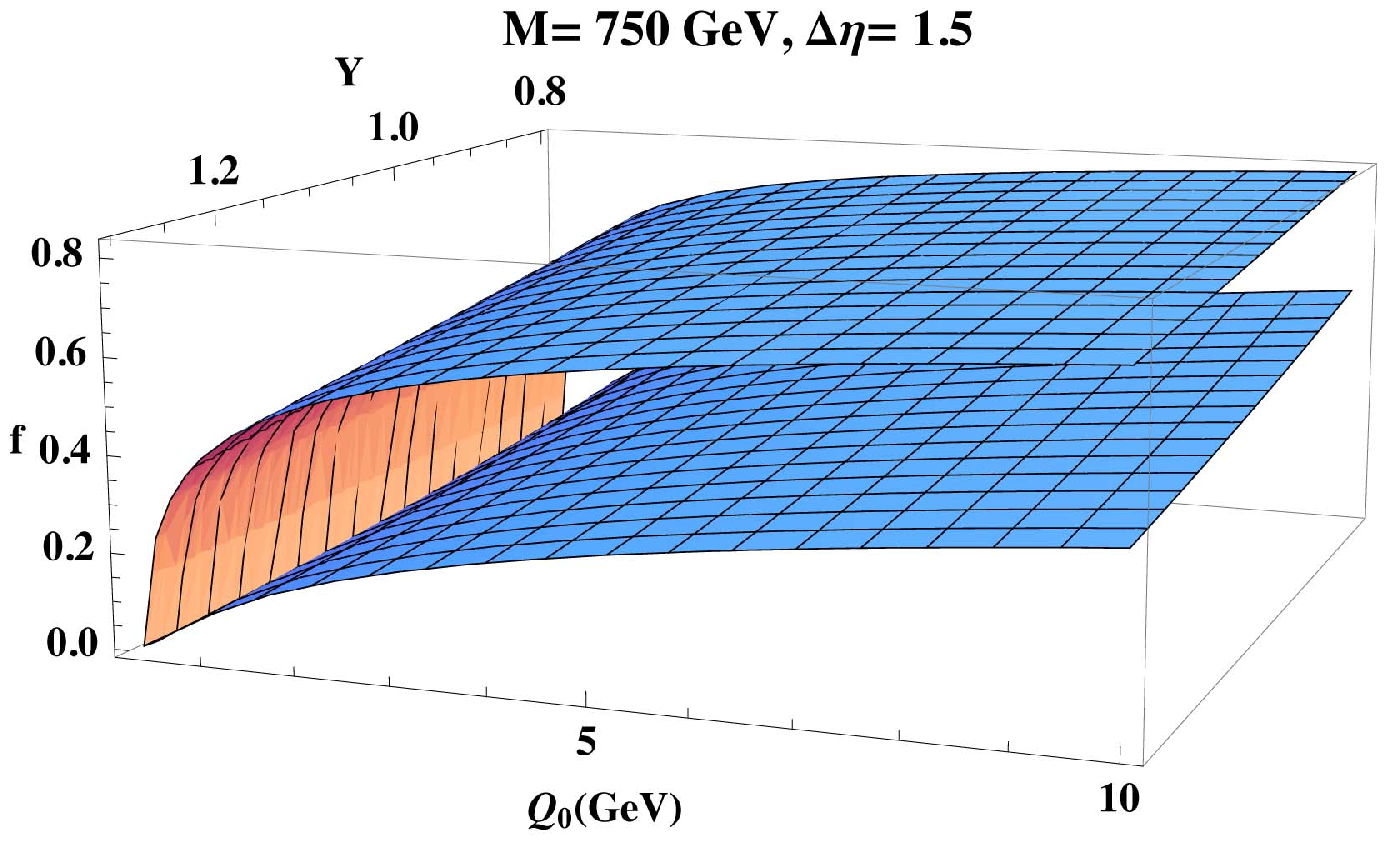}
\label{fig:750f15}
}
\subfigure[]{
\includegraphics[scale=0.545]{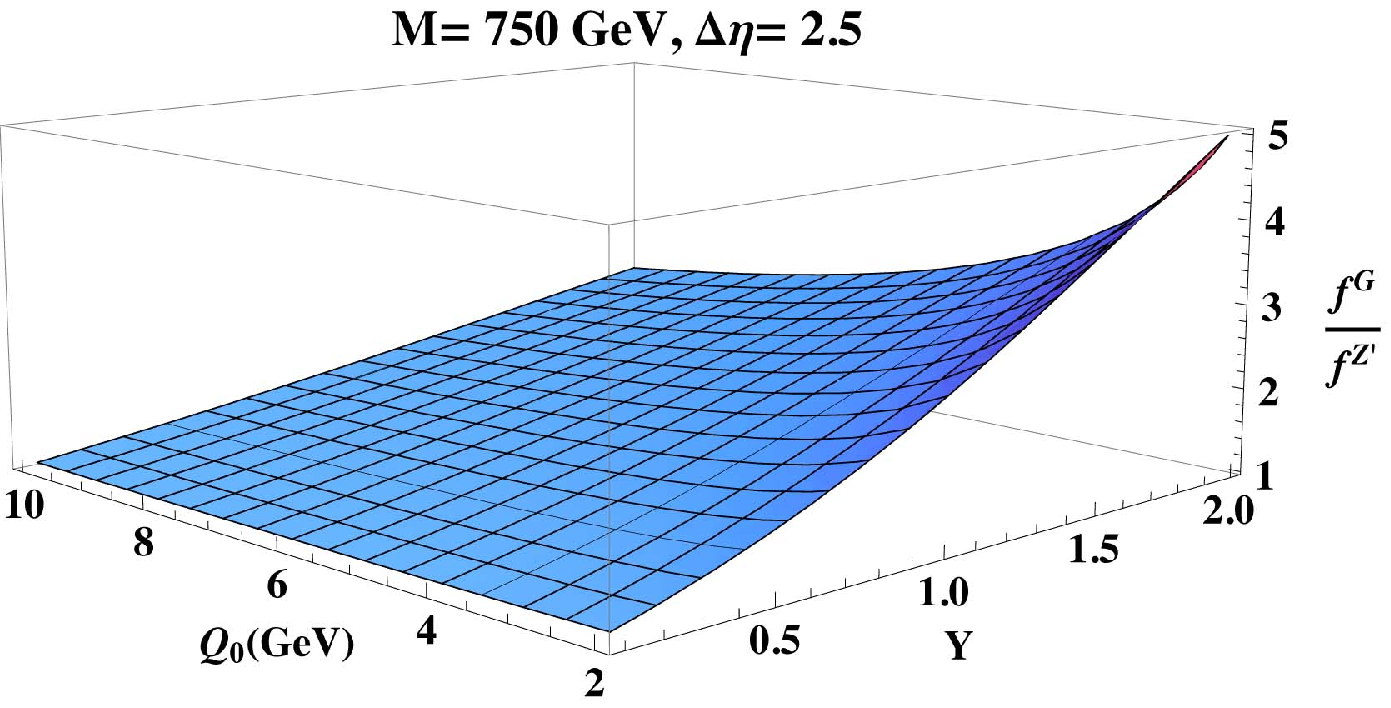}
\label{fig:175025QY}
}
\subfigure[]{
\includegraphics[scale=0.54]{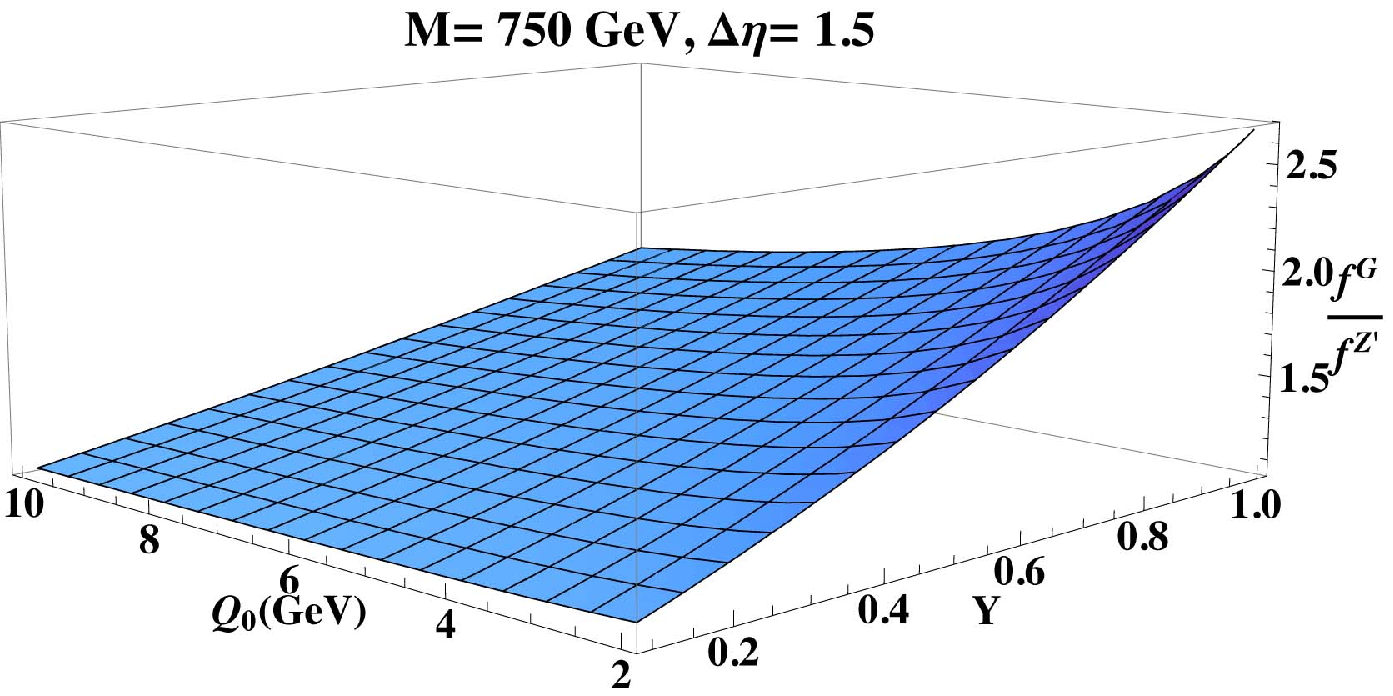}
\label{fig:175015QY}
}
\caption[ad]{
Same as Fig.~\ref{fig:150035}, for $M=750~$GeV, in the allowed region in 
gap size $Y$.
}
\label{fig:75025}
\end{figure}

We can interpret these results as follows. 
As we saw in Eq.~(\ref{eq:ddd}) of Sec.~\ref{sec:sadm}, 
the eigenvectors, $e_1$ and $e_2$, of the soft anomalous dimension matrix 
for $q\bar{q} \to Q\bar{Q}$ are very close to 
a color singlet and a color octet 
for the kinematics and the gap geometry we study in this section. 
For this reason, we have called them 
``quasi-singlet" and ``quasi-octet" eigenvectors, respectively.
Consequently, this leads the hard functions in Eqs.~(\ref{eq:hg}) and (\ref{eq:hz}) 
to retain approximately their color basis dependence 
even after transforming the functions to the diagonal basis.

Following the above property of the hard functions and the eigenvalues, 
shown in Fig.~\ref{fig:real} for $q\bar{q} \to Q\bar{Q}$, 
the quasi-octet component of the hard function for a color octet resonance 
is the leading component, with a corresponding small exponent, $E_{22}^{(\f)}$.
We thus find that the quasi-octet cross section 
$(H^{(\f,LO)}_{G})_{22 } S^{(\f,0)}_{ 22} 
\left[ 
\frac{ \ln \left( \frac{Q_0}{\Lambda} \right) 
}{
 \ln \left( \frac{p_T}{\Lambda} \right) }
 \right]^{E^{(\tf)}_{22}}$ 
is dominant in the partonic gap cross section~(\ref{eq:qct}) 
for an octet resonance~$G$.
The quasi-octet cross section is enhanced 
due to its large overlap with the color-octet basis 
which gives a small exponent $E_{22}<1$.
It is clear that 
this term produces the ``convex" shapes of the dashed gap fraction curves~($f_{gap}$) 
of the log plot on the $x$-axis, as shown 
in Fig.~\ref{fig:fractiont}, and the upper surfaces in Figs.~\ref{fig:150035} and \ref{fig:75025} for top pair production.
These curves rise rapidly at low $Q_0$.

The singlet resonance process, on the other hand, 
has a larger overlap with the quasi-singlet 
component of the hard function. 
The eigenvalue of the quasi-singlet is larger than one, 
and $E_{11}^{(\f)}=(4 / \beta_0) \mbox{Re} \, \lambda^{(\f,1)}_1>1$.
Thus, the quasi-singlet cross section, 
$(H^{(\f,LO)}_{Z'})_{11 } S^{(\f,0)}_{ 11} 
\left[ 
\frac{ \ln \left( \frac{Q_0}{\Lambda} \right) 
}{
 \ln \left( \frac{p_T}{\Lambda} \right) }
 \right]^{E^{(\tf)}_{11}}$, 
 is suppressed at low $Q_0$,
explaining the ``concave" shapes of the solid lines
in Fig.~\ref{fig:fractiont} 
and the lower surfaces in 
Figs.~\ref{fig:150035} and \ref{fig:75025} for top pair production.
These increase slowly at low $Q_0$.
The differential partonic cross sections with respect to $Q_c$ 
are presented in Fig.~\ref{fig:dd2} for octet and singlet resonance processes 
for values $M=1.5~$TeV, $\de=2.5$, and $Y=1.5$. 
The behavior of the exponents corresponds to  
the dominance of the octet resonance process 
at low $Q_c$ in Fig.~\ref{fig:dd2}.

\begin{figure}[fhptb]
\begin{center}
\includegraphics[width=.55\hsize]{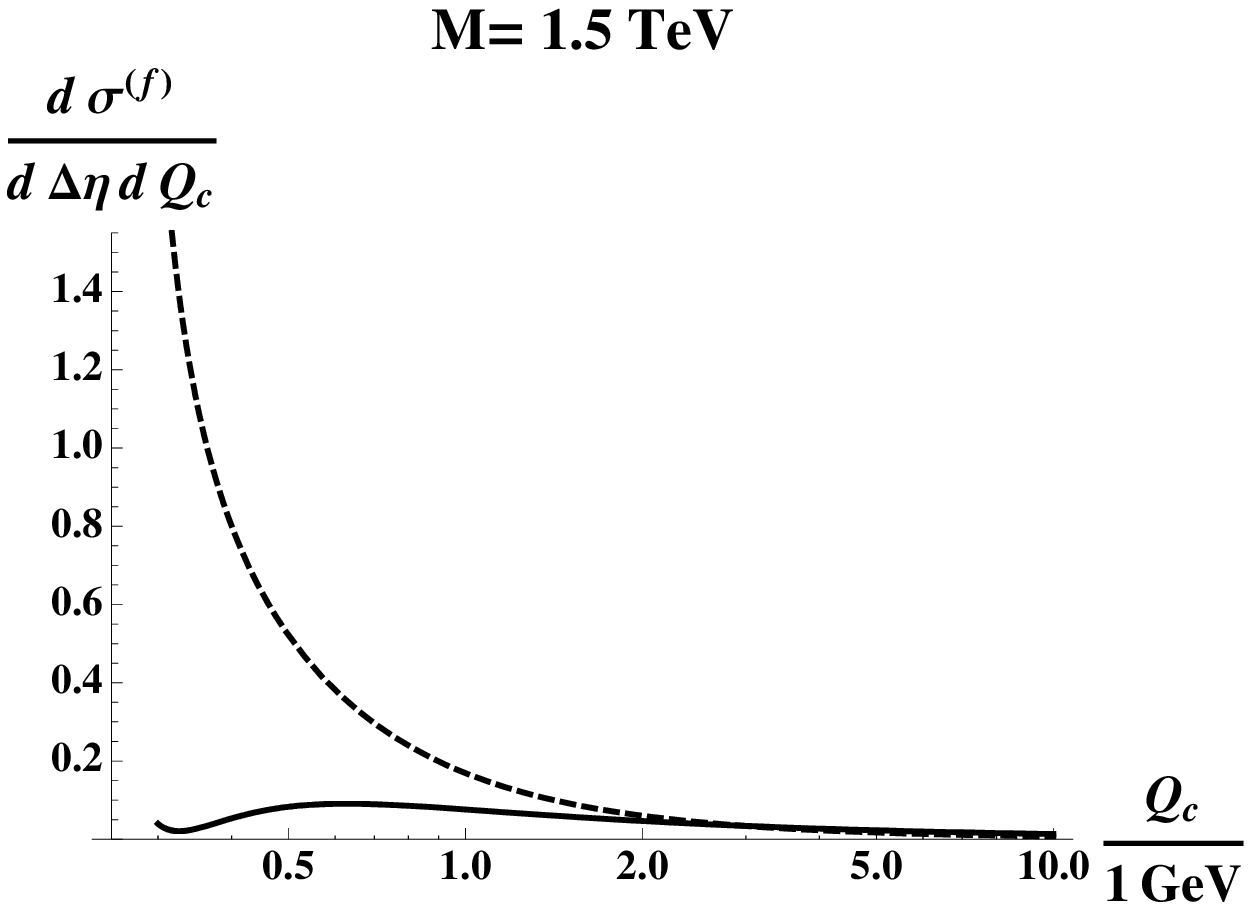}
\end{center}
\caption{
The cross section from the octet resonance~(dashed line) 
and the singlet resonance~(solid line) for $M=1.5~$TeV, $\de=2.5$, and $Y=1.5$. 
Both distributions are normalized to the same area for the range of 
$\Lambda < Q_c < p_T$.}
\label{fig:dd2}
\end{figure}

The gap fraction depends on the quark mass, $m_Q$, 
through the variable $\rho$ in Eq.~(\ref{eq:rho}).
The dependence can be seen in Fig.~\ref{fig:fractiont} by comparing the black curves~(solid and dashed) 
for a top pair with the blue curves~(dotted and dot-dashed) 
for a bottom pair.
For the $b$ quark, 
the gap fractions are slightly smaller at fixed gap threshold energy 
than for a top quark pair for the relatively heavy resonances in Fig.~\ref{fig:fractiont}.
This difference is, however, larger for 
smaller resonance mass, as also shown in Fig.~\ref{fig:fractiont}. 
We note that the quark mass dependence on the gap fraction
for bottom quark pair production
 is negligible 
even for a resonance of relatively low mass, but not for a top pair.
For instance, we obtain $\rho=1.0003$ for 
 bottom pair production via a resonance of $M=550~$GeV with $\de=2$,
while $\rho=3.8200$ for a top pair.
This results in the very different shapes of the black curves compared to the blue curves
in Fig.~\ref{fig:550a}.

The results in this section show
that the measurement of gap fractions can be an effective tool to identify the color content of 
a spin-1 $s$-channel resonance. In the next section, we study gap fractions 
for spin-0 or 2 resonance processes, which contain more partonic channels.

\section{Gap Fractions for Spin-0 or 2 Resonances}
\label{sec:spins}
As we discussed briefly in Sec.~\ref{sec:hard}, 
in several models of new physics, there are 
$s$-channel spin-0 and spin-2 resonances, 
decaying into a heavy quark pair, which can be produced, 
through gluon-induced processes in addition to quark-antiquark reactions.
We readily recognize that 
spin-0 or 2 resonance processes might lead to 
a different pattern of soft gluon radiation 
between the Tevatron and the LHC, 
because of the presence of both quark and 
gluon-induced partonic processes.
As we shall see, PDFs no longer cancel in the leading order gap fraction.
This is distinguished from the analysis in Sec.~\ref{sec:spin1} 
for spin-1 resonances, involving only one partonic channel.
The results in Sec.~\ref{sec:spin1} are energy-independent, at leading order,
applying to both the Tevatron and the LHC.

Let's assume that there are two partonic processes 
that produce a  heavy resonance, 
decaying into a heavy quark pair,
$\mbox{f}_1: q \bar{q} \to V' \to Q\bar{Q}$ 
and 
$\mbox{f}_2: gg  \to V' \to Q \bar{Q}$.
In addition, we assume for simplicity that
the coupling of light quarks is universal.
Recall that the process f$_2$ cannot occur for a spin-1 resonance in general.
Therefore, we consider the octet resonance as spin-0 or 2.
In this case, we obtain the gap fraction, following Eq.~(\ref{eq:fraction}), 
\bea
f_{gap}^{(LO)}  
&=&
\frac{
\int d x_1 d x_2 
\left( 
\sum_{f_1 f_2} \frac{d \hat{\sigma}^{(\tf_1)}(Q_0) }
{d \de \, d M^2 \, d \eta_{V'} } 
\phi_{f_1/A}(x_1) \phi_{f_2/B}(x_2)
+ \frac{d \hat{\sigma}^{(\tf_2)}(Q_0) }{d \de \, d M^2 \, d \eta_{V'}  }  
\phi_{g/A}(x_1) \phi_{g/B}(x_2)
\right)
}
{
 \int d x_1 d x_2 
 \left(  
 \sum_{f_1 f_2} \frac{d \hat{\sigma}^{(\tf_1,LO)}} 
 {d \de \, d M^2 \, d \eta_{V'}   } 
\phi_{f_1/A}(x_1) \phi_{f_2/B}(x_2)
+ \frac{d \hat{\sigma}^{(\tf_2,LO)}}{d \de \, d M^2 \, d \eta_{V'} } 
\phi_{g/A}(x_1) \phi_{g/B}(x_2)
\right)
}
\nn \\
&=&
p^{q\bar{q} \to Q\bar{Q}} f^{q\bar{q} \to Q\bar{Q}}_{gap} 
+ p^{gg \to Q\bar{Q}} f^{gg \to Q\bar{Q}}_{gap} 
\, ,
\label{eq:gapsp}
\eea
where the $p^{\f_i}$ are defined by
\bea
p^{q\bar{q} \to Q\bar{Q}} 
&\equiv&
\frac{
\frac{d \hat{\sigma}^{(\tf_1,LO)}}{d \de  } 
\left( \sum_{f_1 f_2}
\phi_{f_1/A}(x_1) \phi_{f_2/B}(x_2) 
\right) 
}
{
\frac{d \hat{\sigma}^{(\tf_1,LO)}}{d \de  } 
\left( 
\sum_{f_1 f_2} \phi_{f_1/A}(x_1) \phi_{f_2/B}(x_2)   
\right)
+\frac{d \hat{\sigma}^{(\tf_2,LO)}}{d \de  } 
\phi_{f_g/A}(x_1) \phi_{f_g/B}(x_2)
} \, ,
\nonumber \\
p^{gg \to Q\bar{Q}} 
&\equiv&
\frac{
\frac{d \hat{\sigma}^{(\tf_2,LO)}}{d \de  } 
\phi_{g/A}(x_1) \phi_{g/B}(x_2) 
}
{
\frac{d \hat{\sigma}^{(\tf_1,LO)}}{d \de  } 
\left( 
\sum_{f_1 f_2} \phi_{f_1/A}(x_1) \phi_{f_2/B}(x_2)   
\right)
+\frac{d \hat{\sigma}^{(\tf_2,LO)}}{d \de  } 
\phi_{f_g/A}(x_1) \phi_{f_g/B}(x_2)
} \, ,
\eea
and the gap fractions,  $f^{q\bar{q} \to Q\bar{Q}}_{gap}$
and $ f^{gg \to Q\bar{Q}}_{gap} $, are given by
\bea
 f^{q\bar{q} \to Q\bar{Q}}_{gap} (Q_0) 
 &=&
 \frac{\frac{d \hat{\sigma}^{(\tf_1)}(Q_0)}{d \de  }  }
{\frac{d \hat{\sigma}^{(\tf_1,LO)}}{d \de  }  
} \, , \, \, \,\, 
 f^{gg \to Q\bar{Q}}_{gap} (Q_0) 
 =
 \frac{\frac{d \hat{\sigma}^{(\tf_2)}(Q_0)}{d \de  }  }
{\frac{d \hat{\sigma}^{(\tf_2,LO)}}{d \de  }  
} \, ,
\eea
which can be calculated perturbatively, 
following Eqs.~(\ref{eq:qc3}) and (\ref{eq:LO}).
We can interpret the $p^{\f_i}$ as the probability for the process f$_i$ 
among the channels, $p^{q\bar{q} \to Q\bar{Q}}+ p^{gg \to Q\bar{Q}} =1$.

We can study the above gap fraction
for representative choices of $p^{q\bar{q} \to Q\bar{Q}}$ 
and $p^{gg \to Q\bar{Q}}$. 
For the first example,  
we compare the gap fraction for an octet resonance
 with the gap fraction for a singlet resonance through a pure gluon-induced process, 
 $p^{gg \to Q\bar{Q}}=1$ and  $p^{q \bar{q} \to Q\bar{Q}}=0$.
 A pseudo-scalar or a boson-phobic scalar resonance is a good example for the above  resonance process, 
 where gluon-induced processes are dominant and the branching ratio to a top pair 
 can be taken to be unity~\cite{Frederix:2007gi}. 
In general, SUSY models with this feature can be constructed for singlet resonances~\cite{Djouadi:2005gj}, and split SUSY scenarios can explain octet resonances from 
bound states of meta-stable gluinos~\cite{Cheung:2004ad}.

The gap fractions for the pure gluon-induced resonances are  
determined by the soft anomalous dimension matrix 
in Eq.~(\ref{eq:adg}), its eigenvalues and eigenvectors for the process 
$gg \to Q\bar{Q}$. The resulting gap fractions for 
octet and singlet resonances
 are illustrated in Fig.~\ref{fig:puregg2} for $M=1.5~$TeV, 
$\de=2.5$, and $Y=1$.

To study the effect of an admixture of quark- and gluon-induced processes, 
the gap fraction for an octet resonance with 
$p^{gg \to Q\bar{Q}} =0.8$ and $p^{q\bar{q} \to Q\bar{Q}}=0.2$   
is represented in Fig.~\ref{fig:admix} as a dashed curve in comparison to 
the gap fraction for the pure gluon-induced singlet resonance, shown as 
a solid curve in the same figure.  
This quark/gluon admixture  for an octet resonance can occur for 
the reggeon resonances of Ref.~\cite{Perelstein:2009qi}.

\begin{figure}[fhptb]
\centering
\subfigure[]{
\includegraphics[scale=0.62]{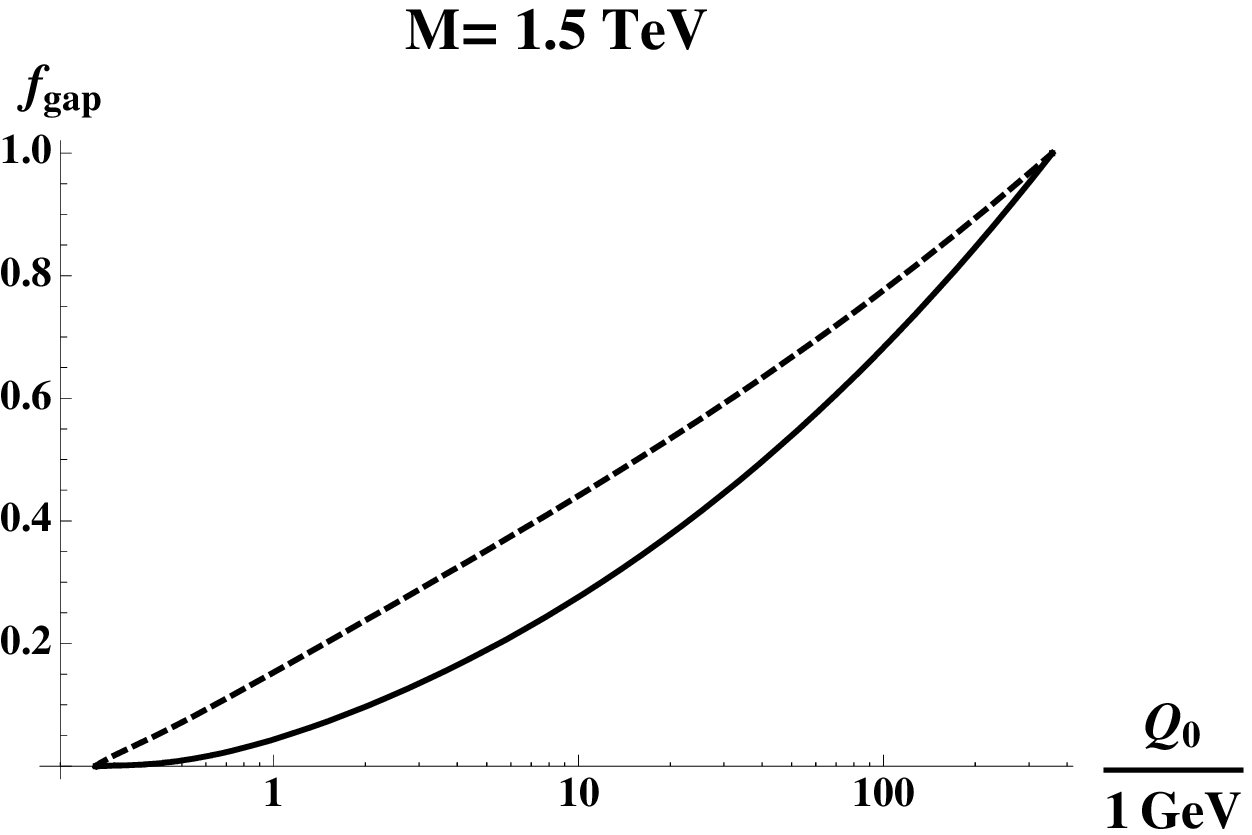}
\label{fig:puregg2}
}
\subfigure[]{
\includegraphics[scale=0.62]{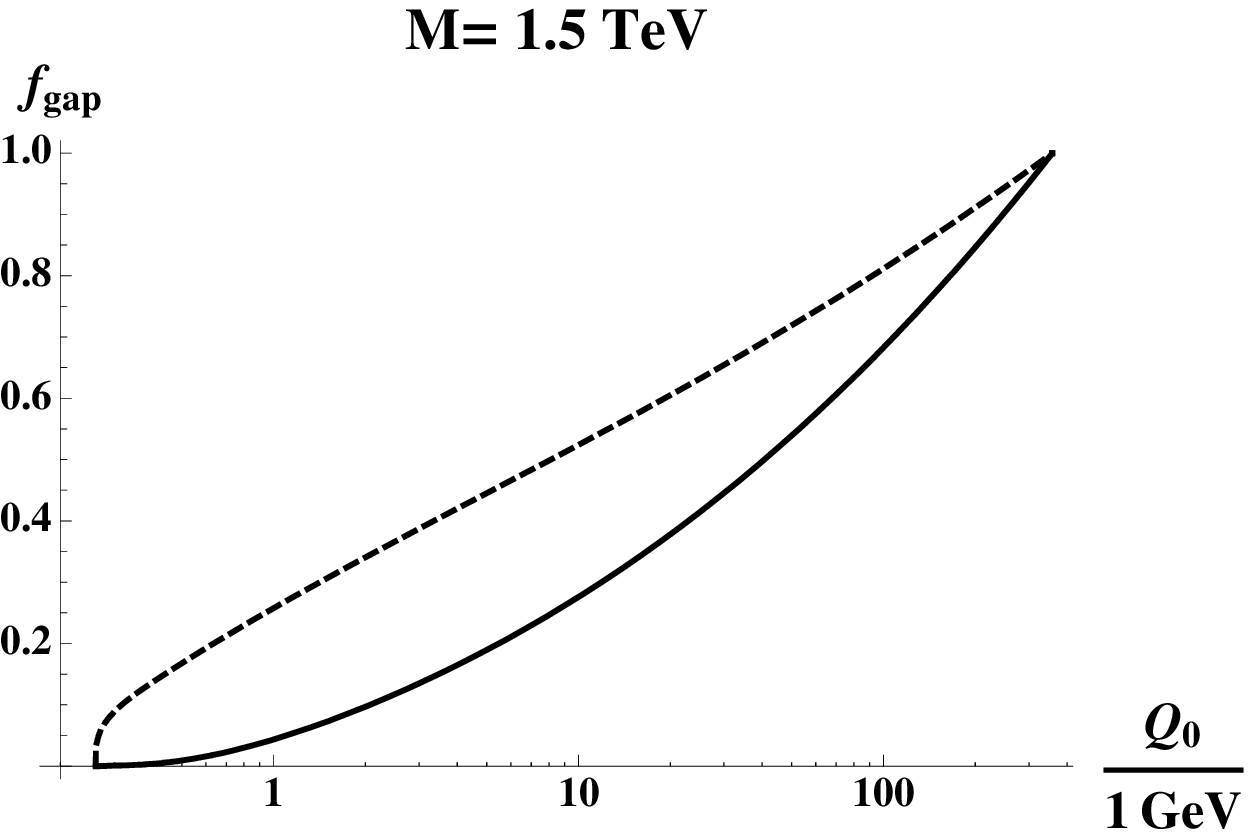}
\label{fig:admix}
}
\caption[ad]{
Gap fraction as a function of gap energy threshold $Q_0$ at $\de=2.5$ with 
$M=1.5~$TeV and $Y=1$. 
In (a) and (b), the solid lines describe the gap fraction through a $Z'$ resonance (color-singlet) for $p^{gg \to Q\bar{Q}}=1$. 
The dashed line describes the gap fraction through a $G$ resonance (color-octet) in (a) for $p^{gg \to Q\bar{Q}}=1$ and in (b) for $p^{gg \to Q\bar{Q}}=0.8$ and $p^{q\bar{q} \to Q\bar{Q}}=0.2$.
}
\label{fig:fractiontl2}
\end{figure}

\begin{figure}[fhptb]
\centering
\includegraphics[scale=0.64]{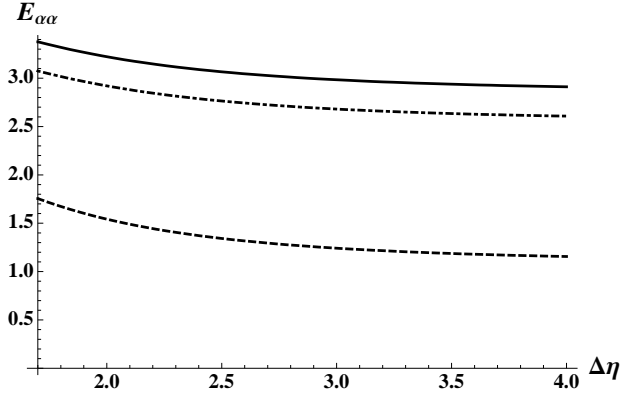}
\caption[ad]{
Plot of the exponents, $E_{\alpha \alpha}^{(\f)}$, 
of the soft anomalous dimension matrix for $gg \to Q\bar{Q}$ 
as a function of $\de$
for $Y=1.5$ and $m_Q=m_t$
with resonance mass $M=1.5~$TeV. 
The solid line identifies the quasi-singlet exponent, 
the dashed and the dot-dashed lines the two quasi-octets.
}
\label{fig:expgluon}
\end{figure}

\begin{figure}[fhptb]
\centering
\subfigure[]{
\includegraphics[scale=0.64]{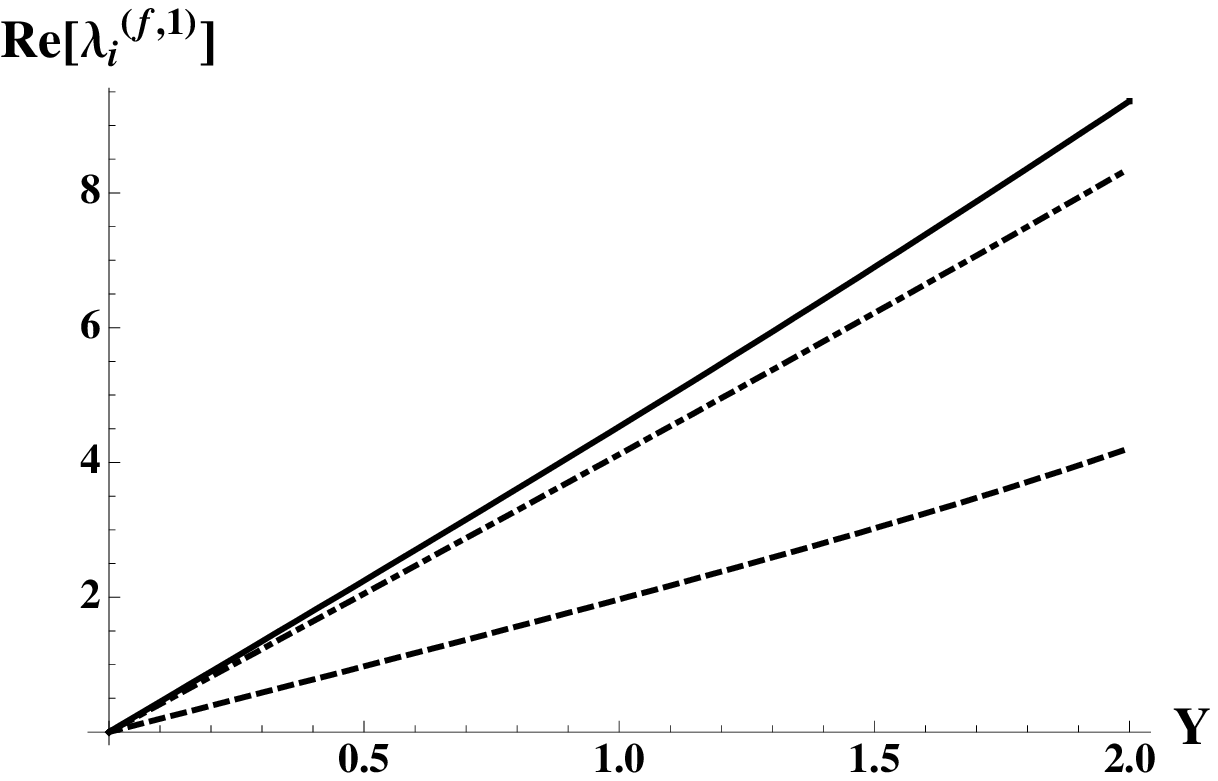}
\label{fig:real2}
}
\subfigure[]{
\includegraphics[scale=0.64]{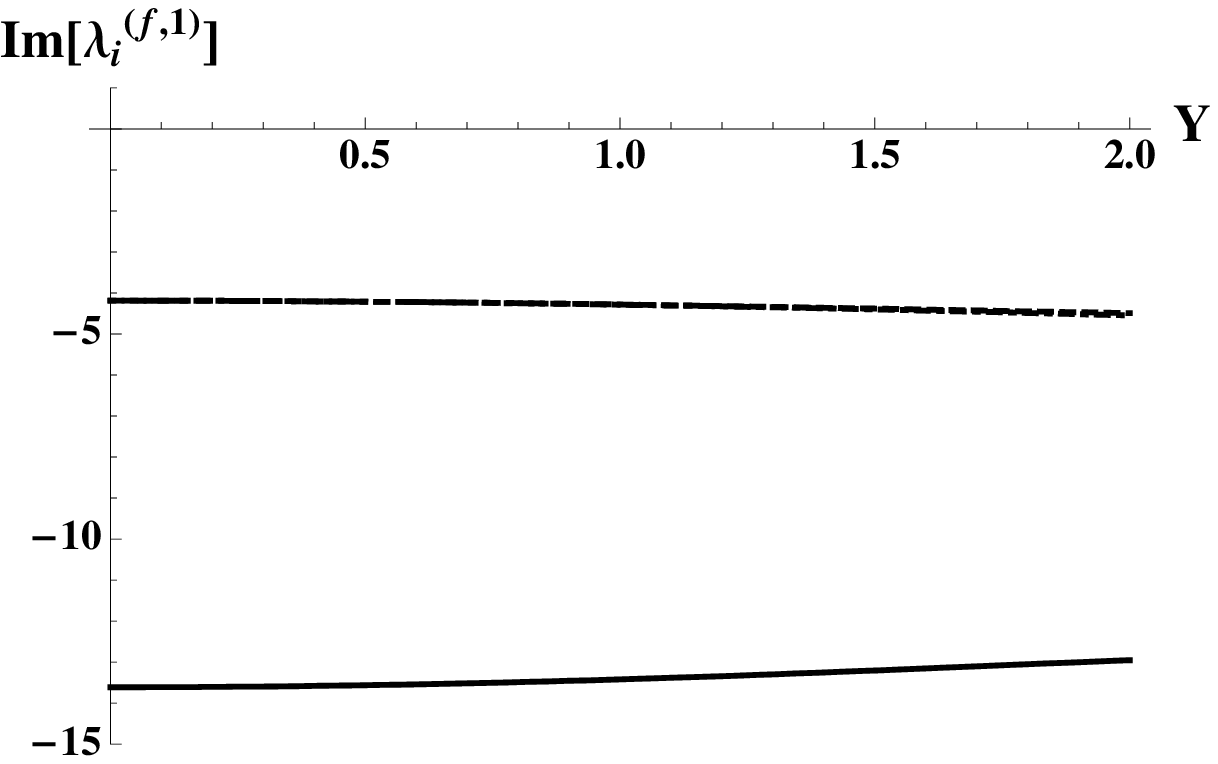}
\label{fig:im2}
}
\caption[ad]{
Plot of the real~(a) and imaginary~(b) parts of 
the eigenvalues of the soft anomalous dimension matrix 
for $gg \to Q\bar{Q}$ of $M=1.5~$TeV, $\de=2.5$ and $m_Q=m_t$.
The solid line identifies the quasi-singlet eigenvalue, 
the dashed and the dot-dashed lines the two quasi-octets.
}
\label{fig:eigeng}
\end{figure}

As observed in the previous section, 
we again see  in Fig.~\ref{fig:puregg2} 
that there is ``less" radiation into the gap region for 
 the color-octet resonance process. 
At the same time, we notice that for the same resonances there is ``more" radiation 
for gluon-induced processes 
than for quark-induced processes.  
This can be seen by comparing solid and dashed curves in Fig.~\ref{fig:fractiont} with the curves in Fig.~\ref{fig:puregg2}. 

Scattered gluons tend to emit more radiation, 
compared to scattered quarks.
We note that the shape of the gap fraction for gluon-induced octet resonances in Fig.~\ref{fig:puregg2} 
is ``linear", while the shape for quark-induced octet resonances in Fig.~\ref{fig:fractiont} is ``convex". 
Following the arguments in the previous section, 
the values of the exponents for gluon-induced processes in Fig.~\ref{fig:expgluon} 
can account for the difference.
The eigenvalues for the process $gg \to Q \bar{Q}$ in Fig.~\ref{fig:eigeng} 
ensure that all exponents
$E_{\alpha \beta}^{(gg \to Q \bar{Q})}$ are larger than one for $Y$ of order unity, 
as shown in Fig.~\ref{fig:expgluon}. 
We note that two quasi-octet eigenvectors, $e_2$ and $e_3$, 
of $\ad^{(gg \to Q\bar{Q})}$ in Eq.~(\ref{eq:adg}), which we recall are determined numerically, 
turn out to approach $e_2$ and $e_3$ in Eq.~(\ref{eq:e1e2e3}), obatined in large $N_c$ limit.  
As a result, both the quasi-octet components, $(H'_G)_{22}$ and $(H'_G)_{33}$, of the octet hard function, written in the basis that diagonalizes $\ad^{(1)(gg \to Q\bar{Q})}$, are important.
 Their exponents, $E_{22}^{(gg \to Q \bar{Q})}$ and $E_{33}^{(gg \to Q \bar{Q})}$,  are shown as the dashed and the dot-dashed lines in Fig.~\ref{fig:expgluon}.
An interesting feature is that one of the quasi-octet exponents is close to the exponent for 
a quasi-singlet.
A semi-numerical expression for the cross section~(\ref{eq:qc3}) for gluon-induced octet resonances is given for $M=1.5~$TeV, $Y=1$, and $\de=1.5$ by
\bea
\frac{ d \hat{\sigma}_G^{(gg\to Q\bar{Q})}}{d \de } 
&=& \sum_{\beta, \gamma} (H_G^{'(\f_2,LO)})_{\beta \gamma}
\, S^{(\f_2,0)}_{\gamma \beta}
\left[ \frac{ \ln \left( \frac{Q_0}{\Lambda} \right) }
{ \ln \left( \frac{p_T}{\Lambda} \right) } \right]^{E^{(\tf)}_{\gamma \beta}} 
\label{eq:semi} 
 \\
&\simeq& h_G' \left( (0.23) \left[ \frac{ \ln \left( \frac{Q_0}{\Lambda} \right) }
{ \ln \left( \frac{p_T}{\Lambda} \right) } \right]^{2.27} + (6.19) \left[ \frac{ \ln \left( \frac{Q_0}{\Lambda} \right) }
{ \ln \left( \frac{p_T}{\Lambda} \right) } \right]^{2.08} 
+(6.33) \left[ \frac{ \ln \left( \frac{Q_0}{\Lambda} \right) }
{ \ln \left( \frac{p_T}{\Lambda} \right) } \right]^{1.17} \right) \, , \nn  
\eea
where the first coefficient, 0.23, of second line in Eq.~(\ref{eq:semi}) is $(H_G^{'(\f_2,LO)})_{11}
 S^{(\f_2,0)}_{11}$, and the other two, 6.19 and 6.33, are 
 $(H_G^{'(\f_2,LO)})_{22}
 S^{(\f_2,0)}_{22}$ and $(H_G^{'(\f_2,LO)})_{33}
 S^{(\f_2,0)}_{33}$, respectively.
 Here $h'_G$ has been defined in Eq.~(\ref{eq:hg2}).
In Eq.~(\ref{eq:semi}), we omit off-diagonal contributions such as $(H_G^{'(\f_2,LO)})_{23}
 S^{(\f_2,0)}_{32}  $, because they contribute less than 2\% of the total.
We observe that each $Q_0$-dependent factor,  
$\left[ 
 \ln \left( \frac{Q_0}{\Lambda} \right) 
/
 \ln \left( \frac{p_T}{\Lambda} \right) 
 \right]^{E^{(gg \to Q \bar{Q})}_{\alpha \beta}}$,
 is thus suppressed, relative to linear behavior in a log plot at low $Q_0$, for all quasi-octet exponents 
 in the partonic cross section~(\ref{eq:semi}).
 This explains why the dashed curve in Fig.~\ref{fig:puregg2}
 increases slowly at low $Q_0$ 
 compared to the dashed curves in Fig.~\ref{fig:1600_1}.

In Fig.~\ref{fig:fractiontl2}, 
 the dashed curve for a pure gluon-induced process in (a) shows a 
 smaller gap fraction than the dashed curve 
 for a mixed quark-gluon process in (b) for given $Q_0$.
As we discussed in the previous section, 
the gap fraction for ${q\bar{q} \to Q\bar{Q}}$ rises rapidly at low $Q_0$.
The term, $f_{gap}^{q \bar{q} \to Q \bar{Q}}$, 
even with the small assumed probability $p^{q \bar{q} \to Q \bar{Q}}=0.2$  thus explains
the difference of the dashed curves  
in Fig.~\ref{fig:fractiontl2}.
We note that the shape of the solid curves for $gg \to Z' \to Q\bar{Q}$ in Fig.~\ref{fig:fractiontl2} 
corresponds to the largest amount of radiation among the resonance processes we have studied in this paper. 
This is due to the large quasi-singlet exponent $E_{11}^{(gg \to Q \bar{Q})}$, compared to other exponents. 

\begin{figure}[fhptb]
\centering
\subfigure[$g g \to Z' \to Q \bar{Q}$]{
\includegraphics[scale=0.8]{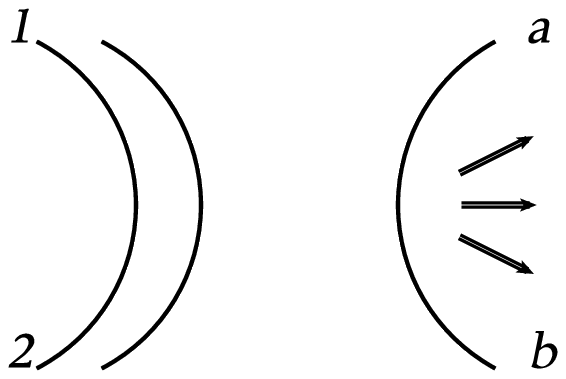}
\label{fig:aggZQQ}
}
\hspace{0.5in}
\subfigure[$g g \to G \to Q \bar{Q}$]{
\includegraphics[scale=0.8]{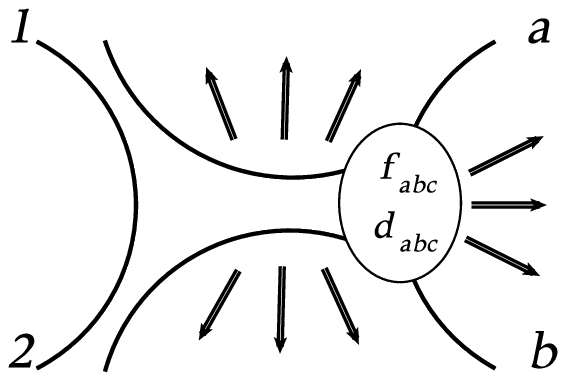}
\label{fig:aggGQQ}
}
\\
\subfigure[$q \bar{q} \to Z' \to Q \bar{Q}$]{
\includegraphics[scale=0.8]{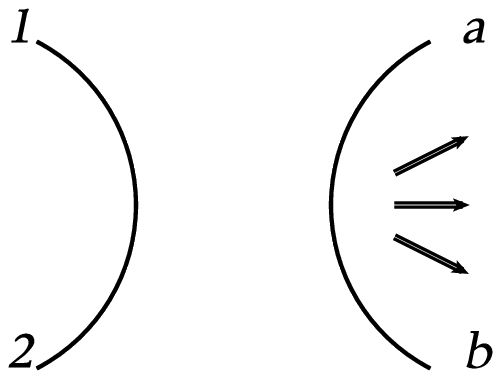}
\label{fig:aqqZQQ}
}
\hspace{0.5in}
\subfigure[$q \bar{q} \to G \to Q \bar{Q}$]{
\includegraphics[scale=0.8]{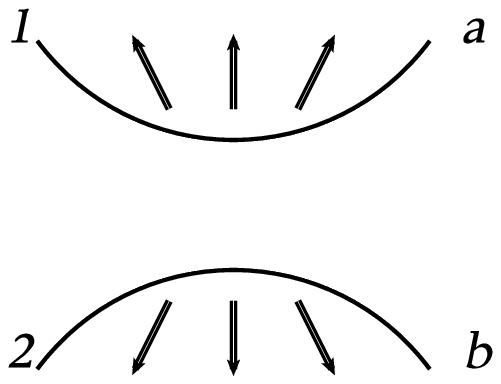}
\label{fig:aqqGQQ}
}
\caption[ad]{
Color dipole configurations for resonance processes, 
$f_1(p_1) + f_2(p_2) \to V' \to Q(p_a) + \bar{Q}(p_b)$. 
Arrows in each figure represent the enhanced direction of radiation.
}
\label{fig:sum}
\end{figure}

The above results, depending on partonic channels 
and the gauge content of resonances, 
can be understood intuitively in terms of color dipole configurations.
In Fig.~\ref{fig:sum}, the color dipole structures and 
the preferred directions of radiation are represented as solid curves and arrows, 
respectively.
Each figure describes the surplus of radiation inside a dipole~\cite{Dokshitzer:1988bq,Dokshitzer:1991wu,Ellis:1991qj}. 
For examples, the dipole configurations in Figs.~\ref{fig:aggZQQ}  
 and \ref{fig:aqqZQQ} 
allow more radiation into the 
$Q\bar{Q}$ region, which overlaps the gap region, 
while the configuration for $q\bar{q} \to G \to Q\bar{Q}$ in Fig.~\ref{fig:aqqGQQ} describes enhanced radiation in the $qQ$ and the $\bar{q} \bar{Q}$ regions that are outside the gap region. 
In Fig.~\ref{fig:aggGQQ}, the color-(anti)symmetric couplings control the dipole configuration, which results in radiation into both the $qQ$ and the $Q\bar{Q}$ regions. 
We have also predicted the least(most) amount of radiation into the gap region for $q \bar{q} \to G \to Q \bar{Q}$~($g g \to Z' \to Q \bar{Q}$). 
These color dipole configurations help explain our results for the patterns of radiation,
associated with the SU(3) color representation of resonances and the partonic channels involved.

At the end of Sec.~\ref{sec:sadm}, 
we argued that for a narrow resonance, generally a gauge singlet, 
there is an additional contribution 
when the radiation energy scale is larger than $\sqrt{\Gamma M}$.
In this limit, we expect that the color dipole configuration is further dominant   
since the process becomes a pure color singlet decay.
This contribution is expected to be absent for an octet resonance 
since the width is in general large.
We point out that this may result in more radiation for a very narrow singlet resonance process 
compared to our results in Secs.~\ref{sec:spin1} and \ref{sec:spins}, 
leading an even more distinguishable difference from an octet resonance process.
The details of this case are left for future work.

\section{Conclusion}

In this paper, we have shown that it may be possible to determine 
the color SU(3) representation of resonances from new physics signals
by analyzing the distribution of soft radiation into a rapidity gap.
The results we have found are based on  
perturbative calculations of factorized partonic cross sections, 
and apply for ordered scales, $\Lambda_{QCD}<Q_c<\sqrt{\Gamma M} \sim p_T$ with energy flow, $Q_c$, resonance mass, $M$, and decay width, $\Gamma$. 
These conditions correspond to a broad width resonance process.
 The gap fractions for octet and singlet resonances, 
which we defined in Sec.~\ref{sec:gapfraction},  
show different radiation patterns 
of energy flow.
To obtain an analytical form of soft gluon emission 
in heavy quark pair production, 
the massive soft anomalous dimension 
matrix for rapidity gap events was introduced.
The results, in general, describe more radiation for singlet than for octet resonances, 
which can be explained in terms of color dipole configurations. 
Especially, for spin-1 resonance production, involving only one partonic sub-process, 
we obtained a relatively large difference in radiation into the gap region 
between color singlet and octet resonances.
The results at leading order do not require a convolution with PDFs, and are the same 
at the Tevatron and the LHC in this case. 
At the end of Sec.~\ref{sec:spins}, 
we discussed the consequences of a narrow width resonance, 
expecting more radiation for a singlet resonance process compared to 
our results based on a broad resonance assumption. 

In principle, following Ref.~\cite{Berger:2001ns},
given any fixed set of final-state partons, 
we may calculate the distribution of energy flow 
into any fixed region of rapidity and azimuthal angle 
rather than the simple rapidity gap we defined in this paper.
This may allow not only 
to illuminate more distinguishable features of color flow, depending on 
the gauge content of resonances, 
but also to shed valuable light on the dynamics of QCD radiation itself.

\section*{Acknowledgements}
I would like to express my gratitude to George Sterman, 
for many conversations and comments on the manuscript, and for all his help.
I am indebted to Gilad Perez and Zuowei Liu for their valuable advice, 
and to Leo Almeida, Rikkert Frederix and Alex Mitov for discussions.
This work was supported in part by the National Science Foundation, PHY-0354776, PHY-0354822 and PHY-0653342.

%%%%%%%%%%%%%%%%%%%%%%%%

\section*{Appendix}
\setcounter{section}{0}
\renewcommand{\thesection}{\Alph{section}}

%%%%%%%%%%%%%%%%%%%%%%%%%%%%
\renewcommand{\theequation}{A.\arabic{equation}}
\setcounter{equation}{0}
%%%%%%%%%%%%%%%%%%%%%%%%%%%%

In general, we represent the four momentum of a particle of mass $m_Q$ as
\bea
p^{\mu}= \left( m_T \cosh \eta, \, p_x, \, p_y \, , m_T \sinh \eta \right) \, ,
\eea
where $\eta$ is rapidity and where $m_T$ is a transverse mass defined by
\bea
m_T^2 = m_Q^2+ p_x^2 + p_y^2 = m_Q^2 + p_T^2 \, .
\eea
In the $\eta$-$\phi$ coordinate system of rapidity and azimuthal angle, 
four momentum $p^{\mu}$ can be written in terms of 
quark mass $m_Q$ and $\phi$ as
\bea
p^{\mu}= \left(\sqrt{p_T^2+m_Q^2} \cosh \eta, \, p_T \sin \phi, \, p_T \cos \phi, \,  \sqrt{p_T^2+m_Q^2} \sinh \eta \right) \, .
\eea
To describe the process in Eq.~(\ref{eq:process}) with the gap geometry, shown in Fig.~\ref{fig:gap} at lowest order, we need five external partons of momenta 
\bea
p_1^{\mu}&=& \frac{\sqrt{\hat{s}}}{2} \left(1,\, 0,\, 0,\, 1\right) \, , \nn \\
p_2^{\mu}&=& \frac{\sqrt{\hat{s}}}{2} \left(1,\, 0,\, 0,\, -1\right) \, , \nn \\
p_a^{\mu}&=&\left(\sqrt{p_T^2+m_Q^2} \cosh \left( \frac{\da}{2}\right), \, 0, \, p_T, \,  \sqrt{p_T^2+m_Q^2} \sinh \left( \frac{\da}{2}\right) \right) \, , \nn \\
p_a^{\mu}&=&\left(\sqrt{p_T^2+m_Q^2} \cosh \left( \frac{\da}{2}\right), \, 0, -\, p_T, \,  -\sqrt{p_T^2+m_Q^2} \sinh \left( \frac{\da}{2}\right) \right) \, , \nn \\
k^{\mu}&=&k_T \left( \cosh y, \, \sin \phi, \, \cos \phi, \, \sinh y \right) \, ,
\eea
where we are working in the partonic center of mass frame, and where we fix azimuthal angles for a quark pair at $\phi_a=0$ and $\phi_b=\pi$.
Here $y$ is rapidity of soft gluon~$k$.


\begin{thebibliography}{99}

\bibitem{Collins:1989gx}
  J.~C.~Collins, D.~E.~Soper and G.~Sterman,
  %``Factorization of Hard Processes in QCD,''
  Adv.\ Ser.\ Direct.\ High Energy Phys.\  {\bf 5}, 1 (1988)
  [arXiv:hep-ph/0409313];
  %%CITATION = 00319,5,1;%%  
  G.~Sterman,
  %``QCD and jets,''
  arXiv:hep-ph/0412013.
  %%CITATION = HEP-PH/0412013;%%
  
  \bibitem{Contopanagos:1997nh}
  H.~Contopanagos, E.~Laenen and G.~Sterman,
  %``Sudakov factorization and resummation,''
  Nucl.\ Phys.\  B {\bf 484}, 303 (1997)
  [arXiv:hep-ph/9604313].
  %%CITATION = NUPHA,B484,303;%%

  
\bibitem{Sen:1981sd}
  A.~Sen,
  %``Asymptotic Behavior Of The Sudakov Form-Factor In QCD,''
  Phys.\ Rev.\  D {\bf 24}, 3281 (1981).
  %%CITATION = PHRVA,D24,3281;%%

\bibitem{Botts:1989kf}
  J.~Botts and G.~Sterman,
  %``Hard Elastic Scattering In QCD: Leading Behavior,''
  Nucl.\ Phys.\  B {\bf 325}, 62 (1989).
  %%CITATION = NUPHA,B325,62;%%
  
  %\cite{Derrick:1995pb}
\bibitem{hera1}
  M.~Derrick {\it et al.}  [ZEUS Collaboration],
  %``Rapidity Gaps between Jets in Photoproduction at HERA,''
  Phys.\ Lett.\  B {\bf 369}, 55 (1996)
  [arXiv:hep-ex/9510012].
  %%CITATION = PHLTA,B369,55;%%
  
  %\cite{Chekanov:2006pw}
\bibitem{hera2}
  S.~Chekanov {\it et al.}  [ZEUS Collaboration],
  %``Photoproduction of events with rapidity gaps between jets at HERA,''
  arXiv:hep-ex/0612008.
  %%CITATION = HEP-EX/0612008;%%
  
  %\cite{Adloff:2002em}
\bibitem{hera3}
  C.~Adloff {\it et al.}  [H1 Collaboration],
  %``Energy flow and rapidity gaps between jets in photoproduction at HERA,''
  Eur.\ Phys.\ J.\  C {\bf 24}, 517 (2002)
  [arXiv:hep-ex/0203011].
  %%CITATION = EPHJA,C24,517;%
  
%\cite{Abachi:1995gz}
\bibitem{tevatron}
  S.~Abachi {\it et al.}  [D0 Collaboration],
  %``Jet Production via Strongly-Interacting Color-Singlet Exchange in
  %$p\bar{p}$ Collisions,''
  Phys.\ Rev.\ Lett.\  {\bf 76}, 734 (1996)
  [arXiv:hep-ex/9509013];
  %%CITATION = PRLTA,76,734;%%
%\cite{Abe:1997ie}
%\bibitem{tevatron2}
  F.~Abe {\it et al.}  [CDF Collaboration],
  %``Dijet production by color-singlet exchange at the Fermilab Tevatron,''
  Phys.\ Rev.\ Lett.\  {\bf 80}, 1156 (1998);
  %%CITATION = PRLTA,80,1156;%%
 %\cite{Abe:1994de}
%\bibitem{Abe:1994de}
  F.~Abe {\it et al.}  [CDF Collaboration],
  %``Observation of rapidity gaps in $\bar{p}p$ collisions at 1.8 TeV,''
  Phys.\ Rev.\ Lett.\  {\bf 74}, 855 (1995).
  %%CITATION = PRLTA,74,855;%%

%\cite{Bjorken:1992er}
\bibitem{Bjorken:1992er}
  J.~D.~Bjorken,
  %``Rapidity gaps and jets as a new physics signature in very high-energy
  %hadron hadron collisions,''
  Phys.\ Rev.\  D {\bf 47}, 101 (1993).
  %%CITATION = PHRVA,D47,101;%%


  
%\cite{Oderda:1998en}
\bibitem{Oderda:1998en}
  G.~Oderda and G.~Sterman,
  %``Energy and color flow in dijet rapidity gaps,''
  Phys.\ Rev.\ Lett.\  {\bf 81}, 3591 (1998)
  [arXiv:hep-ph/9806530].
  %%CITATION = PRLTA,81,3591;%%
  
  %\cite{Berger:2001ns}
\bibitem{Berger:2001ns}
  C.~F.~Berger, T.~Kucs and G.~Sterman,
  %``Energy flow in interjet radiation,''
  Phys.\ Rev.\  D {\bf 65}, 094031 (2002)
  [arXiv:hep-ph/0110004].
  %%CITATION = PHRVA,D65,094031;%%
  
%\cite{Appleby:2003sj}
\bibitem{Appleby:2003sj}
  R.~B.~Appleby and M.~H.~Seymour,
  %``The resummation of inter-jet energy flow for gaps-between-jets  processes
  %at HERA,''
  JHEP {\bf 0309}, 056 (2003)
  [arXiv:hep-ph/0308086].
   
  
 \bibitem{Kyrieleis:2006dt}
   A.~Kyrieleis and M.~H.~Seymour,
  %``The colour evolution of the process q q --> q q g,''
  JHEP {\bf 0601}, 085 (2006)
  [arXiv:hep-ph/0510089].
  %%CITATION = JHEPA,0601,085;%% 

%\cite{Sjodahl:2008fz}
\bibitem{Sjodahl:2008fz}
  M.~Sjodahl,
  %``Color evolution of 2 $\to$ 3 processes,''
  JHEP {\bf 0812}, 083 (2008)
  [arXiv:0807.0555 [hep-ph]].
  %%CITATION = JHEPA,0812,083;%%
 
 %\cite{Sjodahl:2009wx}
\bibitem{Sjodahl:2009wx}
  M.~Sjodahl,
  %``Color structure for soft gluon resummation - a general recipe,''
  arXiv:0906.1121 [hep-ph].
  %%CITATION = ARXIV:0906.1121;%%

   \bibitem{Dasgupta:2001sh}
  M.~Dasgupta and G.~P.~Salam,
  %``Resummation of non-global QCD observables,''
  Phys.\ Lett.\  B {\bf 512}, 323 (2001)
  [arXiv:hep-ph/0104277].
  %%CITATION = PHLTA,B512,323;%%


  \bibitem{Dasgupta:2002bw}
   M.~Dasgupta and G.~P.~Salam,
  %``Accounting for coherence in interjet E(t) flow: A case study,''
  JHEP {\bf 0203}, 017 (2002)
  [arXiv:hep-ph/0203009].
  %%CITATION = JHEPA,0203,017;%%

  
\bibitem{Forshaw:2006fk}  
    J.~R.~Forshaw, A.~Kyrieleis and M.~H.~Seymour,
  %``Super-leading logarithms in non-global observables in QCD,''
  JHEP {\bf 0608}, 059 (2006)
  [arXiv:hep-ph/0604094].
  %%CITATION = JHEPA,0608,059;%%
  %\cite{Forshaw:2008cq}
\bibitem{Forshaw:2008cq}
  J.~R.~Forshaw, A.~Kyrieleis and M.~H.~Seymour,
  %``Super-leading logarithms in non-global observables in QCD: Colour basis
  %independent calculation,''
  JHEP {\bf 0809}, 128 (2008)
  [arXiv:0808.1269 [hep-ph]].
  %%CITATION = JHEPA,0809,128;%%
  %\cite{Keates:2009dn}

  
\bibitem{Keates:2009dn}
  J.~Keates and M.~H.~Seymour,
  %``Super-leading logarithms in non-global observables in QCD: Fixed order
  %calculation,''
  JHEP {\bf 0904}, 040 (2009)
  [arXiv:0902.0477 [hep-ph]].
  %%CITATION = JHEPA,0904,040;%%

  %\cite{Appleby:2002ke}
\bibitem{Appleby:2002ke}
  R.~B.~Appleby and M.~H.~Seymour,
  %``Non-global logarithms in inter-jet energy flow with kt clustering
  %requirement,''
  JHEP {\bf 0212}, 063 (2002)
  [arXiv:hep-ph/0211426].
  %%CITATION = JHEPA,0212,063;%%


%\cite{Forshaw:2009fz}
\bibitem{Forshaw:2009fz}
  J.~Forshaw, J.~Keates and S.~Marzani,
  %``Jet vetoing at the LHC,''
  JHEP {\bf 0907}, 023 (2009)
  [arXiv:0905.1350 [hep-ph]].
  %%CITATION = JHEPA,0907,023;%%
  

  %\cite{Forshaw:2007vb}
\bibitem{Forshaw:2007vb}
  J.~R.~Forshaw and M.~Sjodahl,
  %``Soft gluons in Higgs plus two jet production,''
  JHEP {\bf 0709}, 119 (2007)
  [arXiv:0705.1504 [hep-ph]].
  %%CITATION = JHEPA,0709,119;%%
  
  %\cite{Kulesza:2008jb}
\bibitem{Kulesza:2008jb}
  A.~Kulesza and L.~Motyka,
  %``Threshold resummation for squark-antisquark and gluino-pair production at
  %the LHC,''
  Phys.\ Rev.\ Lett.\  {\bf 102}, 111802 (2009)
  [arXiv:0807.2405 [hep-ph]].
  %%CITATION = PRLTA,102,111802;%%
  
  %\cite{Kulesza:2009kq}
\bibitem{Kulesza:2009kq}
  A.~Kulesza and L.~Motyka,
  %``Soft gluon resummation for the production of gluino-gluino and
  %squark-antisquark pairs at the LHC,''
  arXiv:0905.4749 [hep-ph].
  %%CITATION = ARXIV:0905.4749;%%

  
  %\cite{Frederix:2007gi}
\bibitem{Frederix:2007gi}
  R.~Frederix and F.~Maltoni,
  %``Top pair invariant mass distribution: a window on new physics,''
  JHEP {\bf 0901}, 047 (2009)
  [arXiv:0712.2355 [hep-ph]].
  %%CITATION = JHEPA,0901,047;%%
  
  
  %\cite{Agashe:2006hk}
\bibitem{Agashe:2006hk}
  K.~Agashe, A.~Belyaev, T.~Krupovnickas, G.~Perez and J.~Virzi,
  %``LHC signals from warped extra dimensions,''
  Phys.\ Rev.\  D {\bf 77}, 015003 (2008)
  [arXiv:hep-ph/0612015].
  %\cite{Agashe:2007ki}

\bibitem{Fitzpatrick:2007qr}
  A.~L.~Fitzpatrick, J.~Kaplan, L.~Randall and L.~T.~Wang,
  %``Searching for the Kaluza-Klein Graviton in Bulk RS Models,''
  JHEP {\bf 0709}, 013 (2007)
  [arXiv:hep-ph/0701150].
  %%CITATION = JHEPA,0709,013;%%
  %\cite{Lillie:2007yh}
\bibitem{Lillie:2007yh}
  B.~Lillie, L.~Randall and L.~T.~Wang,
  %``The Bulk RS KK-gluon at the LHC,''
  JHEP {\bf 0709}, 074 (2007)
  [arXiv:hep-ph/0701166].
  %%CITATION = JHEPA,0709,074;%
  
    %\cite{Agashe:2007zd}
\bibitem{Agashe:2007zd}
  K.~Agashe, H.~Davoudiasl, G.~Perez and A.~Soni,
  %``Warped Gravitons at the LHC and Beyond,''
  Phys.\ Rev.\  D {\bf 76}, 036006 (2007)
  [arXiv:hep-ph/0701186].
  %%CITATION = PHRVA,D76,036006;%%
  %\cite{Fitzpatrick:2007qr}
  
  
\bibitem{Agashe:2007ki}
  %%CITATION = PHRVA,D77,015003;%%
  K.~Agashe {\it et al.},
  %``LHC Signals for Warped Electroweak Neutral Gauge Bosons,''
  Phys.\ Rev.\  D {\bf 76}, 115015 (2007)
  [arXiv:0709.0007 [hep-ph]].
  %%CITATION = PHRVA,D76,115015;%%

%\cite{Baur:2008uv}
\bibitem{Baur:2008uv}
  U.~Baur and L.~H.~Orr,
  %``Searching for $t \bar{t}$ Resonances at the Large Hadron Collider,''
  Phys.\ Rev.\  D {\bf 77}, 114001 (2008)
  [arXiv:0803.1160 [hep-ph]].
  %%CITATION = PHRVA,D77,114001;%%

\bibitem{Agashe:2008jb}
  K.~Agashe, S.~Gopalakrishna, T.~Han, G.~Y.~Huang and A.~Soni,
  %``LHC Signals for Warped Electroweak Charged Gauge Bosons,''
  arXiv:0810.1497 [hep-ph].
  %%CITATION = ARXIV:0810.1497;%%

    
    %\cite{Davoudiasl:2009jk}
\bibitem{Davoudiasl:2009jk}
  H.~Davoudiasl, S.~Gopalakrishna and A.~Soni,
  %``Big Signals of Little Randall-Sundrum Models,''
  arXiv:0908.1131 [hep-ph].
  %%CITATION = ARXIV:0908.1131;%%
    
\bibitem{Butterworth:2008iy}
  J.~M.~Butterworth, A.~R.~Davison, M.~Rubin and G.~P.~Salam,
  %``Jet substructure as a new Higgs search channel at the LHC,''
  Phys.\ Rev.\ Lett.\  {\bf 100}, 242001 (2008)
  [arXiv:0802.2470 [hep-ph]].
  %%CITATION = PRLTA,100,242001;%%

  %\cite{Thaler:2008ju}
\bibitem{Thaler:2008ju}
  J.~Thaler and L.~T.~Wang,
  %``Strategies to Identify Boosted Tops,''
  JHEP {\bf 0807}, 092 (2008)
  [arXiv:0806.0023 [hep-ph]].
  %%CITATION = JHEPA,0807,092;%%

%\cite{Kaplan:2008ie}
\bibitem{Kaplan:2008ie}  
  D.~E.~Kaplan, K.~Rehermann, M.~D.~Schwartz and B.~Tweedie,
  %``Top Tagging: A Method for Identifying Boosted Hadronically Decaying Top
  %Quarks,''
  Phys.\ Rev.\ Lett.\  {\bf 101}, 142001 (2008)
  [arXiv:0806.0848 [hep-ph]].
  %%CITATION = PRLTA,101,142001;%%  

%\cite{Almeida:2008yp}
\bibitem{Almeida:2008yp}
  L.~G.~Almeida, S.~J.~Lee, G.~Perez, G.~Sterman, I.~Sung and J.~Virzi,
  %``Substructure of high-p_T Jets at the LHC,''
  Phys.\ Rev.\  D {\bf 79}, 074017 (2009)
  [arXiv:0807.0234 [hep-ph]].
  %%CITATION = PHRVA,D79,074017;%%

%\cite{Almeida:2008tp}
\bibitem{Almeida:2008tp}
  L.~G.~Almeida, S.~J.~Lee, G.~Perez, I.~Sung and J.~Virzi,
  %``Top Jets at the LHC,''
  Phys.\ Rev.\  D {\bf 79}, 074012 (2009)
  [arXiv:0810.0934 [hep-ph]].
  %%CITATION = PHRVA,D79,074012;%%
  
  %\cite{Godfrey:2008vf}
\bibitem{Godfrey:2008vf}
  S.~Godfrey and T.~A.~W.~Martin,
  %``Identification of Extra Neutral Gauge Bosons at the LHC Using b- and
  %t-Quarks,''
  Phys.\ Rev.\ Lett.\  {\bf 101}, 151803 (2008)
  [arXiv:0807.1080 [hep-ph]].
  %%CITATION = PRLTA,101,151803;%%

  %\cite{FileviezPerez:2008ib}
\bibitem{FileviezPerez:2008ib}
  P.~Fileviez Perez, R.~Gavin, T.~McElmurry and F.~Petriello,
  %``Grand Unification and Light Color-Octet Scalars at the LHC,''
  Phys.\ Rev.\  D {\bf 78}, 115017 (2008)
  [arXiv:0809.2106 [hep-ph]].
  %%CITATION = PHRVA,D78,115017;%%
  
  %\cite{Ellis:2009su}
\bibitem{Ellis:2009su}
  S.~D.~Ellis, C.~K.~Vermilion and J.~R.~Walsh,
  %``Techniques for improved heavy particle searches with jet substructure,''
  arXiv:0903.5081 [hep-ph].
  %%CITATION = ARXIV:0903.5081;%%
 
 %\cite{Wang:2008sw}
\bibitem{Wang:2008sw}
  L.~T.~Wang and I.~Yavin,
  %``A Review of Spin Determination at the LHC,''
  Int.\ J.\ Mod.\ Phys.\  A {\bf 23}, 4647 (2008)
  [arXiv:0802.2726 [hep-ph]].
  %%CITATION = IMPAE,A23,4647;%%
  
    %\cite{ArkaniHamed:1998rs}
\bibitem{ArkaniHamed:1998rs}
  N.~Arkani-Hamed, S.~Dimopoulos and G.~R.~Dvali,
  %``The hierarchy problem and new dimensions at a millimeter,''
  Phys.\ Lett.\  B {\bf 429}, 263 (1998)
  [arXiv:hep-ph/9803315].
  %%CITATION = PHLTA,B429,263;%%
  
  %\cite{Randall:1999ee}
\bibitem{Randall:1999ee}
  L.~Randall and R.~Sundrum,
  %``A large mass hierarchy from a small extra dimension,''
  Phys.\ Rev.\ Lett.\  {\bf 83}, 3370 (1999)
  [arXiv:hep-ph/9905221].
  %%CITATION = PRLTA,83,3370;%%
  
  %\cite{Dicus:1994bm}
\bibitem{Dicus:1994bm}
  D.~Dicus, A.~Stange and S.~Willenbrock,
  %``Higgs decay to top quarks at hadron colliders,''
  Phys.\ Lett.\  B {\bf 333}, 126 (1994)
  [arXiv:hep-ph/9404359].
  %%CITATION = PHLTA,B333,126;%%
  
  %\cite{Bernreuther:1997gs}
\bibitem{Bernreuther:1997gs}
  W.~Bernreuther, M.~Flesch and P.~Haberl,
  %``Signatures of Higgs bosons in the top quark decay channel at hadron
  %colliders,''
  Phys.\ Rev.\  D {\bf 58}, 114031 (1998)
  [arXiv:hep-ph/9709284].
  %%CITATION = PHRVA,D58,114031;%%
  
  
  %\cite{Choudhury:2007ux}
\bibitem{Choudhury:2007ux}
  D.~Choudhury, R.~M.~Godbole, R.~K.~Singh and K.~Wagh,
  %``Top production at the Tevatron/LHC and nonstandard, strongly interacting
  %spin one particles,''
  Phys.\ Lett.\  B {\bf 657}, 69 (2007)
  [arXiv:0705.1499 [hep-ph]].
  %%CITATION = PHLTA,B657,69;%%
  

 
 \bibitem{Kidonakis:1996aq}
  N.~Kidonakis and G.~Sterman,
  %``Subleading Logarithms In QCD Hard Scattering,''
  Phys.\ Lett.\  B {\bf 387}, 867 (1996).
  %%CITATION = PHLTA,B387,867;%%
  
  %\cite{Kidonakis:1997gm}
\bibitem{Kidonakis:1997gm}
  N.~Kidonakis and G.~Sterman,
  %``Resummation for QCD hard scattering,''
  Nucl.\ Phys.\  B {\bf 505}, 321 (1997)
  [arXiv:hep-ph/9705234].
  %%CITATION = NUPHA,B505,321;%%
  
  \bibitem{Kidonakis:1998bk}
  N.~Kidonakis, G.~Oderda and G.~Sterman,
  %``Threshold resummation for dijet cross sections,''
  Nucl.\ Phys.\  B {\bf 525}, 299 (1998)
  [arXiv:hep-ph/9801268].
  %%CITATION = NUPHA,B525,299;%%
  
  %\cite{Kidonakis:2009ev}
\bibitem{Kidonakis:2009ev}
  N.~Kidonakis,
  %``Two-loop soft anomalous dimensions and NNLL resummation for heavy quark
  %production,''
  Phys.\ Rev.\ Lett.\  {\bf 102}, 232003 (2009)
  [arXiv:0903.2561 [hep-ph]].
  %%CITATION = PRLTA,102,232003;%%
  
  %\cite{Mitov:2009sv}
\bibitem{Mitov:2009sv}
  A.~Mitov, G.~Sterman and I.~Sung,
  %``The Massive Soft Anomalous Dimension Matrix at Two Loops,''
  Phys.\ Rev.\  D {\bf 79}, 094015 (2009)
  [arXiv:0903.3241 [hep-ph]].
  %%CITATION = PHRVA,D79,094015;%%
  
  %\cite{Becher:2009kw}
\bibitem{Becher:2009kw}
  T.~Becher and M.~Neubert,
  %``Infrared singularities of QCD amplitudes with massive partons,''
  Phys.\ Rev.\  D {\bf 79}, 125004 (2009)
  [arXiv:0904.1021 [hep-ph]].
  %%CITATION = PHRVA,D79,125004;%%
  
  %\cite{Beneke:2009rj}
\bibitem{Beneke:2009rj}
  M.~Beneke, P.~Falgari and C.~Schwinn,
  %``Soft radiation in heavy-particle pair production: all-order colour
  %structure and two-loop anomalous dimension,''
  arXiv:0907.1443 [hep-ph].
  %%CITATION = ARXIV:0907.1443;%%
  
  
   %\cite{Czakon:2009zw}
\bibitem{Czakon:2009zw}
  M.~Czakon, A.~Mitov and G.~Sterman,
  %``Threshold Resummation for Top-Pair Hadroproduction to
  %Next-to-Next-to-Leading Log,''
  arXiv:0907.1790 [hep-ph].
  %%CITATION = ARXIV:0907.1790;%%
  
 %\cite{Ferroglia:2009ep}
\bibitem{Ferroglia:2009ep}
  A.~Ferroglia, M.~Neubert, B.~D.~Pecjak and L.~L.~Yang,
  %``Two-loop divergences of scattering amplitudes with massive partons,''
  arXiv:0907.4791 [hep-ph].
  %%CITATION = ARXIV:0907.4791;%% 

%\cite{Dokshitzer:1988bq}
\bibitem{Dokshitzer:1988bq}
  Y.~L.~Dokshitzer, V.~A.~Khoze and S.~I.~Troian,
  %``COHERENCE AND PHYSICS OF QCD JETS,''
  Adv.\ Ser.\ Direct.\ High Energy Phys.\  {\bf 5}, 241 (1988).
  %%CITATION = 00319,5,241;%%
 
   %\cite{Dokshitzer:1991wu}
\bibitem{Dokshitzer:1991wu}
  Y.~L.~Dokshitzer, V.~A.~Khoze, A.~H.~Mueller and S.~I.~Troian,
  %``Basics Of Perturbative QCD,''
%\href{http://www.slac.stanford.edu/spires/find/hep/www?irn=2556677}{SPIRES entry}
{\it  Gif-sur-Yvette, France: Ed. Frontieres (1991) 274 p. (Basics of)}.

%\cite{Ellis:1991qj}
\bibitem{Ellis:1991qj}
  R.~K.~Ellis, W.~J.~Stirling and B.~R.~Webber,
  %``QCD and collider physics,''
  Camb.\ Monogr.\ Part.\ Phys.\ Nucl.\ Phys.\ Cosmol.\  {\bf 8}, 1 (1996).
  %%CITATION = CMPCE,8,1;%%
 
   \bibitem{Berger:2003iw}  
   C.~F.~Berger, T.~Kucs and G.~Sterman,
  %``Event shape / energy flow correlations,''
  Phys.\ Rev.\  D {\bf 68}, 014012 (2003)
  [arXiv:hep-ph/0303051].
  %%CITATION = PHRVA,D68,014012;%% 
  

 
 %\cite{Banfi:2004nk}
\bibitem{Banfi:2004nk}
  A.~Banfi, G.~P.~Salam and G.~Zanderighi,
  %``Resummed event shapes at hadron - hadron colliders,''
  JHEP {\bf 0408}, 062 (2004)
  [arXiv:hep-ph/0407287].
  %%CITATION = JHEPA,0408,062;%% 
  
  %\cite{Sterman:2005bf}
\bibitem{Sterman:2005bf}
  G.~Sterman,
  %``Energy flow observables,''
  arXiv:hep-ph/0501270.
  %%CITATION = HEP-PH/0501270;%%
  
  %\cite{Berger:2005ct}
\bibitem{Berger:2005ct}
  C.~F.~Berger,
  %``Dijet event shapes as diagnostic tools,''
  Mod.\ Phys.\ Lett.\  A {\bf 20}, 1187 (2005)
  [arXiv:hep-ph/0505037].
  %%CITATION = MPLAE,A20,1187;%%
  
  
    %\cite{Djouadi:2007eg}
\bibitem{Djouadi:2007eg}
  A.~Djouadi, G.~Moreau and R.~K.~Singh,
  %``Kaluza--Klein excitations of gauge bosons at the LHC,''
  Nucl.\ Phys.\  B {\bf 797}, 1 (2008)
  [arXiv:0706.4191 [hep-ph]].
  %%CITATION = NUPHA,B797,1;%%
  

   
    %\cite{Harris:1999ya}
\bibitem{Harris:1999ya}
  R.~M.~Harris, C.~T.~Hill and S.~J.~Parke,
  %``Cross section for topcolor Z'(t) decaying to t anti-t,''
  arXiv:hep-ph/9911288.
  %%CITATION = HEP-PH/9911288;%%
  
  
    %\cite{ArkaniHamed:2002qy}
\bibitem{ArkaniHamed:2002qy}
  N.~Arkani-Hamed, A.~G.~Cohen, E.~Katz and A.~E.~Nelson,
  %``The littlest Higgs,''
  JHEP {\bf 0207}, 034 (2002)
  [arXiv:hep-ph/0206021].
  %%CITATION = JHEPA,0207,034;%%
  
  %\cite{Schmaltz:2005ky}
\bibitem{Schmaltz:2005ky}
  M.~Schmaltz and D.~Tucker-Smith,
  %``Little Higgs Review,''
  Ann.\ Rev.\ Nucl.\ Part.\ Sci.\  {\bf 55}, 229 (2005)
  [arXiv:hep-ph/0502182].
  %%CITATION = ARNUA,55,229;%%
  
  
  
    %\cite{Boersma:2006hw}
\bibitem{Boersma:2006hw}
  J.~Boersma,
  %``Direct search limits on the Littlest Higgs model,''
  Phys.\ Rev.\  D {\bf 74}, 115008 (2006)
  [arXiv:hep-ph/0608239].
  %%CITATION = PHRVA,D74,115008;%%
  
  %\cite{Boersma:2007fd}
\bibitem{Boersma:2007fd}
  J.~Boersma and A.~Whitbeck,
  %``Decays of the Littlest Higgs $Z_{H}$ and the Onset of Strong Dynamics,''
  Phys.\ Rev.\  D {\bf 77}, 055012 (2008)
  [arXiv:0710.4874 [hep-ph]].
  %%CITATION = PHRVA,D77,055012;%%
  
%\cite{Pasechnik:2009bq}
\bibitem{Pasechnik:2009bq}
  R.~S.~Pasechnik, A.~Szczurek and O.~V.~Teryaev,
  %``Elastic double diffractive production of axial-vector \chi_c(1^{++}) mesons
  %and the Landau-Yang theorem,''
  arXiv:0901.4187 [hep-ph].
  %%CITATION = ARXIV:0901.4187;%%
  
%\cite{Hagiwara:2009wt}
\bibitem{Hagiwara:2009wt}
  K.~Hagiwara, Q.~Li and K.~Mawatari,
  %``Jet angular correlation in vector-boson fusion processes at hadron
  %colliders,''
  JHEP {\bf 0907}, 101 (2009)
  [arXiv:0905.4314 [hep-ph]].
  %%CITATION = JHEPA,0907,101;%%
  
  %\cite{Yang:1950rg}
\bibitem{Yang}
  L.~D.~Landau,
  Dokl.\ Akad.\ Nawk., USSR {\bf 60}, 207 (1948);
  C.~N.~Yang,
  %``Selection Rules For The Dematerialization Of A Particle Into Two Photons,''
  Phys.\ Rev.\  {\bf 77}, 242 (1950).
  %%CITATION = PHRVA,77,242;%%
  
  

  %\cite{Perelstein:2009qi}
\bibitem{Perelstein:2009qi}
  M.~Perelstein and A.~Spray,
  %``Tensor Reggeons from Warped Space at the LHC,''
  arXiv:0907.3496 [hep-ph].
  %%CITATION = ARXIV:0907.3496;%%

%\cite{Djouadi:2005gj}
\bibitem{Djouadi:2005gj}
  A.~Djouadi,
  %``The Anatomy of electro-weak symmetry breaking. II. The Higgs bosons in the
  %minimal supersymmetric model,''
  Phys.\ Rept.\  {\bf 459}, 1 (2008)
  [arXiv:hep-ph/0503173].
  %%CITATION = PRPLC,459,1;%%  
  
    
  
  %\cite{Cheung:2004ad}
\bibitem{Cheung:2004ad}
  K.~Cheung and W.~Y.~Keung,
  %``Split supersymmetry, stable gluino, and gluinonium,''
  Phys.\ Rev.\  D {\bf 71}, 015015 (2005)
  [arXiv:hep-ph/0408335].
  %%CITATION = PHRVA,D71,015015;%%
  
  
  
  \end{thebibliography}
  \end{document}